%% file: main.tex
\documentclass[12pt]{article}
\newcommand{\sym}[1]{\ifmmode^{#1}\else\(^{#1}\)\fi}
\usepackage{etoolbox}
\usepackage[utf8]{inputenc}
\usepackage{geometry}
\usepackage{setspace}
\usepackage{amsmath}
\usepackage{amsthm}
\usepackage{amssymb}
\usepackage{amsfonts}
\usepackage{listings}
\usepackage{tcolorbox} 
\usepackage{soul} 
\usepackage{placeins} 
\usepackage{booktabs}
\usepackage[bf]{caption}
\usepackage{subcaption}
\usepackage{multirow} 
\usepackage{makecell}
\usepackage{datetime}
\usepackage{rotating}   % -> \begin{sidewaysfigure}
\usepackage{siunitx}    % -> horizontal alignment at dots in figures. 
\usepackage{caption}    % -> \captionsetup: adds figure caption options
\usepackage{graphicx}   %-> crop figures using \includegraphics[trim=left bottom right top, clip]
\usepackage{float}
\usepackage{adjustbox}
\usepackage{ragged2e}
\usepackage{tabularx}
\usepackage{makecell}
\usepackage{booktabs} % For better-looking tables
\usepackage{siunitx}
\input{export_tsd/macros}

\usepackage{graphicx}
\usepackage{epstopdf}
\usepackage{svg}
\usepackage{xfp}      % floating point arithmetic

\usepackage{xparse}    % For defining new commands with arguments
\usepackage{array}     % For table formatting
\usepackage{xstring}     % For advanced string manipulation
\usepackage{etoolbox}  % For defining column type with breakable lines

\usepackage{filecontents,catchfile}
\usepackage{dsfont}

\usepackage[round,authoryear,comma]{natbib}    % Added 'comma' for \citep citations
\usepackage{xcolor}  % required for color support of links (if not already loaded)
\usepackage[backref=page]{hyperref}

\renewcommand*{\backref}[1]{}
\renewcommand*{\backrefalt}[4]{%
  \ifcase #1 %
    (Not cited.)%
  \else
    (Cited on p.~#2)%
  \fi}
\setcitestyle{authoryear,open={(},close={)},aysep={}}

\usepackage{stackengine,scalerel} %-> Used to create tripple daggers

\newdateformat{monthyeardate}{\monthname[\THEMONTH] \THEYEAR}
\hypersetup{colorlinks=true,allcolors=black}
\renewcommand{\arraystretch}{1.2}

\usepackage{stackengine,scalerel} %-> Used to create tripple daggers
\hypersetup{
    colorlinks=true,
    linkcolor={red},    % The color of internal links (sections, footnotes, etc.)
    citecolor={blue},   % The color of citations
    urlcolor={blue}     % The color of external URLs
}
\renewcommand{\arraystretch}{1.2}

\makeatletter
\patchcmd{\@maketitle}{\LARGE}{\Large}{}{}
\makeatother

\title{
    \textbf{Pay Beliefs and the Amenity-Pay Tradeoff}\thanks{
        \textit{Acknowledgments}:
        This project has received funding from the European Union’s Horizon Europe Research and Innovation Programme under Grant Agreement No. 101043127. We thank seminar participants at University College London, UCLA, UC Santa Barbara, UC Irvine, UC San Diego, HEC Lausanne, ETH Zurich, Rockwool Foundation in Copenhagen, the Nordic Public Policy Symposium (NPPS) in Copenhagen, the 3rd Annual Oslo Labor Workshop, and the 40th EEA Annual Congress for helpful comments. 
    }
}

\author{
    Martin E. Andresen\footnote{
        University of Oslo and IZA.
        E-mail: \href{mailto:m.e.andresen@econ.uio.no}{m.e.andresen@econ.uio.no}
    } 
    \and
    Manudeep Bhuller\footnote{
        University of Oslo, Statistics Norway, CEPR, IZA, CESifo. 
        E-mail: \href{mailto:manudeep.bhuller@econ.uio.no}{manudeep.bhuller@econ.uio.no}
    } 
    \and
    Alfred L{\o}vgren\footnote{
        University of Oslo. 
        E-mail: \href{mailto:alfred.lovgren@econ.uio.no}{alfred.lovgren@econ.uio.no}
    }
}
\date{}

\begin{document}

\maketitle

\begin{center}
%This Version: \date{\monthyeardate\today} \\
This Version: June 2026 \\
\vspace{-1cm}
First Version: May 2026
\vspace{0.5cm}
\end{center}

\begin{abstract}
This paper studies how workers' beliefs about pay shape the tradeoffs between pay and workplace amenities. We design a multi-stage incentivized survey experiment that combines hypothetical choice experiments with elicited beliefs about starting salaries in real jobs and randomly varies the provision of explicit pay information. Although stated preferences imply sizable willingness to pay for workplace amenities, baseline beliefs about salaries in real jobs are systematically biased along two margins: respondents under-predict starting salaries by $18$\% and expect higher-amenity jobs to pay \emph{more}, substantially over-predicting the amenity-pay gradient. A short-term pay information intervention raises mean beliefs about pay in similar jobs by 4\% and reduces belief dispersion by 15\%, but does not alter the perceived amenity-pay slope or the implied tradeoffs in stated choices. Meanwhile, full disclosure of pay for jobs under consideration raises the pay of chosen jobs by about 4\% and recovers willingness-to-pay estimates closely aligned with full-information hypothetical-choice benchmarks. Short-term disclosure thus moves beliefs but not perceived tradeoffs, while persistent disclosure erases biases in pay beliefs and nearly restores the full-information tradeoffs.

\end{abstract}

\vfill
\vspace{-0.1cm}


\thispagestyle{empty}
\clearpage

\onehalfspacing

\setcounter{page}{1}
\section{Introduction}

Pay and non-pecuniary job attributes---or amenities---jointly determine the value of a job. The classical theory of compensating differentials \citep{rosen1986theory, brown1980equalizing} predicts that workers must be compensated with higher pay to accept jobs with worse amenities, so that, in equilibrium, better-amenity jobs should pay less. A growing literature based on hypothetical choice experiments confirms that workers place substantial monetary value on amenities such as workplace flexibility, the option to work from home, and job security \citep{mas2017valuing, wiswall2018preference, maestas2023value}. Yet decades of empirical work using observational data has found limited support for compensating differentials \citep{bonhomme2009}. The gap between sizable stated preferences and the elusive compensating differential found in observational data remains an important puzzle in labor economics.

This paper proposes that an important part of the puzzle resides in workers' \emph{beliefs}. In many labor markets, pay is rarely disclosed up front in vacancy listings (see, e.g., US evidence in \cite{batra2023online}, \cite{audoly24} for Norway, and \cite{caldwell2025bargaining} for Germany), and job seekers must thus form expectations about the salary they will receive in jobs they are considering. If workers systematically misperceive what jobs pay, and if those misperceptions are correlated with the amenities offered in job ads, then their actual choices may not reflect realized compensating differentials, even when their underlying preferences are exactly what hypothetical experiments suggest. The amenity-pay tradeoffs workers \emph{think} they face can differ sharply from those they would face under full information. 
For instance, if workers expect higher-amenity jobs to pay substantially more than they actually do, then even workers with strong preferences for amenities may appear, in observational data, to undervalue amenities relative to pay. In other words, if better amenities and higher pay are already bundled in the menu of jobs perceived by workers, then choosing a higher-amenity job appears to deliver higher pay rather than to require the worker to forgo pay.

We provide direct evidence on this hypothesis by designing and implementing a multi-stage incentivized survey experiment that elicits both stated preferences for workplace amenities and respondent-level beliefs about pay in real job ads, and randomizes the provision of explicit pay information at the job-ad level. The survey was conducted in person with around $1{,}000$ students recruited from the University of Oslo, a major public university in Norway. The survey provides experimental data on $38{,}920$ choice alternatives featuring hypothetical jobs, $19{,}460$ choice alternatives featuring real job ads coupled with elicited pay beliefs and $19{,}460$ choice alternatives featuring real job ads with experimentally manipulated pay information, besides a rich set of observable respondent and job characteristics.

Our survey has four stages. In the first stage of our survey, we collected informed consent and respondent background characteristics. In the next stage, each respondent made choices over $20$ pairs of \emph{hypothetical} jobs in which we experimentally varied pay together with five non-pecuniary job attributes (permanent contract, work from home, shift work, flexible hours, and travel time). To extract as much information as possible from each respondent, we sampled job attributes using the Bayesian Adaptive Choice Experiment (BACE) methodology of \citet{drake22}, which dynamically tailors each new scenario to maximize the expected information gain about the respondent's preferences given their previous answers. We complement these incentivized revealed choices with self-reported subjective willingness to pay for each attribute. The next two stages of the survey are crucial for our analysis. In these stages, respondents made choices between \emph{real} job ads---drawn from a large curated sample of ads recently posted by Norwegian employers in their local labor market---where we randomly varied whether the ads displayed explicit pay information. These features of our survey allow us to (i) elicit respondents' baseline pay beliefs for jobs relevant to their fields of study, and (ii) provide credible information on starting salaries in those jobs, derived from population-level matched employer-employee data on recently hired workers in the corresponding positions. The order of the pay information and no pay information modules in our survey was randomized across respondents, so that comparing pay beliefs and choices in the no pay information modules across treated and control groups identifies the causal effects of having previously been exposed to explicit pay information for similar jobs.

Our first set of findings concerns respondents' \textit{preferences} for workplace amenities. Estimating mixed logit choice models in the spirit of \citet{revelttrain1998, ReveltTrain2000} on data from the hypothetical choice experiments involving hypothetical jobs, we find that respondents are willing to give up $16.1$\% of their pay for a permanent contract, $10.5$\% for the option to work from home, and $6.4$\% for flexible hours, and require compensation of $7.7$\% for shift work and $16.9$\% for an additional hour of daily commuting. These willingness-to-pay (WTP) estimates are quantitatively close to those in a literature targeted at broader populations \citep{ maestas2023value, wiswall2018preference,mas2017valuing}, and to respondents' own self-reported subjective WTPs, suggesting that students at our target institution exhibit standard preferences over the workplace amenities of interest. We also find remarkably similar WTPs from scenarios where respondents faced hypothetical choices between real job ads that disclosed information about starting salaries, providing a pay disclosure benchmark.

Besides respondents' preferences for workplace amenities, we analyzed their \emph{beliefs} about starting salaries. While individuals' beliefs shape their choices, eliciting consistent measures of beliefs is notoriously difficult and raises several methodological challenges \citep{manski2004measuring}. We propose a tractable framework that allows us to flexibly decompose both the mean and the variance of pay beliefs into individual and job components using data on elicited pay beliefs. Our decomposition is structurally similar to the canonical AKM wage decomposition \citep{abowd1999high}, with the important distinction that we apply it to \emph{beliefs} rather than realized wages. Our framework further allows us to quantify the sources of dispersion in pay beliefs, to study the correlates of the pay belief levels and the precision of these beliefs, and apply standard bias-corrections in the estimation of variance-covariance components.

Our second and central set of findings concerns respondents' beliefs about starting salaries and how those beliefs relate to advertised amenities. Three patterns stand out. First, respondents over-attribute pay to advertised amenities: the slope of their beliefs on the composite amenity value of a job is substantially steeper than the analogous slope of actual pay on the same amenities. The wedge is strongest for permanent contracts: respondents expect a permanent job to pay roughly 8.5\% more than an otherwise identical temporary one, while actual starting salaries for permanent jobs in our linked administrative data are only about 2.2\% higher. Analogous wedges are present for shift work and flexible hours. The pattern is robust to occupation-by-sector and respondent fixed effects. Notably, the misperception here is one of \emph{magnitude}, not sign: even where amenities and pay are (weakly) positively correlated in the data, the perceived gradient exceeds the observational one. Under rational Bayesian inference with correctly-specified priors, the two gradients should coincide; the gap we estimate is therefore diagnostic of either an upward-biased prior over the pay-amenity gradient or systematic over-extrapolation from amenity signals \citep{bordalo2016stereotypes, bordalo2018diagnostic}. Second, on average, respondents \emph{underestimate} actual starting salaries---constructed from administrative data on recently hired workers in the corresponding positions---by approximately 18\%, despite being currently or recently employed and approaching graduation and labor market entry themselves. Third, baseline pay beliefs are heavily dispersed even for identical jobs: $\num[round-mode=places,round-precision=0]{\fpeval{100*(\varpersonzero/\vartotzero)}} \%$ of the variance in beliefs is attributable to persistent person components and $\num[round-mode=places,round-precision=0]{\fpeval{100*(\varjobzero/\vartotzero)}} \%$ to job components, with no evidence of sorting between the two---consistent with the random assignment of job ads to respondents. Beliefs also vary systematically with respondent characteristics: male respondents expect starting salaries about 6\% higher than female respondents \citep{Roussille2024,cortes2023}, and respondents anchor their beliefs in their own current salary with a pass-through of approximately 0.12 \citep{jager2022worker}.

These findings on beliefs speak directly to the elusive patterns of compensating wage differentials often found in observational data. Even before workers act on their preferences, their beliefs attach \emph{higher} expected pay to better amenities. In the spirit of \citet{hwang1998hedonic}, this is consistent with augmenting---rather than compensating---differentials when search frictions are present, since amenities can serve as a signal of unobserved firm productivity and hence pay. It also helps rationalize why hypothetical choice experiments, which strip away these informational signals by construction, recover sizable compensating differential preferences, while observational hedonic regressions of pay on amenities do not. While full-information models of rent-sharing in monopsonistic labor markets \citep{card2018firms} can generate a positive association between pay and advertised amenities, the systematic \emph{wedge} between perceived and observational amenity-pay gradients is instead consistent with the presence of biased beliefs. Our evidence on the slope wedge thus isolates a specific information channel behind the lack of compensating differentials in observational data.

Our third set of findings concerns the causal effects of explicit pay information provision. Our experimental design allows us to uncover the impacts of two distinct information treatments: a short-term pay information intervention (``learning treatment'') and a persistent pay transparency intervention (``full information treatment''). The learning treatment seeks to capture the effects of providing information on starting salaries for some jobs on future pay beliefs and preferred choices among similar jobs. The full information treatment instead seeks to capture the effects of having information on posted starting salaries on all choice alternatives that are under consideration by the agent in an ongoing search process. 

Crucially, our learning treatment separates the ads for which pay is disclosed from the ads for which beliefs are subsequently elicited: treated respondents first see explicit pay information on one set of ten job ad pairs, and then state their pay beliefs over a \emph{separate} set of ten ad pairs in the subsequent stage, with both sets drawn from the same underlying pool of high-skilled service-sector vacancies. We thus identify how prior exposure to pay information on related---but not identical---ads spills over to beliefs about new ads. We find that this pay information exposure raises respondents' mean pay beliefs by about 4\%---closing roughly one-fifth of the average 18\% baseline bias relative to actual starting salaries---reduces the variance of pay beliefs by about 15\%, and shrinks the probability mass at both ends of the belief distribution, lowering the prevalence of both negative and positive biases.

Our evidence shows that while the learning treatment neither affects future job choices nor alters the valuations of workplace amenities implied by these choices, the full information treatment increases selectivity towards high-paying jobs with meaningful impacts on the implied willingness to pay for workplace amenities. A short-term information intervention can thus shift the level and dispersion of beliefs but is unable to break the link by which respondents infer that better-amenity jobs are also better-paid in our experimental setting. Meanwhile, the full information treatment represents a more sustained or salience-enhancing pay-transparency intervention that yields shifts in stated job choices, with implied WTPs that are closely aligned with the WTPs derived from stylized full-information hypothetical choice experiments. 
Our evidence thus contrasts what a full-scale pay-transparency policy could achieve with what short-term information interventions can be expected to deliver.

Our paper contributes to several literatures. First, we contribute to a literature that uses hypothetical choice experiments to recover preferences over amenities and pay \citep{mas2017valuing, maestas2023value, wiswall2018preference, drakeetal2023, lewandowski2022, lewandowski2023, datta2019,nagleretal2022, caldwell2026firm}, complemented by contemporaneous field-experimental work that recovers revealed preference willingness-to-pay from job seekers' clicks and applications on real online job boards \citep{beerli2026}. Closest in spirit to our methodological approach, \citet{caldwell2026firm} embed discrete choice experiments in a large-scale survey of German workers, use randomized wage variation across hypothetical offers from real firms to recover money-metric \emph{firm-level} amenity valuations, and link these valuations to administrative records and firm-level employer-review data. We extend this literature in two directions: by combining hypothetical choices with elicited beliefs over real job ads linked to administrative starting-salary data, and by embedding a randomized pay-information intervention within both a hypothetical-choice and a real-ads choice design. To our knowledge, this is the first survey experiment to \emph{jointly} elicit pay beliefs and incentivized hypothetical choices over the same set of real job ads, allowing us to study how information about pay reshapes both beliefs and the amenity-pay tradeoff in stated choices. Following \cite{drake22}, we also use recent advances in Bayesian adaptive sampling methods to dynamically tailor choice scenarios to our survey respondents.

Second, we contribute to a rapidly growing literature on workers' beliefs and expectations \citep{mueller2022expectations,manski2004measuring, Zafar2011, Zafar2013, WiswallZafar2015,conlon2018labor, conlon2021major, jager2022worker, caldwell2025search,caplin2025subjective, hvidberg2023social, balleer2026biased,blesch2023biased}. More broadly, \cite{haaland_jel_2023} and \cite{stantcheva_2024} discuss the importance of measuring beliefs and review recent work in economics exploiting information provision experiments to study belief updating, including the use of open-ended surveys \citep{haaland_jel_2025}. Closest in spirit to our evidence on systematic underestimation of starting salaries are \citet{conlon2021major}, finding that first-year university students underestimate average salaries across major fields, and \citet{christensen2026relative}, showing that recent Danish university graduates underestimate the pay gaps between jobs that are relevant for their education. While much of this literature focuses on beliefs about own future wages, unemployment risk, or wage offers in search, we provide new evidence on the \emph{joint} beliefs about pay and amenities, exploiting the fact that we observe many respondents' beliefs about the same actual jobs. Methodologically, our two-way fixed-effects model of beliefs provides a tractable framework for decomposing the systematic and idiosyncratic components of beliefs data, complementing the extensive literature in labor economics that decomposes sources of dispersion in wages \citep{abowd1999high}.

Third, our findings speak directly to the long-standing literature on compensating differentials \citep{brown1980equalizing, hwang1998hedonic, bonhomme2009, sorkin2018ranking, sockin2022, audoly24, beerli2026, caldwell2026firm}. Recent empirical evidence on the relationship between firm-level amenities and pay in actual labor markets is mixed: \citet{audoly24} and \citet{beerli2026} document positive correlations between advertised amenities and pay in Norwegian and Swiss vacancy data, respectively, while \citet{caldwell2026firm} estimate firm-specific amenity valuations using discrete choice experiments over real firms in Germany and find that these valuations are approximately orthogonal to firm wage premia. We adopt the text-analysis approach of \citet{audoly24} to detect amenities in vacancy texts and link our job ads to revealed-preference measures of employer quality \citep{abowd1999high, sorkin2018ranking, bagger2019empirical}. We show that in the \emph{beliefs} of our respondents, the association between perceived pay and advertised amenities is robustly positive and quantitatively strong. Our evidence thus provides support for a belief-based mechanism behind augmenting differentials \citep{hwang1998hedonic} and helps rationalize why hedonic regressions of pay on amenities so often produce wrong-signed or null estimates.

Fourth, we contribute to the literature on pay transparency and information frictions in labor markets
\citep{cullen2024pay,cullenecma2024,caldwell2025search,arnold2024, Roussille2024,jager2022worker,Balgova2025,beerli2026,jalal2026}. Much of this literature studies the effects of pay disclosure on outcomes such as wage compression, gender gaps, or vacancy filling. Recent field experiments provide evidence complementary to ours. \citet{Balgova2025} randomize the inclusion of pay information in real vacancy postings on a job board in Ethiopia and find that exposure substantially reduces the absolute error in applicants' pay beliefs. \citet{beerli2026} randomize information on wages and benefits in Swiss online job boards and trace the effects on click and application behavior. \citet{jalal2026} mandates pay disclosure in job ads in a field experiment with Pakistan's largest job-search platform, and shows that revealing pay nearly doubles women's applications to large firms and reverses the gender gap in directed search. Our study complements these experiments by allowing us to (i) jointly elicit pay beliefs and hypothetical choices over a curated sample of real ads linked to administrative starting salaries, (ii) decompose the structure of pay beliefs along individual and job dimensions, and (iii) assess how a pay-information intervention reshapes not only beliefs but also the amenity-pay tradeoff in stated choices.\footnote{An extensive non-experimental literature studies how information on posted wages affects recruitment outcomes \citep[e.g.,][]{faberman2018}, including recent studies examining the wage elasticity of vacancy duration \citep{mueller2024}, applications \citep{azar2022,banfi2019highwage}, and hires \citep{bassier2022, hirsch2022}, with \citet{datta2024monopsony} exploiting variation in whether vacancy postings make the wage top-up explicit to estimate recruitment-wage elasticities of around 6, roughly twice the separation-wage elasticity. We add to this literature by providing experimental evidence on how pay information provision affects beliefs about the amenity-pay gradients, even prior to actual job choices.} The persistence of positive amenity-pay belief associations under pay disclosure suggests that short-term information interventions on a subset of ads may not be sufficient to fully correct workers' systematic misperceptions of the amenity-pay tradeoff.

The remainder of the paper is organized as follows. Section~\ref{sec:design_data} describes the survey design, the construction of hypothetical and real choice sets, and the recruitment of respondents. Section~\ref{sec:wtp} presents our choice model for estimating preferences for workplace amenities and presents evidence on these. Section~\ref{sec:beliefs} outlines how we estimate mean and precision of pay beliefs and presents evidence on these beliefs, how they decompose into person- and job components and how they relate to job characteristics. Section~\ref{subsec:jobchoices} reports the impact of the pay information intervention on beliefs and on the amenity-pay tradeoff. Section~\ref{sec:conclusion} concludes.

\section{Survey Design}\label{sec:design_data}
This section describes the survey experiment we designed and details on its implementation. We provide an overview of the different stages of the survey, describe how we designed the hypothetical choice sets, and how we recruited respondents and collected their survey responses. We also document key characteristics of respondents in our data collection.

\subsection{Overview of the Survey}\label{subsec:design}

Our survey experiment has four main objectives. First, we want to quantify the tradeoffs between pay and workplace amenities by estimating respondents' willingness to pay for specific amenities. Second, we want to measure respondents' beliefs about pay in real jobs, and, by relating these beliefs to revealed-preference measures of job quality, actual pay and advertised amenities provide evidence on the correlates of beliefs. Third, we aim to study how exposure to explicit pay information in job ads affects respondents' beliefs about pay in other similar jobs posted in the labor market around the same time. Fourth, we want to assess how exposure to pay information in job ads affects respondents' stated choices and how they trade off pay against amenities. Our survey experiment was pre-registered in the AEA RCT registry with trial number \textit{AEARCTR-0013426} \citep{BhullerAEA2024}.

\begin{figure}[t!]
    \begin{center}
    \caption{The Survey Design.}
    \includegraphics[width=\linewidth]{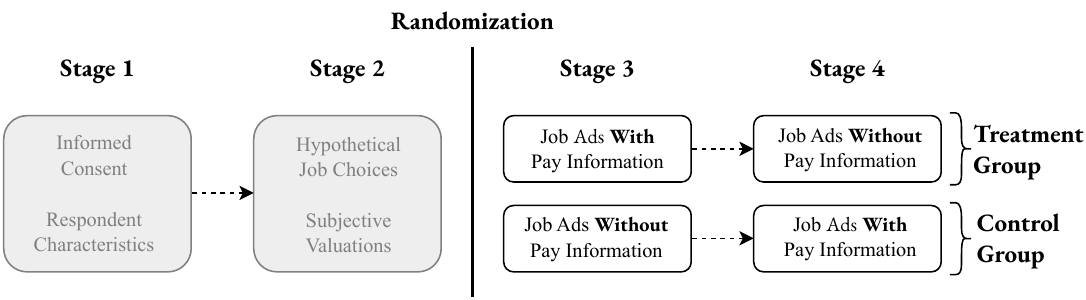}
    \label{fig:surveyDesign}
    \vspace{-1cm}
    \end{center}
    %\par \scriptsize{\emph{Notes:} The figure provides a schematic overview of the four stages in our survey experiment.}
\end{figure}

Figure \ref{fig:surveyDesign} provides a schematic overview of our survey in four stages. In Stage 1, we gather background information about respondents and their informed consent for participating. After completing this, respondents are directed to Stage 2, where we present each respondent with a sequence of 20 hypothetical job choice scenarios. Each scenario has two job alternatives and the respondent is asked to state their preferred choice among these two alternatives, as illustrated in Appendix Figure \ref{fig:appendix_hypothetical_jobs}. Each alternative provides explicit information on monthly pay and five non-pecuniary job attributes, namely whether the job is permanent or temporary, whether the job provides the possibility to work flexible hours, whether the job involves shift work, whether the job has the option to work from home, and finally, the travel time to work. Using information on respondents' preferred choices and the choice sets they faced in each of the 20 hypothetical job choice scenarios, we can infer their willingness to pay for (or accept) each of the five non-pecuniary job attributes.
 
Notably, the elicitation of preferences for workplace amenities in Stage 2 is done based on hypothetical job choices, where respondents have full information about pay and non-pecuniary attributes and these attributes are randomly manipulated by design. Additionally, the survey clearly stated that all jobs are full time jobs, and we informed the respondents that other than the stated differences, the jobs were identical in all other respects (see Appendix Figure \ref{fig:information_sheet}). These aspects limit concerns related to omitted variables bias, unobserved choice sets and information frictions, which are common identification challenges in studies that take a revealed-preference approach using observational data. Following the hypothetical choice scenarios, we also ask respondents to state their own, i.e., subjective, willingness to pay for (or accept) workplace (dis)amenities, by asking them how much of their wage they would be willing to give up (must be compensated) to get a job with a particular amenity (to accept a job with a particular disamenity), all else the same (see Appendix Figure \ref{fig:information_sheet_subj}). Based on the information collected in Stage 2, we can thus compare respondents' subjective valuations to the willingness to pay estimates derived from their hypothetical choices.

The next phase is a crucial part of our survey design. In each of the next stages, every respondent faces 10 job choice scenarios, where, as earlier, each scenario has two job alternatives and the respondent is asked to state their preferred choice among these alternatives. Importantly, however, in Stages 3-4, respondents no longer face purely hypothetical jobs, but make hypothetical choices between real job ads. These ads were recently posted on a major Norwegian job board in the local labor market, as shown in Appendix Figure \ref{fig:appendix_pay_inf_noinf}.

As respondents transition from Stage 2 to Stage 3 in our survey, they are randomized into different modules. As shown in Figure \ref{fig:surveyDesign}, while each respondent must complete both survey modules, the order in which they complete these modules is randomized. The treated respondents (upper block) first complete job ads \textbf{with} explicit information about pay in Stage 3 (Appendix Figure \ref{fig:appendix_pay_inf_noinf}, Panel (a)), before moving on to job ads \textbf{without} such pay information in Stage 4 (Appendix Figure \ref{fig:appendix_pay_inf_noinf}, Panel (b)). The control respondents (lower block) complete the same two modules, but in the opposite order. Notably, the job ads that are shown in both modules are sampled from the same underlying sample of job ads, where we experimentally vary the provision of explicit pay information. This feature ensures that whether or not respondents view explicit pay information is randomly assigned, for otherwise similar job postings. In both modules, respondents are asked which job they prefer. Additionally, in the module that does not contain explicit pay information, respondents are asked to provide their belief (i.e., best guess) about pay in each job alternative. The respondents were required to complete these fields before proceeding.\footnote{Prior to randomizing respondents to the different sequences of job ads modules--jobs ads with and without pay information--we also randomized the sequence of the hypothetical job choice module and the job ads modules, such that around one-half of the respondents completed the job ads modules before the hypothetical job choice module. As we do not exploit this randomization, we omit this feature from Figure \ref{fig:surveyDesign} for brevity. Notably, the two job ads modules were ``tagged together'' in a connected block, so once a respondent had started on their first job ads module, the next module they faced was always the other job ads module.}

We use the experimental design in Stages 3-4 to shed light on several aspects related to pay beliefs and the role of pay information in job ads. First, using the control respondents that were randomly assigned to job ads without pay information in Stage 3, we can document the distribution of baseline pay beliefs  (e.g., the systematic dispersion in pay beliefs across respondents facing similar jobs), and, by relating these beliefs to revealed-preference measures of job quality, actual pay and advertised amenities, provide evidence on the correlates of pay beliefs. Next, by comparing the respondents' stated pay beliefs in modules without explicit pay information, across the treated group (in Stage 4) and the control group (in Stage 3), we can study how repeated exposure to pay information affects future beliefs about pay in other similar jobs (i.e., learning effects on beliefs). Notably, in this comparison, we exploit the fact that the treated group has already (in Stage 3) been exposed to job ads with explicit pay information. Finally, we can evaluate the impact of pay information on respondents' stated choices and trade-offs between pay and amenities. Specifically, by comparing the respondents' choices in modules without explicit pay information, across the treated group (in Stage 4) and the control group (in Stage 3), we can study how repeated exposure to pay information affects choices (i.e., learning effects on job selection).

\subsection{Hypothetical Choice Sets}\label{subsec:sampling}

We now describe how we sampled job attributes and real job ads in order to design the hypothetical choice sets that respondents faced in our survey experiment.

\paragraph*{Sampling of Job Attributes}

As shown in Figure \ref{fig:surveyDesign}, our survey comprises a hypothetical choice experiment where respondents choose between hypothetical jobs in Stage 2. A common approach to design choice sets in such experiments is to randomly sample jobs from a pre-specified population of jobs characterized by a given joint distribution of pay and non-pecuniary job attributes. However, as respondents in our survey experiment face multiple scenarios of hypothetical job choices, a random sampling procedure would be inefficient as it ignores what the researcher may have learned about a respondent's preferences based on their choices in an earlier scenario when sampling the next set of job alternatives. 

We follow the methodology developed by \citet{drake22} for the sampling of job attributes. Specifically, we perform dynamic efficient sampling of job attributes using Bayesian adaptive choice experiments (BACE) methodology. Rather than pre-specifying a fixed set of job attributes, the BACE algorithm draws the values of pay and non-pecuniary characteristics for each choice alternative across the 20 scenarios that respondents face in Stage 2 in a dynamic manner to select the choice scenario that is expected to be the most informative about the respondent's preferences, given their previous answers. The hypothetical jobs are thus selected by the BACE algorithm, with the goal of maximizing the expected information gain about individual preferences. Essentially, BACE uses Bayesian updating to infer from the respondent's past choices the set of job attributes that are expected to maximize information gain about the respondent's preferences. Especially in settings where respondents can be expected to credibly respond to only a limited number of choice alternatives (as in our survey experiment), the BACE algorithm is shown to outperform pure random sampling of alternatives and achieve large efficiency gains already with a relatively small number of choice scenarios. We provide further details on the BACE methodology in Appendix \ref{sec:appendix_wtp}. 

A general worry in any hypothetical choice experiment is related to respondent inattention. Some respondents may simply browse through the survey, answering more or less in a random fashion, in order to finish the survey and receive the gift card. The BACE methodology allows us to investigate this, as it dynamically produces scenarios embedded with attention checks in the hypothetical choice experiment. In these scenarios, one of the jobs dominates the other on at least one pay or amenity attribute, without being worse on all others, so that attentive respondents should never select the dominated job. We exploit responses to these attention checks for robustness in our empirical analyses by restricting to attentive respondents or weighing down responses consistent with inattentive behavior.

\paragraph*{Sampling of Job Ads}

In Stages 3-4, respondents are faced with choices between real job ads. For this part of our survey, we collected job ads from a leading Norwegian online job board. We focused on a subset of job ads that were posted by employers based in Oslo or the surrounding labor market regions over a six-month period from June to December 2023. We have access to the full text of each of the posted ads that we could utilize as part of our survey experiment. Notably, the majority of job ads posted in Norway do not contain explicit pay information, and there is no legal requirement to disclose such information in job ads.\footnote{\cite{audoly24} find that less than 10\% of ads posted between 2021 and 2024 mention a salary number or bracket, while around 25\%  of ads mention a collective bargaining agreement (CBA). Taken together, around 30\%  of job ads feature some information about the actual pay level or CBA-related information.} This feature aids our design as we aim to collect respondents' beliefs about pay in different jobs, and experimentally manipulate the provision of information on expected pay to evaluate respondents' hypothetical choices under alternative informational settings.

Our initial sample consisted of $11,393$ job ads for full-time positions. We limit our analysis to jobs ads with a valid establishment identifier, where we could credibly assign a 4-digit occupational code based on the textual information on job title, role or position stated in the ad, and where we could link information on recently hired workers in respective positions. Using salary information for recently hired workers, we constructed a measure of the expected monthly starting salary for each job posting. Imposing these restrictions, we retain $5,017$ job ads, which is about 44\% of the initial sample. 

Next, we filtered the remaining sample of job ads further, with the aim of selecting job ads that are most relevant for our target population (see Section \ref{subsec:recruitment}). Specifically, we retained a subset of high-skilled service sector job ads, such as advisors and professionals in consulting, sales, teaching, finance, public administration, logistics, and legal occupations. We further retained job ads with an expected monthly starting salary between \$2,000 and \$8,000, to avoid jobs that are outliers in the pay distribution.\footnote{According to \cite{Statistics_Norway_2023}, the median pre-tax monthly pay in Norway was \$5,060 in 2023.} Moreover, we dropped job ads posted by staffing agencies. As indicated above, we also dropped a small share of ads that explicitly provided information on compensation (e.g., salary level) as part of the publicly posted information. After these restrictions, the relevant sample retains $1,234$ job ads posted by around 500 unique employers, which we use in Stages 3-4 of our survey experiment. 

A novel feature of our job ads data is that we have establishment identifiers for the posting employer, and can thus relate salary information on recently hired workers in these establishments to respondents' pay beliefs about starting salaries in respective positions. Besides salary information, we are also able to link job ads to alternative revealed-preference measures of employer quality. Following \cite{audoly24}, we use Norwegian population level employer-employee data between 2021 and 2024 to estimate alternative revealed-preference measures of employer quality for the set of strongly connected employers.\footnote{Within a strongly connected set of employers, each employer has at least one worker moving in from another employer in this set and one worker moving to another employer in this set during the years 2021-2024. To precisely estimate these employer-level parameters, we use a \textit{k}-means clustering algorithm \citep{bonhomme2019distributional} as in \cite{audoly24}. The algorithm partitions employers with similar observable characteristics in clusters, increasing the number of movers per group and improving estimation precision. The clustering and employer quality estimation is performed for the strongly connected set of establishments, using Norwegian population-level employer-employee data between 2021 and 2024. On average, each cluster of employers consists of 50 unique establishments. See \cite{audoly24} for further estimation details.} Using this methodology, we estimate alternative employer quality measures, such as the \textit{overall employer value} of \cite{sorkin2018ranking} and the \textit{poaching index} of \cite{bagger2019empirical}, as well as \textit{employer pay premiums} from AKM-style wage equations \citep{abowd1999high}. We are able to link these measures to $1,028$ of the $1,234$ job ads posted by $417$ employers.

Appendix Table \ref{tab:jobAdsCollection} shows the characteristics of job ads that we retained at each stage. As compared to the initial sample of job ads, our final sample of relevant job ads has higher shares of permanent jobs and jobs offering the possibility to work from home or have flexible hours, and a lower share of jobs involving shift work, and a similar sectoral composition. Using the text-analysis methodology developed by \cite{audoly24}, we further processed the vacancy texts to classify non-pecuniary job attributes that are publicly posted in our sample of job ads. For instance, this allows us to identify textual mentions of a job having ``good colleagues'', ``central or beautiful location'', providing ``on the job training'', or other ``minor perks'', such as parking in the premise, access to canteen, or a sports facility.

As described in Section \ref{subsec:design}, at Stages 3-4 of our survey experiment, we aim to measure respondents' baseline pay beliefs and study how their job choices depend on the provision of pay information in job ads. Notably, for each of the $1,234$ job ads retained in our final sample of relevant job ads, we do have credible information on the expected starting salary based on monthly full-time equivalent base pay for recently hired workers in these positions, as recorded in high quality administrative matched employer-employee data. In our experimental design, we exploit this feature by constructing hypothetical choice scenarios that embed an information intervention that we create through manipulation of job ad texts that respondents see prior to stating their preferred choices. Specifically, in job scenarios with pay information, as shown in Panel (a) of Appendix Figure \ref{fig:appendix_pay_inf_noinf}, we inform respondents about the expected starting salary associated with each alternative. By contrast, in job scenarios without pay information, as shown in Panel (b) of Appendix Figure \ref{fig:appendix_pay_inf_noinf}, there is no explicit information about pay, and respondents are instead asked to provide their pay beliefs. 

To avoid contamination of our experimental design, we are careful in the design and sampling of job ads that are shown to respondents. To ensure comparability of responses across the different survey modules in Stages 3-4, we always draw job ads from the same underlying sample of $1,234$ job ads. As already discussed, we focus solely on job ads that do not provide explicit pay information, as otherwise we would not be able to elicit baseline pay beliefs and estimate the effects of pay information provision in a credible manner. As each respondent faces ten job scenarios with two choice alternatives in both Stage 3 and 4, i.e., they face a sequence of up to 40 job ads, we sample job ads without replacement within respondent to avoid contaminating learning effects. This feature ensures that the same job ad is not displayed twice by the same respondent across either survey module. However, the same job ad could be viewed by multiple respondents, which allows us to study whether there is systematic dispersion in pay beliefs across respondents that view the same jobs.

In order to ensure comparability between the choice alternatives in each scenario, we further draw pairs of job ads within sector-by-occupation cells. This means that we ask respondents to compare a private sector job to another private sector job, and vice versa for public jobs, and similarly, compare jobs from the same occupational code. This feature of our design allows us to study how signals provided in job ad texts related to workplace amenities and other job-specific attributes matter for respondents' pay beliefs and stated job choices, and abstract from sectoral and occupational differences across jobs. In Stages 3-4, each respondent faces ten choice scenarios in each stage, where they are provided five choices between pairs of public sector jobs and five choices between pairs of private sector jobs, selected in a random order. In each scenario, jobs are sampled from the same occupation. We selected five occupational codes that we considered relevant for our target group. The sampling of job ads is done such that all respondents face two scenarios from each occupational code, in a random order, in each stage. The randomization of occupational codes is pairwise, such that the first two scenarios will show job ads in the same occupation, then the next two scenarios will show ads from the next randomly drawn occupation, and so on.

\subsection{Recruitment and Data Collection}\label{subsec:recruitment}

We now provide information about the recruitment of respondents, the data collection procedure, and provide summary statistics covering key respondent characteristics.

\paragraph*{Recruitment}
We targeted our survey to students enrolled at the University of Oslo, which is a large public university in Norway, with a total of 26,505 students enrolled as of Autumn 2024. We aimed to recruit students from the Faculty of Social Sciences, which is one of the eight major divisions at the University. A total of 4,900 students (i.e., 18.5\%) were enrolled at the Faculty as of Autumn 2024, across a diverse set of study programs, including anthropology, economics, political science, psychology, and sociology. We targeted our survey to students enrolled in 43 different courses, which represent a diverse set of study programs at the Faculty, across 9 waves of data collection carried out in different months. Our main criteria for selecting these courses were: (i) each course was expected to have a large enough mass of enrolled students, (ii) the courses reflected the heterogeneity in study programs at the Faculty, (iii) avoid overlap in the student pools enrolled in the different courses (avoid double counting), and (iv) the availability of auditoriums or rooms used for the survey.

We conducted a series of in-class surveys among students enrolled in the selected courses between April 2024 and October 2025. As part of our initial planning and selection of courses to target, we had informed the course instructor about our survey (all instructors gave us their approval) and reserved the auditorium or instruction room used for the lecture for an additional one-hour slot right after the normal lecture schedule. The normal lecture sequence in all courses at the Faculty consists of a 45-minute instruction session, followed by a 15-minute break, and finally another 45-minute instruction session. During the break, we invited all students attending the lecture sequence through an in-person announcement. We informed the students that they could participate in our survey by remaining seated in the same classroom after the lecture ended and that they would be rewarded with a \$20 gift card for the university cafeteria upon completion of the survey.\footnote{The gift card was valued at NOK 200, which amounts to around \$20 (1 USD is approximately 10 NOK).} We also informed the students that the survey would take around 30 minutes to complete, and that the processing and use of personal information would be based on their informed consent, in compliance with the European General Data Protection Regulation (GDPR) regulation.

Across the 43 courses that we selected, there were a total of around 2,800 students (based on our head count) attending the lectures when we made our announcement. We will refer to this group of students attending lectures that heard our announcement as the target population for our survey. This group corresponds to more than 50\% of all students enrolled at the Faculty. While we do not have access to background characteristics for all students in this target population, we believe our targeted group is broadly similar to the population of students enrolled at the Faculty. If anything, as our survey was implemented through a classroom announcement, we expect to have targeted students that are more likely to be present in class during lectures (e.g., more likely to be full-time student) than the average.

\paragraph*{Data Collection}
As described in Section \ref{subsec:design}, our survey contains several experimental and nonstandard features (e.g., dynamic selection of alternatives, randomization in different survey modules, visualization of both pictures and texts). To implement the survey, we built a dedicated website to ensure that we met the technical requirements of our design and to comply with GDPR regulations. This further allowed us to have complete control of the survey workflow from the point of survey entry to submissions at each stage.

At the start of our survey, all students that were present in the classroom after the lecture were provided a quick response (QR) code, which they could use to load the survey onto their mobile device or computer. Alternatively, they could type in the provided hyperlink to the survey website. Notably, we designed the survey such that it could automatically adapt resolution and other technical features to mobile or desktop view. As shown in Figure \ref{fig:surveyDesign}, Stage 1 of our survey involved signing an informed consent, using a national personal identification system equipped with two-factor authentication, besides filling out some background personal characteristics. Participation in our survey was voluntary. 

Across the various survey waves, a total of 1,091 students were present in the classroom after the lecture (i.e., indicating their intention to participate in the survey), suggesting a gross response rate close to 40\%.\footnote{While the same student could in principle attend more than one of the 43 courses we targeted, we were able to limit this by carefully selecting courses that we believed represented different fields and study programs (e.g., bachelors or masters class) with little overlap. Our initial head counts indicated a total of 2,814 students present in the lectures during the announcements, with 1,127 students indicating their intention to participate in the survey, i.e., an initial response rate of 40\%. Among the 1,127 students in our head count, four respondents never entered into our electronic survey, reducing the number of participating respondents to 1,123. Furthermore, among these 1,123 respondents, about 32 had already completed the survey in a previous course (i.e., 2.8\% of respondents were counted more than once), reducing the number of unique participating respondents to 1,091. We only use the first completed response for each respondent.} The response rate in our survey thus compares favorably to response rates of 11\%-26\% reported in recent surveys targeted at broader populations (see, e.g., \cite{caldwell2025search} and \cite{Humlum2025}). Only 20 out of these 1,091 respondents did not sign our informed consent form, while another 8 did not fill out the survey form on background characteristics. We thus retain 1,063 respondents, whom we refer to as our initial sample of respondents. Potential reasons for not participating or giving consent could, for instance, be that (i) the student had already scheduled other plans for their evening, as we conducted most surveys at 4-5 pm after the regular lecture sessions, or simply that (ii) the student was on exchange and thus had limited knowledge of the Norwegian language or context or lacked a Norwegian personal ID, which is required to use the national personal identification system and thus to fill out the consent form.

\begin{table}[t!]
%\centering
\caption{Respondent Characteristics and Survey Completion.}
\label{tab:attrition}
\input{export_tsd/attrition.tex}
\vspace{0.5em}
\par \scriptsize{
\emph{Notes:} Column 1 shows mean values of background characteristics for all respondents who signed the informed
consent form and completed the survey form on background information in Stage 1, with standard deviations in brackets. Columns 2-4 shows the difference from this mean for individuals who finished various modules of the survey, with standard errors in parenthesis. Column 2 includes respondents who completed hypothetical choice experiments in Stage 2 (required for estimation of willingness to pay for workplace amenities), Column 3 includes respondents who completed the pay information experiments in Stages 3-4, and Column 4 includes respondents who completed all stages of the survey. $^{*} p<0.1$, $^{**} p<0.05$, $^{***} p<0.01$.
}
\end{table}

\paragraph*{Respondent Characteristics}
As shown in Table \ref{tab:attrition}, Column 1, the average age of our respondents is around 22, around 30\% are male, and 10\% had an immigrant background. Virtually all respondents had a paid employment contract in the past. Two-thirds were also currently employed, but only about 37\% had ever held a full-time job, suggesting that most respondents had worked part-time besides studying. The average respondent is in the second year of their study program. By comparison, the average worker entering the sample of strongly connected set of employers, which we use for the estimation of revealed-preference measures of employer quality as in \cite{audoly24}, was aged around 40, with a male share of 50.4\% and an immigrant share of 17.4\%. As expected, our survey respondents are thus substantially younger, with around 70\% being female, and less likely to have an immigrant background, reflecting the population of individuals enrolled in higher education.

Notably, we recorded submissions at each of the four stages of our survey, meaning that we could retain responses from the consent form in Stage 1 and the hypothetical choice experiments in Stage 2, even though the respondent later on decided to drop out of the survey. As shown in Table \ref{tab:attrition}, a total of 973 respondents completed all stages of the survey, i.e., we have a net completion rate of 91.5\% among the initial group of 1,063 respondents. Only about 2\% of students who signed the consent form dropped out of the survey before participating in the hypothetical choice experiment (in Stage 2), while another 6\% dropped out before completing the pay information experiment involving choices between real job ads (in Stages 3-4). Although we do not have knowledge of why a student would drop out of the survey, we can use information recorded on the consent form and their background characteristics to test for the correlates of survey attrition. In Table \ref{tab:attrition}, Columns 2-4, we test for whether the differences in average characteristics across respondents that gave consent and those who completed hypothetical choice experiment and pay information experiment, are statistically significant. We find no evidence indicating that respondents in the initial sample who later drop out are statistically different along their observable characteristics from the respondents that completed all modules. If anything, respondents with immigrant background are slightly less likely to complete the modules featuring real job ads, possibly due to limited knowledge of the Norwegian language and/or limited experience in the Norwegian labor market. However, we cannot reject the test for whether all observable characteristics are jointly significant.

\section{Preferences for Workplace Amenities}\label{sec:wtp}

In this section, we start by describing the specification of the choice model we use to estimate preferences over pay and workplace amenities. We then provide our estimates of willingness to pay that capture respondents' preferences for specific workplace amenities, across choice scenarios involving hypothetical jobs and real job ads, respectively, and compare them to a systematic review of comparable estimates from the  literature using hypothetical choice experiments and subjective willingness to pay measures reported by the respondents. Finally, we provide evidence on how respondents' hypothetical choices align with revealed-preference based rankings of jobs derived from actual labor market choices for the population at large.

\subsection{Choice Model}

To estimate respondents' willingness to pay (WTP) for workplace amenities, we estimate a mixed logit choice model, following the methodology in \cite{revelttrain1998, ReveltTrain2000} for repeated choices. Let the utility of job $j$ for individual $i$ be:
\begin{equation}
U_{ij} = u_{ij}(w_{j},z_{j})+\epsilon_{ij}=\theta w_{j}+\pi_iz'_{j}+\epsilon_{ij}, \label{eq_utility}
\end{equation}

where $w_{j}$ is the pay offered in job $j$, measured as the log of pre-tax monthly salary, $z'_{j}$ is a vector of non-pecuniary job attributes (i.e., amenities), and $\pi_i$ is a vector of parameters that captures the preferences for these attributes, which are allowed to be individual-specific. By assumption, the idiosyncratic utility components are independently and identically distributed, $\epsilon_{ij}\overset{\mathrm{i.i.d.}}{\sim} \text{Gumbel}(0,1)$, where the scale parameter is normalized to 1.

As described in Section \ref{sec:design_data}, we experimentally design choice sets, varying both the offered pay $w_{j}$ and non-pecuniary attributes $z'_{j}$. The vector $z'_{j}$ contains indicators for \textit{permanent contract}, \textit{work from home}, \textit{shift work}, and \textit{flexible hours}, and a measure of \textit{travel time}. A permanent job does not have a predetermined contract end date, while the counterpart in our design is a temporary one-year contract with the possibility for extension at the end of the contract period. Work from home (WFH) jobs allow the employee to work partly from home, while the counterpart is a job that does not have any WFH possibility. Shift work jobs involve some pre-scheduled and predictable evening and/or weekend shifts (but not night work), while the counterpart is a job that does not involve any form of shift or night work. Jobs offering flexible working hours allow the employee to decide parts of their work time schedule themselves, while the counterpart is a job that requires fixed working hours from 8:30 am to 4:00 pm, with no possibility of flexible hours on the part of the employee. Travel time is reported as daily commuting time in hours (i.e., units of 30 minutes each way).

In our survey experiment, each choice scenario contains two alternatives, and each respondent faces a series of choice scenarios $t=\{1,2,...,T\}$. To capture this structure of choice sets, we denote a given choice scenario $t$ by individual $i$ as $it$. Each scenario consists of two alternatives, $j(it)$ and $j'(it)$. We define the outcome $y_{it}$ as a dummy for individual $i$ stating that alternative $j$ is preferred over $j'$ in scenario $it$, i.e., $j(it) \succ j'(it)$. Recognizing that $y_{it} = 1 \Leftrightarrow U_{ij(it)} - U_{ij'(it)} > 0$ and using properties of the Gumbel distribution, we can express the probability that individual $i$ prefers alternative $j$ over alternative $j'$ as:

\begin{equation}
Pr(y_{it}=1|\pi_i,\theta)=\frac{e^{\theta(w_{j(it)}-w_{j'(it)})+\pi_i(z'_{j(it)}-z'_{j'(it)})}}{1+e^{\theta(w_{j(it)}-w_{j'(it)})+\pi_i(z'_{j(it)}-z'_{j'(it)})}} \label{eq_prob}
\end{equation}

The expression for the choice probability resembles the standard logit choice probability, with the exception that the amenity preferences $\pi_i$ are individual-specific. While our primary objective is not to identify preference heterogeneity, the mixed logit model relaxes the restrictive assumptions of standard logit models, specifically the independence of irrelevant alternatives (IIA) property and the requirement of identical preferences across individuals, and provides a better fit than standard choice models. \cite{McFaddenTrain2000} demonstrate that any discrete choice model derived from a random utility model can be approximated to any desired degree of accuracy by the mixed logit model. Moreover, the fact that our survey experiment features a series of repeated choices for each individual aids the identification of individual-specific choice parameters \citep{revelttrain1998}.

\paragraph{Estimation}

In the estimation of the mixed logit choice model, we assume that preference parameters $\pi_i$ are normally distributed. We obtain choice probabilities by integrating the logit probability over the distribution of random coefficients, which we approximate using simulated maximum likelihood. Conditional on a realization of $\pi_i$, choice probabilities take the logit form as in \eqref{eq_prob}. Following \cite{revelttrain1998, ReveltTrain2000}, we recover individual-level preference parameters by computing the posterior mean of the random coefficients given observed choices. Specifically, for each individual $i$ and non-pecuniary attribute $k$, we compute $\hat{\pi}_{i}^k=\mathbb{E}_i(\pi_{i}^k \mid S_{i})$, where $S_i \equiv (y_{i1},y_{i2},...,y_{iT})$ denotes the sequence of observed choices for individual $i$.\footnote{The simulated draws of $\pi_i$ are weighted by the likelihood of observed choices for each draw \citep{hole2007}.} This procedure admits a Bayesian interpretation: The estimated population distribution of random coefficients serves as an empirical Bayes prior over individual-level parameters, while the observed sequence of choices $S_i$ provides the likelihood. These individual-level choice parameters thus combine information from the estimated population distribution of choice parameters and each respondent’s observed choices. To account for the fact that each respondent faced multiple choice scenarios in our survey design, we cluster the standard errors associated with our estimates at the respondent level.

\paragraph{Willingness to Pay}

To summarize an individual $i$'s preference for the non-pecuniary attribute $k$, we calculate the willingness to pay as $WTP_{i}^k=\mathbb{E}_i(\pi_{i}^k \mid S_{i})/\hat{\theta}=\hat{\pi}_{i}^k/\hat{\theta}$, where $\hat{\theta}$ is the estimated choice parameter associated with the pay component. Analogously, we calculate the average WTP for attribute $k$ across our survey respondents by taking the ratio of the estimated average preference parameter $\hat{\bar{\pi}}^k$ for attribute $k$ and the pay coefficient $\hat{\theta}$. The average WTPs can be interpreted as the (approximate) percentage change of monthly pay an average individual is willing to give up (if positive) or must be compensated (if negative) to accept an otherwise identical job with an additional unit of an amenity. 

\subsection{Evidence}

We now provide our evidence on individuals' willingness to pay (WTP) for workplace attributes. Before considering the WTPs estimated from our choice model, we present WTPs derived from estimates reported in a range of existing studies that use hypothetical choice experiments in Table \ref{tab:willingness_to_pay}, Column 1. Specifically, we performed a targeted literature review focusing on studies providing WTP estimates for at least one of the five non-pecuniary job attributes that featured in our survey experiment.\footnote{Appendix Table \ref{tab:wtpLitA} provides an overview of the estimates of WTPs found in the literature, while in Appendix Table \ref{tab:wtpLitB}, we report the most relevant estimate from each study that corresponds to the non-pecuniary job attributes in our survey. We took several steps to construct estimates of WTPs that are comparable to the estimates that we construct using our data. For instance, in some of the studies included in our literature overview, WTPs were estimated in levels of pay, and in such cases, we scale the estimates by the average or baseline pay to be comparable to our model specification. We also construct a weighted average of the relevant estimates across all studies, using the sample size in each study as weights.} Across the studies we considered, the typical estimates of WTPs reflect strong preferences for workplace amenities. On average, these estimates imply that respondents are willing to sacrifice 17.3\% of their pay for a permanent position, are willing to pay 13\% for jobs offering the possibility to work from home and 18.4\% for jobs offering the possibility to work flexible hours. Similarly, respondents must be compensated 12.7\% higher pay for jobs involving shift work and 15\% for jobs that have an hour longer daily commute (i.e., 30 minutes longer each way).

Next, we present respondents' self-reported subjective WTPs for each non-pecuniary job attribute. Specifically, we obtained subjective WTPs by asking each respondent how much they would be willing to give up (require) of pay (increase) to obtain (avoid) a given amenity (disamenity), all else equal. To fix ideas, we asked respondents to imagine they had a job paying NOK 50,000 per month and asked them to state their subjective valuation in levels, which we later scaled by NOK 50,000 to construct relative WTP measures.\footnote{NOK 50,000 per month amounts to around \$5,000 per month (1 USD is approximately 10 NOK).} In Table \ref{tab:willingness_to_pay}, Column 2, we report the average subjective WTP for each workplace amenity reported by our respondents, after truncating at the 95\% percentile.\footnote{As shown in Table \ref{tab:attrition}, our final sample consists of 973 respondents. However, 8 of these did not fill out the fields eliciting subjective WTPs. Nevertheless, we retain their responses from all other fields in our survey.} The subjective WTPs are quantitatively similar to those reported in the existing literature, although these estimates rely on different methodologies and target populations. Notably, our respondents report lower WTPs for jobs providing flexible hours or work from home options, as compared to the typical estimates in the literature. This could, for instance, reflect differences across labor markets in the notions of what workplace attributes involving the possibility to work flexible hours or work from home typically entail. Further, our respondents report that they must be compensated more to work in shift work jobs. Nevertheless, the subjective WTPs are broadly in line with the literature. 

\begin{table}[t!]
%\centering
\caption{Willingness to Pay for Workplace Amenities.}
\label{tab:willingness_to_pay}
\input{export_tsd/willingness_to_pay.tex}
\vspace{0.5em}
\par \scriptsize{
\emph{Notes:} This table provides alternative estimates of willingness to pay (WTP) for five non-pecuniary workplace attributes. The counterpart of a permanent job is a temporary one-year contract with the possibility for extension at the end of the contract period. Work from home (WFH) jobs allow the employee to only work partly from home, where the intensity of WFH in days or hours is not explicitly specified, and the counterpart is a job that does not have any WFH possibility. Shift work jobs involve some pre-scheduled and predictable evening and/or weekend shifts (but not night work), where the intensity of shift work in days or hours is not explicitly specified, and the counterpart is a job that does not involve any shift or night work. Jobs offering flexible working hours allow the employee to decide their work time schedule themselves beyond the requirement to be present at the workplace during core hours from 9:00 am to 2:30 pm, while the counterpart is a job that requires fixed working hours from 8:30 am to 4:00 pm, with no possibility of flexible hours on the part of the employee. Travel time is reported in daily commuting time in hours (i.e., in units of 30 minutes each way). Column 1 provides the average WTP estimates from our literature review, as detailed in Appendix Table \ref{tab:wtpLitB}. Column 2 provides the average subjective WTP measures that were self-reported by our respondents in the questionnaire shown in Appendix Figure \ref{fig:information_sheet_subj}. Columns 3-5 provide alternative WTP estimates that we estimated based on mixed logit models for the final sample of respondents [$N$=$973$], where we use choice data collected using hypothetical jobs in Stage 2 (Column 3), using real job ads in the pay information modules in Stages 3-4 (Column 4), and from the combination of hypothetical jobs in Stage 2 and real job ads in Stages 3-4 (Column 5). Columns 3-5 provide the average expected WTP parameters estimated using the methodology in \cite{revelttrain1998, ReveltTrain2000}, while Appendix Figure \ref{fig:wtp_density} provides the distribution of individual-specific expected WTPs that correspond to Column 5, and Appendix Table~\ref{tab:robustness_wtp} provides robustness to correlated preferences and respondent inattention. Standard errors are reported in parenthesis, and clustered at the respondent level in Columns 2-5. $^{*} p<0.1$, $^{**} p<0.05$, $^{***} p<0.01$.
}
\end{table}

Next, we report WTP estimates from the choice model \eqref{eq_utility}-\eqref{eq_prob}, using data from our hypothetical choice experiments. In Table \ref{tab:willingness_to_pay}, Column 3, we provide the average WTPs that we estimated using respondents' hypothetical choices from Stage 2, where we collected information on preferred choices across hypothetical job alternatives. As described in Section \ref{subsec:sampling}, we experimentally designed these jobs, varying pay and job attributes across alternatives, while informing respondents that these choice alternatives were identical in all other respects, in order to limit the role of omitted job attributes in our choice model. This provides us with a total of 38,920 choice alternatives, as we have data on choices from $T=20$ choice scenarios with two alternatives each for every respondent (i.e., $973 \times 20 \times 2 = 38,920$). On average, we find that individuals are willing to give up 16.1\% of their salary for having a permanent job, 10.5\% for having the possibility to work from home, 6.4\% for the possibility to work flexible hours, and must be compensated with 7.7\% higher pay to accept a job that involves shift work and 16.9\% for a job with an hour longer daily commute (i.e., 30 minutes longer each way). All of these estimates are highly statistically significant and have the expected signs. With the exception of shift work, where we find lower WTP estimates from the choice model in Column 3 as compared to the subjective WTP in Column 2, the WTPs derived from hypothetical choices and the subjective WTPs are also fairly well aligned.

Further, our survey experiment featured experimentally designed choice sets where we asked respondents to state their preferred choices across multiple pairs of real job ads, which we randomly sampled within sector-by-occupation cells from a targeted sample of job ads recently posted in their local labor market featuring high-skilled service sector jobs. Using respondents' hypothetical choices across real job ads from the modules where respondents were provided explicit information on the expected starting salary in each job, i.e., the choice data collected in Stage 3 for the treated group and in Stage 4 for the control group in Figure \ref{fig:surveyDesign} (see details in Section \ref{subsec:sampling}), we can estimate choice models where respondents' choices depend on the expected starting salary and the advertised workplace amenities in job ads featured in their hypothetical choice sets. This provides us with a total of 19,460 choice alternatives, as we have data on choices from $T=10$ choice scenarios for every respondent, where each choice scenarios featured two real job ad alternatives (i.e., $973 \times 10 \times 2 = 19,460$).

In the choice models where we use data on choices between real job ads to infer WTPs, we focus on the same set of workplace amenities as in the choice models we estimated using data on hypothetical jobs. Specifically, we use the text-analysis approach from \cite{audoly24} to construct binary indicators for whether each of these job attributes was advertised in the job ad texts. Although job ad texts do not provide personalized information on travel times for each person-job pair, they do provide information on the location of each workplace, which respondents could potentially use to infer the expected travel times to each workplace from their residence location. As we collected information on respondents' residence location and the location of workplace associated with each job ad, we are able to construct measures of travel times across the  job alternatives that each respondent faced.\footnote{Specifically, when estimating the choice model \eqref{eq_prob} using respondents' choices between ads, we include travel times measured at the \textit{job-person} level. For notational ease, we still refer to the amenity vector as $z_{j}$.} 

The experimental choice setting that features choices between real job ads with explicit pay information resembles the experimental setting with choices between hypothetical jobs, with the caveats that (i) respondents' choices between pay and advertised workplace amenities depend on how they perceive and process signals about pay and non-pecuniary attributes listed in job ad texts, and (ii) job ads may advertise additional job attributes that we do not include in our choice models. Due to these reasons--the potential non-salience of workplace attributes in job ads and the possibility of omitted job attributes--we may not expect the WTPs we estimate using respondents' choices between real job ads and between hypothetical jobs to be fully aligned.\footnote{In Appendix~\ref{app_choice_omitted}, we clarify how to interpret WTP estimates from choice models with omitted attributes.} In Table \ref{tab:willingness_to_pay}, Column 4, we report the average WTPs estimated using respondents' hypothetical choices across real job ads. Interestingly, for three out of five workplace attributes--permanent job, shift work and flexible hours--we are unable to reject the equality of average expected WTPs that we estimated using respondents' choices between hypothetical jobs in Column 3 and across real job ads in Column 4. Notably, however, we find no evidence that respondents prefer jobs offering the possibility to work from home based on their choices between job ads. Moreover, respondents' distaste for travel time has around half the magnitude as compared to the estimate using hypothetical jobs, likely reflecting that job ad texts provide noisy signals about expected commuting distances.

Finally, in Table \ref{tab:willingness_to_pay}, Column 5, we estimate WTPs in the pooled set of hypothetical and real job scenarios. We now use a larger sample with choice data, exploiting several portions of our survey experiment, and find WTP estimates that are highly statistically significant and have the expected signs. Notably, despite the fact that our survey targeted currently enrolled students, the WTPs for this group are broadly similar to the estimates found in the existing literature on hypothetical choice experiments. We find somewhat larger WTPs for permanent contracts and shift work, and smaller WTPs for flexible hours (Column 5), as compared to the WTPs reported in the literature (Column 1).

In the later parts of our analysis, we will use the WTP estimates in Table \ref{tab:willingness_to_pay}, Column 5, along with information on attributes advertised in job ad texts, to construct composite measures of the job amenity values. This specification pools the largest available sample of choice observations within respondents and therefore delivers the most precise individual-specific WTP parameters. In Appendix Table~\ref{tab:robustness_wtp}, we provide robustness to allowing cross-attribute correlation in the preferences for non-pecuniary job attributes in the choice model, besides alternative checks for respondent inattention, finding very similar results.

While the WTP estimates reported in Table \ref{tab:willingness_to_pay}, Columns 2-5, provide averages across our sample of survey respondents, we also construct individual-specific WTP parameters using the methodology in \cite{revelttrain1998, ReveltTrain2000}. In Appendix Figure \ref{fig:wtp_density}, we show the distributions of individual-specific expected WTPs that we estimate for each of the five non-pecuniary workplace attributes. The model specification corresponds to the one used for the estimates shown in Table \ref{tab:willingness_to_pay}, Column 5, which reports the average WTPs across all respondents. We note that there is sizable heterogeneity in WTPs across respondents.

\begin{figure}[t!]
    \centering
    
    \caption{Rank-Rank Slopes: Hypothetical Choices vs. Market-Level Revealed Preferences.}
    \label{fig:rank_rank}
    
    \begin{subfigure}[t]{0.48\textwidth}
        \centering
        \includegraphics[width=\linewidth]{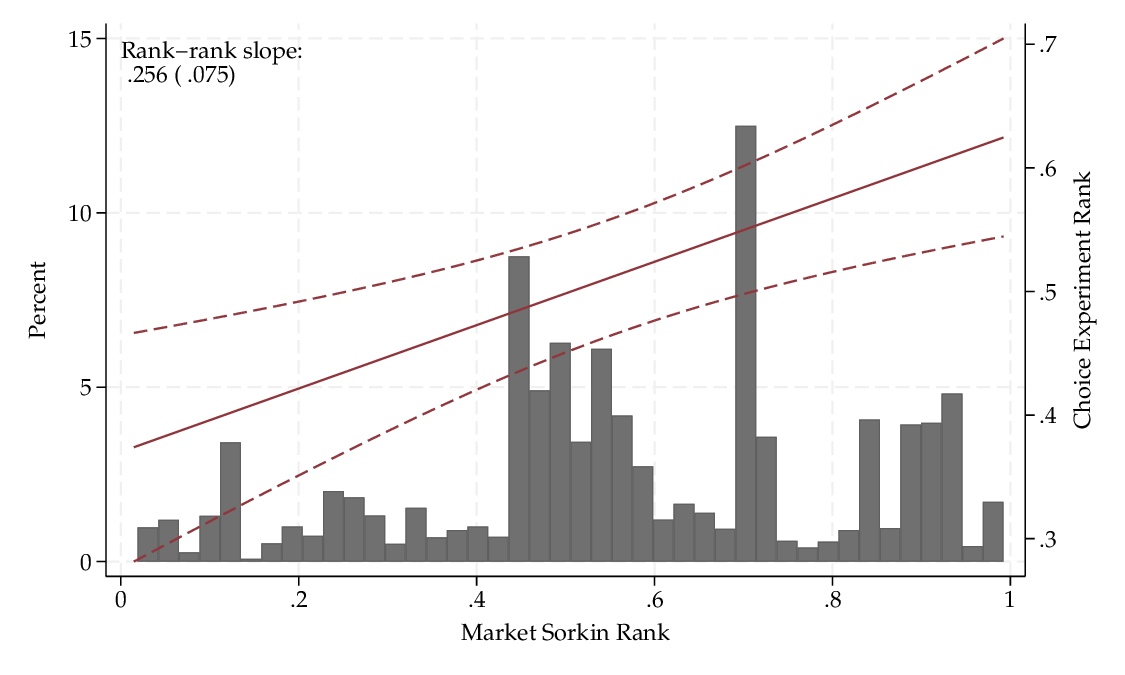}
        \caption{Sorkin Value.}
        \label{fig:first}
    \end{subfigure}
    \hfill
    \begin{subfigure}[t]{0.48\textwidth}
        \centering
        \includegraphics[width=\linewidth]{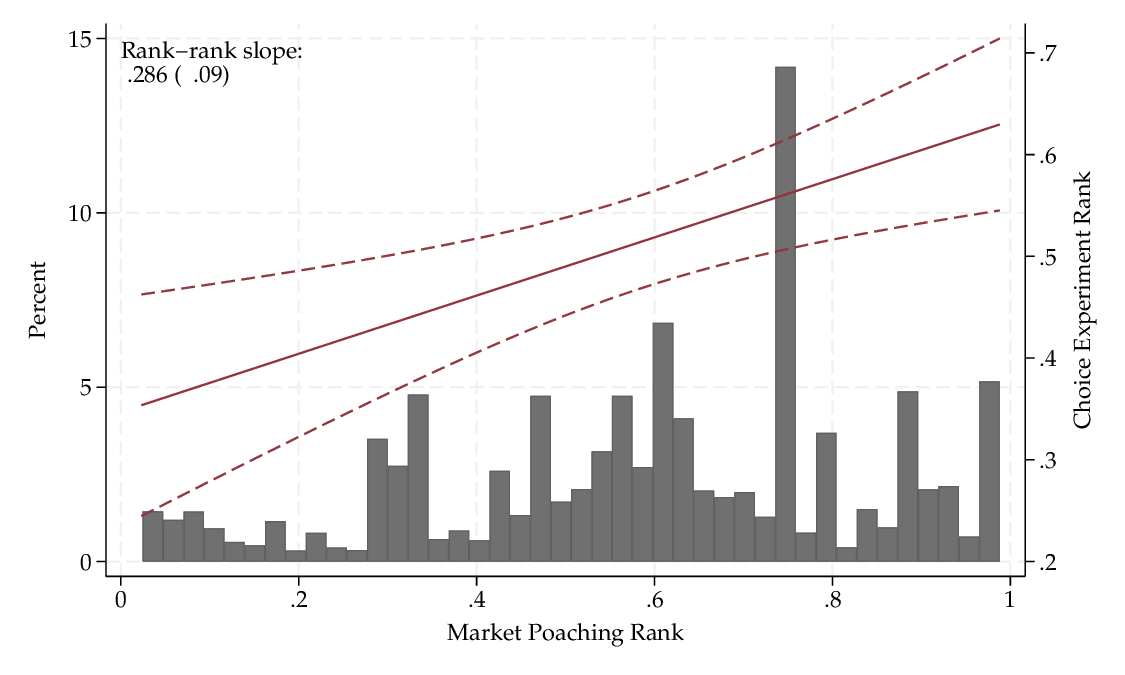}
        \caption{Poaching Index.}
        \label{fig:second}
    \end{subfigure}

        \vspace{0.5em}
    \begin{minipage}{1\textwidth}
        \scriptsize
        \textit{Notes:} This figure shows the rank-rank correlations between the rankings of real job ads derived from respondents' hypothetical choices in our survey experiment (right $y$-axis) and the market-level revealed-preference based ranks of jobs derived from the actual labor market choices for the population at large ($x$-axis). The \textit{choice-experiment ranks} are derived from logit choice models with indicators for each real job ad, estimated separately by occupation and sector, using the estimated choice probabilities. The market-level \textit{revealed-preference ranks} are estimated using matched employer-employee data following the approach in \cite{audoly24}, and derived from the Overall Employer Value of \cite{sorkin2018ranking} in panel (a) and the Poaching Index of \cite{bagger2019empirical} in panel (b), respectively. The rank-rank correlations are estimated from regressions of  the ``choice experiment rank'' as the dependent variable and the ``market-level revealed-preference rank'' as the explanatory variable, while controlling for occupation by sector fixed effects. Standard error are constructed by product bootstrap over individual and job using 500 repetitions, accounting for the statistical uncertainty in the estimation of choice probabilities and the associated ``choice experiment rank'' assigned to each real job ad.
    \end{minipage}

\end{figure}

\paragraph{Ranking of Jobs} The above evidence shows that our estimates of respondents' preferences for specific workplace amenities based on their hypothetical choices are comparable to the estimates found in the literature using hypothetical choice experiments and subjective willingness to pay measures reported by the respondents. We now consider how respondents’ hypothetical choices align with actual labor market choices for the population at large. Crucially, we observe establishment identifiers for the employer posting each job ad, and can follow actual job flows across employers in the population level employer-employee data. 

To study the alignment between the ranking of jobs implied by respondents' hypothetical choices and the ranking of jobs implied by actual job flows for the population of workers at large, we take a simpler modeling approach that abstracts from preference heterogeneity. Specifically, we estimate logit choice models on indicators for each job ad, separately by occupation-sector cells. This provide job-specific choice probabilities, allowing us to construct a common ranking of job ads within occupation-sector cells across respondents. Notably, we use the same choice data as in Table \ref{tab:willingness_to_pay}, Column 4, to estimate the job ranks, but unlike the choice model \eqref{eq_utility}-\eqref{eq_prob} specified earlier, we do not decompose job values in pay and amenity components.\footnote{By comparison, \cite{caldwell2026firm} use discrete choice experiments with randomized wage variation to recover money-metric firm-level amenity valuations for a large sample of German workers.} For comparisons to revealed-preference measures of employer quality, we use matched employer-employee data as in \cite{audoly24} to estimate the Overall Employer Value of \cite{sorkin2018ranking} and the Poaching Index of \cite{bagger2019empirical}.

Figure \ref{fig:rank_rank} provides our evidence on the alignment between the \textit{choice-experiment ranks} of real job ads derived from the hypothetical choices made by our survey respondents and the corresponding market-level \textit{revealed-preference ranks} derived from actual job flows observed in the matched employer-employee data. We obtain rank-rank slopes by regressing choice experiment rank on market rank, controlling for occupation by sector. For both revealed-preference measures of employer quality, we find positive associations with the ranks derived from our survey experiment, with rank-rank regression slopes of around 0.25-0.3. We interpret these correlations as indicate that our respondents' implicit ranking of real job ads is positively aligned with the rankings implied by actual labor market choices for the population at large. Possible reasons for these rankings to not be fully aligned (i.e., correlation of one) include the presence of preference heterogeneity and attenuation bias due to statistical noise in our measurement of ranks.

\section{Pay Beliefs}\label{sec:beliefs}

Our survey experiment has components where we elicited respondents' beliefs about starting salaries in real jobs. We now use information on what respondents expected the starting salaries to be in the various real job ads to construct measures of their systematic pay beliefs. Specifically, to measure respondents' \textit{baseline} pay beliefs, we use information provided by the control group in Stage 3 (see Figure \ref{fig:surveyDesign}), where they viewed real job ads without explicit information on the expected starting salary in each job. Unlike the treated group that was exposed to real jobs with explicit pay information, the control group did not see such information until a later stage of the survey. By virtue of the randomization, the control group's responses in Stage 3 provide measures of the baseline beliefs held by our respondents.

\subsection{Measurement}

We refer to what a respondent expects the starting salary in a job to be after having viewed an ad posted for this job as their \textit{posterior mean pay belief}. We refer to it as \textit{posterior}, as it is their belief after updating according to the public information in the job ad, and as \textit{mean} since their belief about the pay in a given job is uncertain and thus follows a distribution. Formally, we denote respondent $i$'s posterior mean pay belief for job $j$ as $\mathbb{E}_i[\tilde{W}_{j}]$.\footnote{To ease notation, we let respondent $i$'s information set be implicit. Thus, one can interpret $\mathbb{E}_i[\tilde{W}_{j}]:=\mathbb{E}_i[\tilde{W}_{j}|\mathbb{I}_i]$ as capturing what respondent $i$ expects starting salary to be in job $j$, given $i$'s information set $\mathbb{I}_i$.} In practice, in our survey, respondent $i$ views an ad for job $j$, and reports a measure of the expected monthly starting salary in this job in levels (reported in bins of 500 Norwegian Kroner).\footnote{As noted in Section \ref{sec:design_data}, our survey was conducted between April 2024 and October 2025, and the job ads we sampled were posted in the second half of 2023. To facilitate comparability, we throughout deflate measures of the actual wages for hired workers and respondents' beliefs about starting salaries to December 2023 Norwegian Kroner, using Statistics Norway's wage growth index. Over the period of our survey, wage growth was at approximately 9\%, exceeding the 5\% growth in the consumer price index.} One can view this as a measure of what a job seeker expects a job to pay during the early phase of job search, while considering options in the labor market (e.g., by browsing publicly posted job ads). The ad texts used in this part of our survey provide publicly posted signals about workplace amenities and other non-pecuniary job attributes in the job, but by construction do not reveal explicit information on the starting salary. Respondents may hold prior knowledge about starting salaries in various jobs, and additionally use information listed in job ads to form expectations about starting salaries. In Appendix \ref{sec:appendix_belief_model}, we provide a stylized Bayesian model of pay belief formation for settings where job ads typically do not provide explicit pay information. Our experimental setup seeks to mimic such a setting.

In most of our empirical analysis, we focus on the logarithm of the respondent's posterior mean belief. Let $\mathbb{E}_i[\tilde{w}_{j}] \equiv \mathbb{E}_i[\log \tilde{W}_{j}]$. From Jensen's inequality, $\mathbb{E}_i[\tilde{w}_{j}] < \log \mathbb{E}_i[\tilde{W}_{j}]$, so the logarithm of the posterior mean belief reported in levels overstates the true posterior mean belief in logs. Intuitively, this arises from the uncertainty about pay levels. If the posterior distribution of $\tilde{W}_{j}$ is dispersed, log-concavity implies that low realizations reduce $\mathbb{E}_i[\tilde{w}_{ij}]$ more than high realizations increase it. The difference $\log \mathbb{E}_i[\tilde{W}_{j}] - \mathbb{E}_i[\tilde{w}_{j}]$ increases in the posterior variance of $\tilde{W}_{j}$ --- it is larger the greater is the uncertainty of pay beliefs.

To infer respondents' posterior mean beliefs in logs, $\mathbb{E}_i[\tilde{w}_{j}]$, from measures of their reported beliefs in levels, $\mathbb{E}_i[\tilde{W}_{j}]$, we must correct for the bias stemming from posterior variance. If log posterior pay beliefs are normally distributed, then $\mathbb{E}_i[\tilde{w}_{j}]=\log \mathbb{E}_i[\tilde{W}_{j}] - \tfrac12\,\mathrm{Var}_i(\tilde{w}_{j})$. More generally, this relationship arises as a second-order approximation. The approximation is accurate when higher-order moments (skewness, kurtosis, etc.) are small, so that terms beyond the second order are negligible. In the following, we impose structural assumptions that allow us to flexibly model pay beliefs in logs.\footnote{In theory, one could  elicit mean beliefs in logs, but this presumes that respondents are able to perform log-transformations during the survey, which is unreasonable. Another alternative is to elicit information on the precision or variance of respondents' beliefs, which is also known to be extremely challenging. Higher order moments are sensitive to framing \citep{DelavandeRohwedder2008}, and even when first order beliefs are stable, second-order beliefs are much noisier and sensitive to elicitation format \citep{Zafar2011,Zafar2013} and are troubled with measurement error and cognitive simplification \citep{WiswallZafar2015}. For these reasons, we instead opt for estimating the posterior variance, and assess sensitivity of our results to this choice.} We allow both the posterior mean and the posterior variance of pay beliefs to vary flexibly across respondents and jobs:

\begin{align}
\mathbb{E}_i[\tilde{w}_{j}] &= \alpha_i + \gamma_j + \tilde{\varepsilon}_{ij}, &\mathbb{E}(\tilde{\varepsilon}_{ij}\mid \alpha_i,\gamma_j)=0, \label{eq:meanvariancemodel}\\
\mathrm{Var}(\tilde{\varepsilon}_{ij}\mid \alpha_i,\gamma_j)&= \exp\!\left(\rho_i + \psi_j\right).
\label{eq:meanvariancemodel2}
\end{align}
In Equation \eqref{eq:meanvariancemodel}, we decompose respondent $i$'s posterior mean belief about log pay in job $j$ into an individual component $\alpha_i$ (e.g., capturing persistent optimism or pessimism) and a job component $\gamma_j$ (e.g., capturing shared pay belief differentials across jobs). The residual $\tilde{\varepsilon}_{ij}$ represents idiosyncratic deviations around this additive structure. This structure resembles the classical AKM wage structure with two-way heterogeneity \citep{abowd1999high}, with the distinction that we impose this additive structure on individuals' posterior mean beliefs. Notably, our model does not feature interactions between individual and job components. Next, Equation \eqref{eq:meanvariancemodel2} models heteroskedasticity in belief dispersion, allowing the conditional variance of the idiosyncratic deviations $\tilde{\varepsilon}_{ij}$ to vary systematically by individual-specific $\rho_i$ and job-specific $\psi_j$ components. The exponential link guarantees positivity and implies a multiplicative heteroskedasticity structure. In this framework, belief uncertainty is captured as the conditional variance of log belief deviations around the individual–job mean structure.

\paragraph{Estimation}
We jointly estimate the two-equation model \eqref{eq:meanvariancemodel}-\eqref{eq:meanvariancemodel2} using a fixed-point iterative algorithm: We first estimate the mean equation \eqref{eq:meanvariancemodel}, predict residuals $\hat{\tilde{\varepsilon}}_{ij}^2$, then estimate the variance equation \eqref{eq:meanvariancemodel2}, predict the conditional variance and update the implied posterior mean in logs using the Jensen correction, and iterate this procedure until convergence. In particular, the variance equation \eqref{eq:meanvariancemodel2} is estimated in a Poisson regression of the squared residuals $\hat{\tilde{\varepsilon}}_{ij}^2$ on individual- and job-fixed effects. To account for the uncertainty in this procedure, we bootstrap all subsequent regressions that use beliefs as outcomes jointly with this procedure using a product bootstrap over individuals and jobs \citep{CameronGelbachMiller2011}.\footnote{The software package \texttt{belieffit} for Stata, available on \href{https://github.com/martin-andresen/belieffit}{Github}, implements the fixed-point algorithm and allows the user to flexibly model beliefs, do product bootstrap for inference, and run auxiliary regressions.}

Our model permits the following decomposition, similar to the AKM framework:
\begin{align}
\label{eq:flex}\underbrace{\mathrm{Var}(\tilde{w}_{ij})}_{\footnotesize\shortstack{\text{total} \\ \num[round-mode=places,round-precision=3]{\vartotzero} [100 \%]}}&=
\underbrace{\mathrm{Var}(\alpha_i)}_{\footnotesize\shortstack{\text{person component} \\ \num[round-mode=places,round-precision=3]{\varpersonzero} [\num[round-mode=places,round-precision=0]
    {\fpeval{100*(\varpersonzero/\vartotzero)}} \%]}}+
\underbrace{\mathrm{Var}(\gamma_j)}_{\footnotesize\shortstack{\text{job component} \\ \num[round-mode=places,round-precision=3]{\varjobzero} [\num[round-mode=places,round-precision=0]
    {\fpeval{100*(\varjobzero/\vartotzero)}} \%]}}+
\underbrace{2\mathrm{Cov}(\alpha_i,\gamma_j)}_{\footnotesize\shortstack{\text{sorting} \\ \num[round-mode=places,round-precision=3]{\covzero} [\num[round-mode=places,round-precision=0]
    {\fpeval{100*(\covzero/\vartotzero)}} \%]}}+
\underbrace{\mathrm{Var}(\tilde{\varepsilon}_{ij})}_{\footnotesize\shortstack{\text{unexplained} \\ \num[round-mode=places,round-precision=3]{\varunexpzero} [\num[round-mode=places,round-precision=0]
    {\fpeval{100*(\varunexpzero/\vartotzero)}} \%]}}
\end{align}

A common concern in estimating these variance components is that statistical noise may bias the variance of job and person components \citep{https://doi.org/10.3982/ECTA16410}. We apply an empirical Bayes shrinkage procedure to estimate the variance-covariance components.\footnote{The software package \texttt{ebayes\_shrink} for Stata, available on \href{https://github.com/martin-andresen/ebayes_shrink}{Github}, performs Empirical Bayes shrinkage, using a method-of-moments estimator.} In practice, we use a method-of-moments estimator that subtracts the average estimated precision of the job and person components and their covariance (see, e.g., \cite{Morris1983}), using the precision of the job and person components from the product bootstrap described above.\footnote{The estimation of variance components based on AKM-style wage regressions faces similar challenges in observational panel data settings when few individuals change jobs. A common approach to address this limited mobility bias is to apply the leave-one-out estimator of \cite{https://doi.org/10.3982/ECTA16410}. Note, however, that limited mobility bias is not an issue in our setting as each respondent is exposed to a large number of job ads in our survey experiment. We instead apply empirical Bayes shrinkage using the method-of-moments estimator, which unlike the  \cite{https://doi.org/10.3982/ECTA16410} estimator, allows clustering by respondent and job.} 

Having corrected the posterior means for the variance bias, we also estimate regression models of posterior means and variance on job or person characteristics to understand the correlates of beliefs. For all these models, we use a product bootstrap that accounts for the first step uncertainty from estimating the posterior mean and variance. As we randomized job ads to respondents and respondents to treatment, we do not need job- or person-level fixed effects for identification in these models, but can include them for robustness and to reduce residual variance, when the parameters of interest are not perfectly multicollinear.

\subsection{Evidence}
We now present results from a decomposition of pay beliefs in Equation \eqref{eq:meanvariancemodel} using data on pay beliefs that we elicited for the control group in Stage 3. The main results from this decomposition are reported below each variance component in Equation \eqref{eq:flex}. First, it is reassuring to see that there is no evidence of sorting, which supports our randomization. Next, we find that $\num[round-mode=places,round-precision=0]{\fpeval{100*(\varjobzero/\vartotzero)}} \%$ of the variance in pay beliefs can be attributed to job-specific components, capturing common  pay beliefs across jobs shared by respondents.\footnote{By comparison, the bias-corrected estimates of the variance components of AKM wage equations reported in \cite{BonhommeEtAl2023} for Norway suggest that around 12\% of the variance of log annual earnings can be attributed to firm effects, while the sorting component explains about 11–17\%.} Further, $\num[round-mode=places,round-precision=0]{\fpeval{100*(\varpersonzero/\vartotzero)}} \%$ of the variance in beliefs can be attributed to respondent-specific components, suggesting that there are systematic differences in pay beliefs across respondents for the same job. The remaining component is attributed to idiosyncratic deviations (i.e., the unexplained part).

\begin{table}[t!]
\begin{center}
\caption{Correlates of Pay Beliefs: Respondent Background Characteristics.}
\label{tab:baselinebeliefs_p}
\renewcommand{\arraystretch}{0.95}
\input{export_tsd/baselinebeliefs_mean_p.tex}
\renewcommand{\arraystretch}{1}
\end{center}
{\scriptsize \emph{Notes:} This table shows the coefficients from regressions of respondent $i$'s posterior mean pay belief for job $j$ in logs, $\mathbb{E}_i[\tilde{w}_{j}]$, on respondent characteristics. Column (2) adds controls for job ad fixed effects, Column (3) adds indicators for the survey batch, capturing the time period and study course that respondent participated in during the survey, while Column (4) adds both sets of fixed effects. Standard errors are reported in parentheses, estimated by two-way cluster bootstrap over respondent and job ad, accounting for the estimation of posterior variance. Posterior mean beliefs in logs have been corrected for uncertainty bias (see Section \ref{sec:beliefs}), and estimated using the two-equation model \eqref{eq:meanvariancemodel}-\eqref{eq:meanvariancemodel2}. $^{*} p<0.1$, $^{**} p<0.05$, $^{***} p<0.01$.
}
\end{table}

We now consider how respondents' background characteristics relate to their pay beliefs. In Table \ref{tab:baselinebeliefs_p}, we provide regression coefficients from alternative specifications with each respondent's posterior mean pay beliefs in logs, $\mathbb{E}_i[\tilde{w}_{j}]$, as the dependent variable and a vector of respondents' background characteristics as explanatory variables. Across the alternative specifications, we find that two respondent characteristics are highly predictive of their pay beliefs: (i) gender and (ii) salary in current job. On average, male respondents expect starting salaries to be around 6\% higher, as compared to female respondents. This finding may reflect that there exist systematic gender differentials regarding salary expectations for similar jobs \citep{Roussille2024}, which do not necessarily reflect actual gender pay gaps. The total full-time equivalent monthly salary in the latest job is positively correlated with pay beliefs, with a pass-through rate of around 0.12. This may reflect that the current salary anchors respondents' pay expectations \citep{jager2022worker}. Moreover, Appendix Table \ref{tab:baselinebeliefs_var_p} shows that the variance of respondents' beliefs is negatively related to current salary (10\% significance level). This suggests that respondents that currently earn more not only have higher pay expectations, but also more precise beliefs. Unlike the current salary, there does not appear to be a robust association between the amenities in the current job and the mean or variance of respondents' pay beliefs, nor past or current work-experience or years spent studying.

While the above evidence could be interpreted as consistent with the findings in \cite{Roussille2024} and \cite{jager2022worker}, there are important differences in the measurement of pay beliefs across studies. In particular, we asked about the starting salaries that a respondent expects various jobs to pay in general, independent of who gets hired, rather than about their own expected salary in these jobs. Our evidence thus suggests that female respondents expect starting salaries in the market to be lower \textit{in general} when compared to male respondents, while high earners expect starting salaries to be higher \textit{in general}. This is a subtle difference from the measure of \cite{Roussille2024} that female respondents expect to earn less than male respondents, and from that of \cite{jager2022worker} where high earners expect higher outside option wages than low earners. Specifically, our measure does not capture respondents' beliefs about outside options, and should instead be interpreted as a measure of general pay beliefs, which are likely less influenced by perceptions of own-ability and job selection.\footnote{For instance, a worker with a high wage might perceive this wage level as a signal of their own ability and thus expect to receive high wage in other jobs, resulting in a positive association between current salary and beliefs about outside options. Similarly, job choices are endogenous, so asking a worker what they expect to make \textit{elsewhere} will naturally be influenced by their decision to enter the current job.}

\begin{figure}[t!]
    \begin{center}
    \caption{Correlates of Pay Beliefs: Job Quality, Actual Pay, and Amenity Value.}
    \label{fig:baseline_beliefs_bias}
    \vspace{0.25em}

    \begin{subfigure}[t]{0.48\textwidth}
        \centering
        \includegraphics[width=\linewidth]{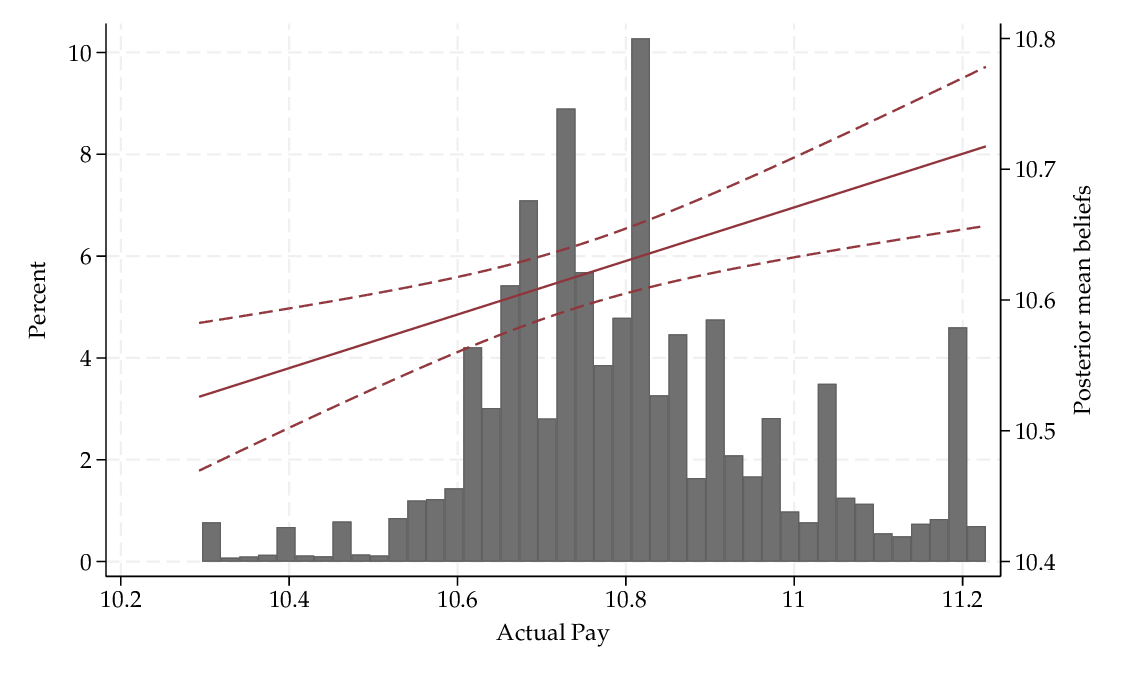}
        \caption{Actual Pay}
        \label{fig:baseline_beliefs_biasA}
    \end{subfigure}
    %\hfill
    \begin{subfigure}[t]{0.48\textwidth}
        \centering
        \includegraphics[width=\linewidth]{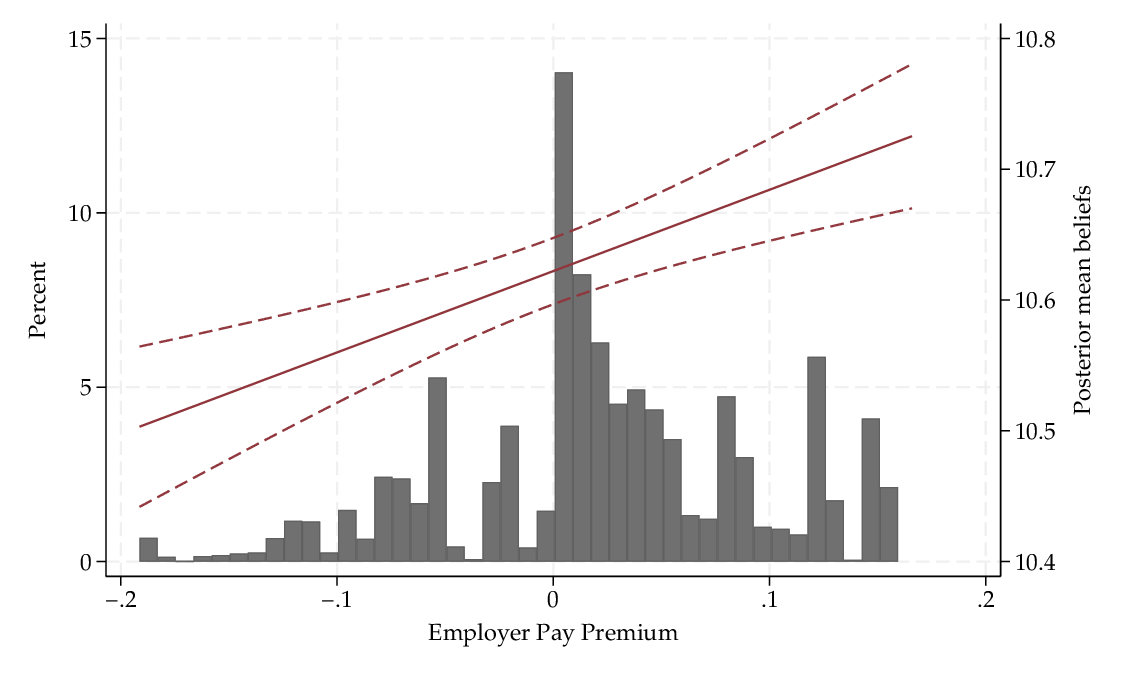}
        \caption{Employer Pay Premium}
        \label{fig:baseline_beliefs_biasB}
    \end{subfigure}

    \vspace{0.5em}

    \begin{subfigure}[t]{0.48\textwidth}
        \centering
        \includegraphics[width=\linewidth]{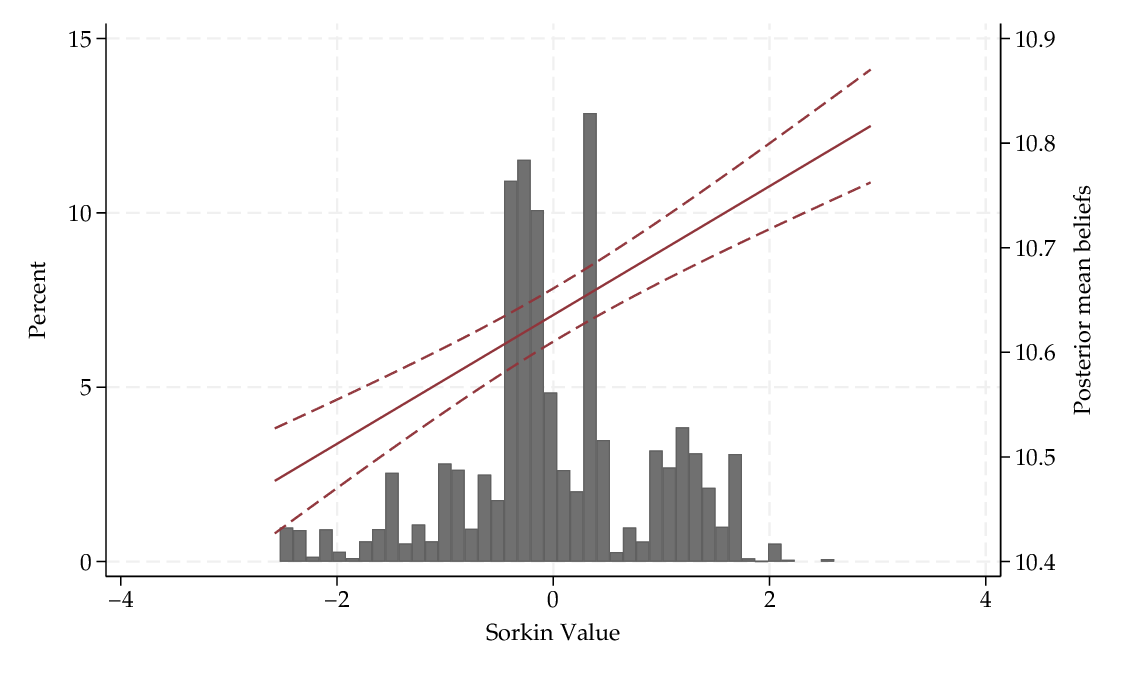}
        \caption{Sorkin Value}
        \label{fig:baseline_beliefs_biasC}
    \end{subfigure}
    %\hfill
    \begin{subfigure}[t]{0.48\textwidth}
        \centering
        \includegraphics[width=\linewidth]{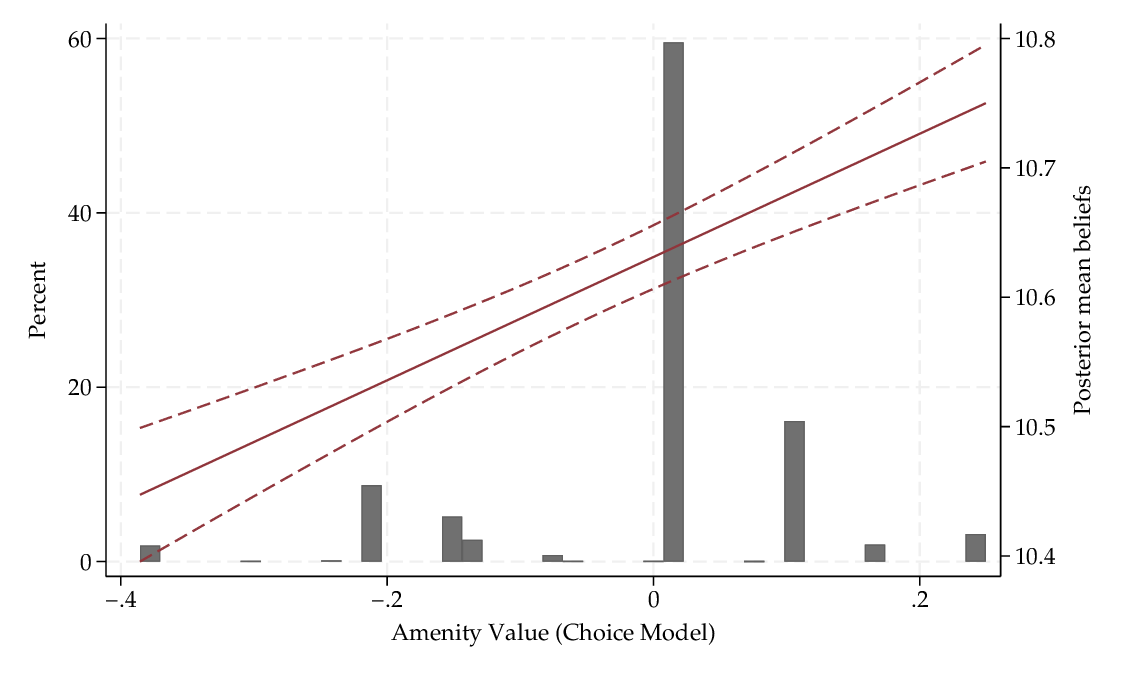}
        \caption{Amenity Value}
        \label{fig:baseline_beliefs_biasD}
    \end{subfigure}
    \end{center}
    \vspace{-0.25em}
    
    \scriptsize{
    \emph{Notes:} This figure shows the distributions (gray histograms) of job quality measures; panel (a) shows actual pay, as measured by the starting salaries for recently hired workers in the position listed in each job ad, panels (b)-(c) show employer pay premium and employer Sorkin value associated with the employer posting the ad, respectively, estimated using matched employer-employee data \citep{audoly24}, and panel (d) shows amenity value associated with each job ad, calculated using the WTPs listed in Table \ref{tab:willingness_to_pay}, Column (5), and information on mentions of workplace amenities detected using the text-analysis approach of \cite{audoly24}. The densities shown in percent along the left y-axis correspond to the value of each job quality measure shown along the x-axis, using the sample of job ads that were shown to the control group in Stage 3 of the survey; see Figure \ref{fig:surveyDesign}. In each subplot, we also show a linear projection (red line) based on a regression of posterior mean pay belief in logs, $\mathbb{E}_i[\tilde{w}_{j}]$, on each job quality measure, with the associated 95\% confidence intervals. Beliefs have been corrected for uncertainty bias; see Section \ref{sec:beliefs}; as estimated using the two-equation model \eqref{eq:meanvariancemodel}--\eqref{eq:meanvariancemodel2}, and the confidence intervals account for this using a product bootstrap over respondents and jobs. Observations outside the 1st and 99th percentile of each distribution are dropped.
    }
\end{figure}

Having established that respondents' gender and current pay correlate strongly with their pay beliefs, we now turn to how beliefs are associated with job characteristics present in the ad. In Figure \ref{fig:baseline_beliefs_bias}, we report the distributions of alternative job characteristics for the sample of job ads viewed by our respondents, along with the linear projection of their posterior mean pay beliefs in logs, $\mathbb{E}_i[\tilde{w}_{j}]$, on each of these measures. In Panel A, we plot the relationship between the posterior mean pay belief and the actual pay, where we measure the latter as the starting pay for recently hired worker(s) in the position(s) listed in the job ad. As expected, the slope of the linear projection shown in the red line indicates that respondents' pay beliefs are positively associated with actual pay. Further, in Panel B, we provide associations between respondents' pay beliefs and employer pay premiums, which we estimated using population-level matched employer-employee data, closely following the approach in \cite{audoly24}. Employer pay premiums capture systematic pay differentials across firms posting the job ads in our survey, where the pay differentials captured by fixed worker-specific components are purged \citep{abowd1999high}. We again find a strong association between respondents' beliefs about pay in jobs and employer pay premiums. Put simply, survey respondents expect high-paying jobs to pay more.

Next, in Panel C, we consider a standardized job quality measure formalized by \cite{sorkin2018ranking}. Intuitively, this approach uses workers' job-to-job transitions to infer a revealed-preference measure of job quality. Notably, the Sorkin measure uses job flows rather than pay premiums, capturing systematic differences across jobs in the overall valuation. We estimate the Sorkin job value measure using population-level matched employer-employee data as in \cite{audoly24}, and provide associations between this measure and respondents' pay beliefs. We find that this job quality measure is also positively associated with respondents' pay beliefs, consistent with the earlier evidence that relies solely on pay information.

In Panel D, we instead use information on workplace amenities that were explicitly posted by employers in the job ad texts viewed by respondents. Specifically, we use the text-analysis methodology of \cite{audoly24} to detect mentions of workplace amenities in each job ad and use the willingness-to-pay estimates from our hypothetical choice model in Table \ref{tab:willingness_to_pay}, Column 5, to construct a composite measure of amenity value associated with each job.\footnote{For the construction of the composite amenity value, we do not use information on travel time as, unlike the other four non-pecuniary job attributes, it depends also on the respondent's residential location.} Respondents' pay beliefs are also positively related to this `posted' amenity value. This suggests that mentions of workplace amenities inform respondents' pay expectations. Specifically, respondents expect jobs offering better amenities to also pay higher salaries.

The finding that pay beliefs are positively correlated to advertised amenities may appear to be at odds with standard models of compensating differentials, where one may expect workers to be willing to accept lower pay for better-amenity jobs. Notably, however, the associations uncovered here reflect respondents' beliefs, which may not necessarily align with the actual associations between pay and amenities. In Appendix~\ref{sec:appendix_belief_model}, we provide a stylized Bayesian model of belief formation that allows us to pin down what we should expect this comparison to look like under rational inference: under correctly-specified priors, the slope of beliefs on advertised amenities should equal the OLS slope of actual pay on the same amenities. We return to the empirical comparison of these two slopes below. The presence of search frictions can, however, lead to augmenting rather than compensating differentials \citep{hwang1998hedonic}. Further, as emphasized by \cite{audoly24}, the texts of job ads may not contain the total amenity value associated with jobs, and there may still exist compensating differentials with respect to the non-advertised (``intrinsic'') amenities.

Further, in Table \ref{tab:baselinebeliefs}, we show that the patterns we reported above are robust to controlling flexibly for occupation-by-sector fixed effects (corresponding to the position listed in the job ad) in Column 2, respondent fixed effects in Column 3, or both in Column 4. In Panel A, we show that the positive association between respondents' pay beliefs and the Sorkin job value measure also carries over to another flow-based job quality measure--the poaching index--formalized by \cite{bagger2019empirical}. In Table \ref{tab:baselinebeliefs}, Panel C, we show that our conclusions are unchanged if we use alternative WTP measures, as found in the literature or directly reported by the respondents (Table \ref{tab:willingness_to_pay}, Columns 1-2, respectively), to construct the composite job amenity value, as opposed to using the WTPs estimated in our choice model (Table \ref{tab:willingness_to_pay}, Column 5). In Panel D, we show associations between respondents' pay beliefs and mentions of workplace amenities from job ad texts, finding that respondents expect higher salaries in permanent jobs and jobs offering the possibility to work flexible hours, while they expect lower salaries for jobs involving shift work. These results confirm the positive associations between mean pay beliefs and composite job amenity values reported in Panel C. While the evidence thus far relies on the model \eqref{eq:meanvariancemodel}-\eqref{eq:meanvariancemodel2} to correct for uncertainty bias, we provide robustness to alternative specifications in Appendix Table~\ref{tab:robustness_baselinebeliefs}, such as using measures of expected pay in levels or not correcting for the posterior variance. We also provide robustness to alternative checks for respondent inattention. Our findings remain unchanged across all specifications. Moreover, as shown in Appendix Table \ref{tab:baselinebeliefs_var}, Panels C-D, the variance of respondents' pay beliefs is negatively related to amenities. This suggests that not only are signals about workplace amenities in job ad texts associated with respondents' pay levels, these signals also correlate with the precision of their beliefs.

\begin{table}
 \begin{center}
\caption{Correlates of Pay Beliefs: Job Quality, Actual Pay, and Amenity Value.}
\label{tab:baselinebeliefs}
\renewcommand{\arraystretch}{0.95}
\input{export_tsd/baselinebeliefs_mean.tex}
\renewcommand{\arraystretch}{1}
\end{center}
{\scriptsize \emph{Notes:} This table shows the coefficients from regressions of respondent $i$'s posterior mean pay belief for job $j$ in logs, $\mathbb{E}_i[\tilde{w}_{j}]$, on job characteristics, as listed in each row header. Panels A-C provide coefficients from separate regressions for each job measure, while Panel D provides coefficients from a joint regression including all workplace (dis)amenities as explanatory variables (controlling for travel time). Column (2) adds controls for respondent fixed effects, Column (3) adds controls for occupation-by-sector fixed effects, corresponding to the position listed in the job ad, while Column (4) adds both sets of fixed effects. Standard errors are reported in parentheses, estimated by two-way cluster bootstrap over respondent and job ad, accounting for the estimation of posterior variance. 
Posterior mean beliefs in logs have been corrected for uncertainty bias (see Section \ref{sec:beliefs}), and estimated using the two-equation model \eqref{eq:meanvariancemodel}-\eqref{eq:meanvariancemodel2}. $^{*} p<0.1$, $^{**} p<0.05$, $^{***} p<0.01$.
}
\end{table}

\begin{table}
\begin{center}
\caption{Pay Beliefs and Actual Pay: Levels, Gaps and Correlates.}
\label{tab:bias}
\renewcommand{\arraystretch}{0.95}
\input{export_tsd/bias.tex}
\renewcommand{\arraystretch}{1}
\end{center}
{\scriptsize \emph{Notes:} The first row in Columns (1)-(2) of Panel A reports the mean values across respondent-job ad pairs of each respondent $i$'s posterior mean pay belief for job $j$ in logs, $\mathbb{E}_i[\tilde{w}_{j}]$. See the notes below Table \ref{tab:baselinebeliefs} for details on the construction of this measure. Columns (3)-(4) instead report the mean values of the actual full-time equivalent starting salaries (i.e., the actual pay) for recently hired workers in the positions corresponding to those posted in the job ads in job $j$ viewed by respondent $i$. The second row in Panel A reports the mean gap in these measures within each choice scenario, sorted by the job with the highest actual pay versus the lowest actual pay. Columns (1)-(2) of Panels B-C report the coefficients from regressions of respondent $i$'s posterior mean pay belief for job $j$ on job amenity (value) measures, as listed in each row header. These coefficients are identical to those reported in Table \ref{tab:baselinebeliefs}, Columns (1) and (3), Panels C-D. Columns (3)-(4)  of Panels B-C instead report the coefficients from regressions of the actual pay on the same set of explanatory variables, estimated in the same set of job ads (as seen in the survey to reflect the population of jobs in the survey). Standard errors are reported in parentheses, estimated by two-way cluster bootstrap over respondent and job ad, accounting for the estimation of posterior variance for the variables in Columns (1)-(2), and by regular standard errors clustered by job in columns (3)-(4).. $^{*} p<0.1$, $^{**} p<0.05$, $^{***} p<0.01$.
}
\end{table}

We now return to the positive associations between respondents' beliefs about starting salaries in the job ads they viewed and workplace amenities listed in the corresponding ad text, and ask whether these associations are reflected in actual starting salaries for recently hired workers in the positions corresponding to the job ads. As shown in Table \ref{tab:bias}, Panel B, Column 3, we find some evidence that the amenity values associated with advertised workplace amenities in job ad texts are also positively associated with the actual pay. However, these associations are much weaker than those found using respondents' pay beliefs in Columns 1-2, and become statistically insignificant when we control for occupation-by-sector fixed effects in Column 4. Turning to the individual workplace amenities in Panel C, we find that jobs involving shift work also appear to pay less. Taken together, these findings echo the lack of evidence of compensating differentials often found in observational studies \citep{bonhomme2009}, which may reflect the strong presence of search frictions that lead to augmenting as opposed to compensating differentials \citep{hwang1998hedonic}, or simply reflect that jobs differ along other dimensions correlated with pay (i.e., omitted variables bias). Importantly for our interpretation, however, the slope of beliefs on advertised amenities is not just larger than the slope of actual pay on advertised amenities, it is in itself inconsistent with rational Bayesian inference: as we show in Appendix~\ref{sec:appendix_belief_model}, under rational inference with correctly-specified priors the two slopes should coincide. The gap we estimate is therefore evidence of either an upward-biased prior over the pay-amenity gradient or systematic over-extrapolation from amenity signals (or both), independent of whether the underlying gradient in actual pay is positive, zero, or negative.

Finally, we note that respondents expect starting salaries in the job ads they viewed to be on average around 18\% lower than the actual starting salaries of workers recently hired in these jobs. This average ``level bias'' in respondents' pay beliefs is reflected in the difference between the mean values reported in the first row of Panel A in Table \ref{tab:bias}, across Columns 1-2 and 3-4, respectively. Consistent with \cite{christensen2026relative}, we find that respondents substantially underestimate differences in pay between the highest and lowest paying jobs within occupation-sector-level choice scenarios. As reported in the second row of Panel A in Table \ref{tab:bias}, while the average actual highest-to-lowest pay gap is about 21\%, there is virtually no pay belief gap. The large difference between pay beliefs and actual pay indicates that there is misperception about the pay levels and gaps among our survey respondents. The ``level bias'' could also reflect pay anchoring, as most respondents are relatively young and currently employed in jobs that pay substantially lower than high-skilled service sector jobs that were included in our survey.\footnote{Notably, the average current full-time equivalent monthly salary (in logs) of employed respondents in our survey is 10.34, which is substantially lower than the corresponding salary of 10.81 for recently hired workers in the jobs viewed by respondents in our survey. This difference in current salary and actual pay in viewed jobs, combined with the pass-through from current pay on pay beliefs at 0.116 reported in Table \ref{tab:baselinebeliefs_p}, implies 5.5\%-points lower pay beliefs, accounting for 30\% of the ``level bias'' in respondents' pay beliefs.} This also indicates that despite being partly attached to the labor market early in their career, many respondents have misperceptions about the pay levels in jobs that are likely relevant to them given their fields of study.

\section{The Effects of Pay Information in Job Ads}\label{subsec:jobchoices}

In the previous section, we provided evidence on pay beliefs for a random subsample of our survey respondents--the control group--finding that these respondents on average expected starting salaries in the job ads they viewed to be lower than the actual starting salaries of workers recently hired in these jobs. Our evidence also indicated systematic dispersion in pay beliefs across respondents, with males and those with higher current salary stating higher pay beliefs. Further, respondents expected jobs offering better amenities to be high-paying, overstating the positive associations found in observational data. These findings indicated the presence of misperceptions about starting salaries among our survey respondents. We now seek to understand how the provision of explicit pay information in job ads can shape respondents' beliefs about pay levels in other similar jobs. In particular, we study whether alternative pay information interventions can affect the dispersion in pay beliefs, and in turn, job choices and the tradeoffs between pay and workplace amenities.

\subsection{Pay Information Intervention}\label{subsec:intensity}

Our experimental design allows us to uncover the impacts of two distinct information treatments: a short-term pay information intervention (``learning treatment'') and a persistent pay transparency intervention (``full information treatment''). The learning treatment seeks to capture the effects of providing information on starting salaries for some jobs on future pay beliefs and preferred choices among other similar jobs. The full information treatment instead seeks to capture the effects of having information on posted starting salaries on all choice alternatives that are under consideration by the agent in an ongoing search process. To study the impacts of these treatments, we use responses from the final two stages of our survey. For the learning treatment, we focus on stages where respondents stated their beliefs about starting salaries in different jobs as well as their preferred job alternatives, comparing responses from the treatment and control groups. For the treatment group, pay beliefs were elicited in Stage 4, while for the control group, pay beliefs were elicited in Stage 3. Notably, before we elicited pay beliefs for the treatment group, these respondents had already seen a series of job ads with explicit information about pay in Stage 3. We utilize this feature of our survey design--randomization of the sequence of survey modules--to study the effects of the learning treatment. For the full information treatment, we instead use information on the treated group's preferred job alternatives in Stage 3 where they received explicit pay information, and compare the treated group's choices to those made by the control group that lacked explicit pay information. We refer to Section \ref{subsec:design} for further details on the design.

\begin{table}
\begin{center}
\caption{Randomization Tests.}
\label{tab:randomization}
\vspace{-0.3cm}
\renewcommand{\arraystretch}{0.92}
\input{export_tsd/randomization.tex}
\renewcommand{\arraystretch}{1}
\end{center}
\vspace{-0.3cm}
\par \scriptsize { \emph{Notes:} This table shows randomization tests of respondent and job characteristics across our treatment- and control groups, defined based on the sequence of exposure to pay information, as illustrated in Figure \ref{fig:surveyDesign}. Columns (1)-(2) show means and standard deviations for the control- and treatment group. Column (3) shows the difference in means, with standard errors in parenthesis, while column (4) shows coefficients and standard errors from a regression of treatment indicator on each characteristic. The final two rows provide F-statistic and p-value for the joint significance of all characteristics in Panels A-B.}
\end{table}

\paragraph{Balance Tests} We start by providing evidence that the randomization of respondents in the treatment and control groups--effectively exposing some respondents to explicit pay information earlier than others--worked as intended. Table \ref{tab:randomization} provides results from randomization tests, where we assess balance across the treatment and control groups in terms of the respondents' background characteristics (Panel A) and the characteristics of job ads (Panel B) that they viewed. The treatment and control groups are balanced on most of these characteristics, and we are unable to reject the null hypothesis on their joint significance.\footnote{The respondents used on average about 10.6 minutes on each of the stages involving job ads. In Appendix Table \ref{tab:rob-timeuse}, we show that the time spent on each stage does not vary across the treatment and control groups, once we control for the sequence and type of module. These regressions use responses from Stages 3-4 for both treatment and control groups, effectively allowing us to control for the sequence and type of module.}

The randomization design facilitates comparisons of responses across the treatment and control groups in Stages 3-4 in our survey, and implies that such comparisons are informative about the impacts of our pay information interventions. In several analyses that follow, we perform such comparisons by either splitting our estimation sample based on an indicator for treatment status (e.g., when decomposing the variance of pay beliefs for each group using Equations \eqref{eq:meanvariancemodel}-\eqref{eq:meanvariancemodel2}) or by performing estimations where we fully interact all parameters with this indicator (e.g., when estimating choice models using Equation \eqref{eq_prob}). Notably, the results from these analyses admit interpretations as standard treatment-control comparisons.

\paragraph{Treatment Intensity} 
In some of the analyses that follow, we also study the effects of pay information intervention on respondent-level outcomes in a regression framework. The core of our pay information treatment was to show treated respondents the actual starting salaries of recently hired workers when they were asked to rank job ads. The information on actual salaries, however, was nominally fixed to the salary rates in December 2023, when we designed our survey experiment, and not adjusted for wage growth across the different survey waves that were carried out between April 2024 and October 2025. This effectively implies that the intensity of pay information treatment diminished over time.

To account for the variation in treatment intensity across the different survey waves, we construct a treatment intensity measure that allows us to scale the treatment. Notably, the scaled treatment effect estimates rely not only on standard treatment-control comparisons, but also utilize variation in treatment intensity \textit{within} the treatment group. Over the period of our survey, wage growth was approximately 9\%. This wage growth implies that the real wages displayed in our job ads gradually declined over time across the different survey waves, as illustrated by the blue line in Appendix Figure \ref{fig:intensity}. Meanwhile, using the same wage growth deflator on respondents' pay beliefs, we find no trend in growth in posterior mean pay beliefs across survey waves, as illustrated by the flat red line in Appendix Figure \ref{fig:intensity}.\footnote{This evidence is consistent with respondents reporting their current pay beliefs in real time, while the actual pay levels displayed in job ads in our survey were nominally fixed to December 2023 salary rates.} To account for the variation in pay information treatment over time, we thus construct a measure of treatment intensity, $S_{t}=1-\frac{k_t}{\bar{w}-\bar{b}}$, where $\bar{w} \equiv  \overline{\log W_j}$  is the average actual pay in job ads shown to the treatment group (December 2023 salary rates) and $\bar{b} \equiv \overline{\mathbb{E}_i[\tilde{w}_{j}]}$ is the average value of posterior pay belief in the control group (deflated to December 2023). We define $k_t \equiv  \log K_t $ as the general wage index relative to December 2023. The treatment intensity $S_{t}$ measures how much our pay information treatment depreciated due to aggregate wage growth, relative to the initial disparity between the average displayed salary in job ads and the average deflated value of posterior mean pay beliefs for the control group.

To estimate the effects of pay information treatment on respondents' pay beliefs, we estimate the following regression model that allows for variable treatment intensity:

\begin{equation}
\label{eq:intensity}
y_{ij} = \alpha + \beta T_i\times S_{t(i)} + e_{ij},
\end{equation}

where $T_i$ is the randomized treatment indicator, $S_{t(i)}$ is the measure of treatment intensity (as defined above) that varies by the time of survey $t(i)$ across respondents. Depending on the regressions we run, $y_{ij}$ denotes either respondent $i$'s posterior mean pay belief, $\mathbb{E}_i[\tilde{w}_{j}]$, or the variance of pay belief, $\mathrm{Var}_i[\tilde{w}_{j}]$, for job $j$. Notably, the treatment intensity $S_{t(i)}$ can vary across treated respondents that entered our survey at different times $t(i)$, but the variation in $S_{t(i)}$ across $t$ comes solely from the aggregate wage growth index $k_t$. We estimate standard errors by two-way product bootstrap over respondent and job ad, which accounts for the uncertainty in the estimation of posterior mean and variance of pay beliefs.

\subsection{Learning Effects on Pay Beliefs}\label{subsec:treat_belief}

\begin{figure}[t!]
\caption{The Effects of Pay Information Intervention on Variance of Pay Beliefs.}
\includegraphics[width=0.9\textwidth]{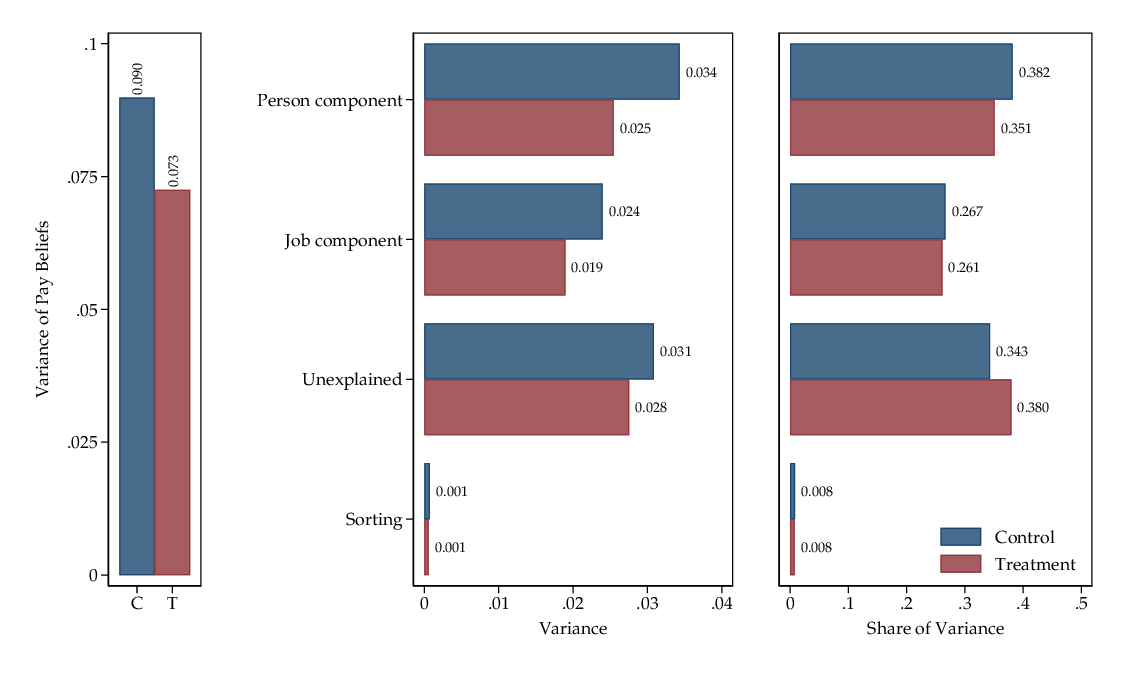}
\label{fig:variancedecomp}
\vspace{-0.5cm}
\par \scriptsize{ \emph{Notes:} This figure shows a decomposition of the variance components of respondents' posterior mean pay beliefs, separately for the treated (red) and control (blue) groups, using data on pay beliefs collected in Stage 4 for the treatment group and in Stage 3 for the control group (see Figure \ref{fig:surveyDesign}). Respondents' posterior mean pay beliefs are estimated using the two-equation model \eqref{eq:meanvariancemodel}-\eqref{eq:meanvariancemodel2}. The decomposition relies on the additive structure in Eq. \eqref{eq:meanvariancemodel}, with the variance components listed in Eq. \eqref{eq:flex}.}
\end{figure}

We start by providing evidence on how the pay information intervention affected the dispersion of pay beliefs, i.e., the learning treatment. Using the additive structure in Equations \eqref{eq:meanvariancemodel}-\eqref{eq:meanvariancemodel2}, we decompose the variance of pay beliefs in components that can be attributed, respectively, to respondents and jobs, as in Equation \eqref{eq:flex}, separately for the treatment and control groups. Figure \ref{fig:variancedecomp} shows the results from this variance decomposition. The left panel shows that the overall variance of pay beliefs is lower in the treatment group than the control group, indicating that pay beliefs are more aligned after the pay information intervention. The middle panel shows that there is lower variance in components associated with persistent person and job characteristics, as well as in the unexplained component. Taken together, this evidence indicates that there is less systematic dispersion in pay beliefs among treated respondents, and their pay beliefs are more aligned for similar jobs. In the right panel, we show that it is the person component that has the largest relative reduction in dispersion.

\begin{figure}[tb]
    \begin{center}
    \caption{The Distributional Effects on Posterior Mean Pay Beliefs.}
    \includegraphics[width=0.7\textwidth]{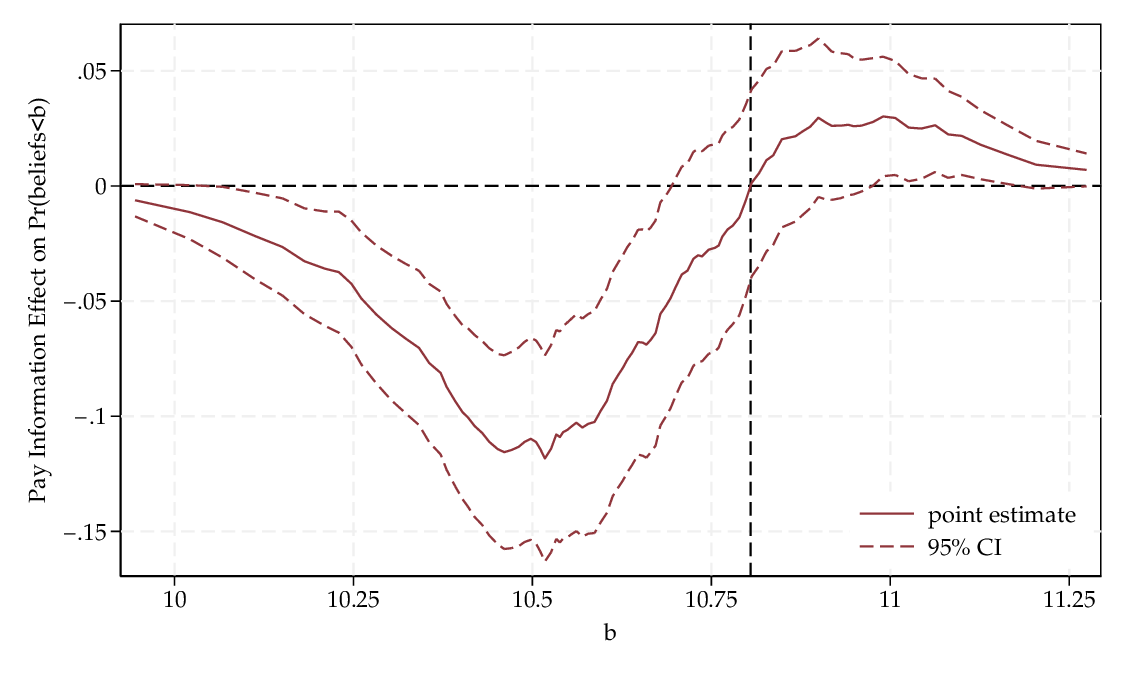}
    \label{fig:distributionEffect}
    \end{center}
    \vspace{-1em}    
    \par \scriptsize{
\emph{Notes:} This figure shows coefficient estimates from a series of estimations of the regression model in Eq. \eqref{eq:intensity} using indicators, $1[\mathbb{E}_i[\tilde{w}_{j}]<b]$, defined for each value of $b$ that spans the empirical distribution of pay beliefs as outcomes. Respondents' posterior mean pay beliefs, $\mathbb{E}_i[\tilde{w}_{j}]$, adjust for the uncertainty bias and are estimated using the two-equation model in Eqs. \eqref{eq:meanvariancemodel}-\eqref{eq:meanvariancemodel2}. Shaded lines show the joint 95\% confidence interval.}
\end{figure}

Next, we study the impacts of pay information intervention across the distribution of pay beliefs. As reported in the first row of Table \ref{tab:bias}, we recall that the average posterior mean pay belief among control respondents was at 10.63 (in logs), while the average actual starting salary of workers recently hired was 10.81 (in logs), indicating that the average control respondent expected salaries to be 18\% lower than what the actual starting salaries turned out to be. However, within this group, some respondents expected salaries to be higher (``positive bias'') while others expected salaries to be lower  (``negative bias'') than the actual salaries. To study how the pay information intervention affects the distribution of pay beliefs, we estimate the regression model \eqref{eq:intensity} using a series of indicators as binary outcomes, $1[\mathbb{E}_i[\tilde{w}_{j}]<b]$, defined for each value of $b$ that spans the empirical distribution of pay beliefs. Figure \ref{fig:distributionEffect} plots estimates of the resulting $\beta$ coefficients from these regressions, along with the associated 95\% confidence intervals, for different values of $b$ shown along the horizontal axis.

To help interpret the evidence shown in Figure \ref{fig:distributionEffect}, we add a vertical line at $b=10.81$, which reflects the average value of the salary shown in pay information intervention to treated respondents. The point estimates illustrated in this figure indicate that the probability $p_b \equiv \Pr\!\left[\mathbb{E}_i[\tilde{w}_{j}]<b\right]$ that a treated respondent holds relatively low pay beliefs (``negative bias'') reduces substantially, as compared to the average actual salary. This reduction is clearly visible across the range of values that span $b \in \left[10.1,10.7\right]$, with the strongest reduction at around $b\approx10.6$, which is close to the average posterior mean pay belief among control respondents. Meanwhile, the information intervention also lowers the probability that a respondent holds relatively high pay beliefs (``positive bias''), as compared to the average actual salary. Notably, across the range of values that span $b \in \left[10.9,11.2\right]$, the point estimates shown in Figure \ref{fig:distributionEffect} are clearly positive, indicating that the probability $p_b$ goes up, so the complement probability $1-p_b \equiv \Pr\!\left[\mathbb{E}_i[\tilde{w}_{j}]\geq b\right]$ is lower. This evidence suggests that the information intervention reduced the dispersion in pay beliefs at both ends of the belief distribution, lowering both negative and positive biases in respondents' pay beliefs.

\begin{table}
\begin{center}
\caption{The Effects of Pay Information Intervention on Respondents' Pay Beliefs.}
\label{tab:effects_beliefs}
\renewcommand{\arraystretch}{0.95}
\input{export_tsd/effects_beliefs.tex}
\renewcommand{\arraystretch}{1}
\end{center}
\vspace{-0.5cm}
{\par \scriptsize \emph{Notes:} This table shows the coefficients from regressions of respondent $i$'s posterior mean pay belief for job $j$ in logs, $\mathbb{E}_i[\tilde{w}_{j}]$, in Columns (1)-(3), and respondent $i$'s posterior variance of pay belief for job $j$ in logs, $\mathrm{Var}_i[\tilde{w}_{j}]$, in Columns (4)-(6), on information treatment (see Equation \eqref{eq:intensity}). The estimation sample consists of both treatment and control groups. Panel A reports the baseline treatment effects and associated control group mean values for the dependent variable, Panel B shows interaction effects between the information treatment and respondent gender and current pay, and Panel C further by the amenity value associated with each job ad, calculated using the WTPs from Table \ref{tab:willingness_to_pay}, respectively, from the literature in Columns (1), from the subjective elicitation in Column (2), or from the choice model in Column (5), and information on mentions of workplace amenities detected using the text-analysis approach of \cite{audoly24}. Columns (2) and (5) add controls for job ad fixed effects, while Columns (3) and (6) further add indicators for the survey batch, capturing the time period and study course that respondent participated in during the survey, and controls for respondent background characteristics, as listed in Table \ref{tab:baselinebeliefs_p}. Standard errors are reported in parentheses, estimated by two-way cluster bootstrap over respondent and job ad, accounting for the estimation of posterior variance. Posterior mean beliefs in logs have been corrected for uncertainty bias (see Section \ref{sec:beliefs}), and estimated using the two-equation model \eqref{eq:meanvariancemodel}-\eqref{eq:meanvariancemodel2}. $^{*} p<0.1$, $^{**} p<0.05$, $^{***} p<0.01$.
}
\end{table} 

As noted earlier, the average control respondent expected starting salaries to be 18\% lower than what the actual salaries turned out to be. Consistent with this disparity, we find that the information intervention raised the mean value of respondents' pay beliefs, besides reducing the dispersion in pay beliefs. In Table \ref{tab:effects_beliefs}, Panel A, Columns 1-3, we show respondents' posterior mean pay beliefs increase by around 4\%, which measured relative to the baseline average ``level bias'' in pay beliefs implies an increase by almost 22\%. Earlier, in Table \ref{tab:baselinebeliefs_p}, we showed that among the control group, male respondents held almost 6 percent higher mean pay beliefs and the coefficient on respondents' current pay was around 0.12 and statistically significant. In Table \ref{tab:effects_beliefs}, Panel B, Columns 1-3, we test for interaction effects between pay information treatment and these respondent characteristics. The interaction terms for males and respondents with higher current pay are negative--indicating that males and higher-current-pay respondents update their beliefs less, but neither interaction is statistically significant. Similarly, in Table \ref{tab:baselinebeliefs}, we showed that among the control group, respondents on average expect jobs with better amenities to also pay higher salaries. In Table \ref{tab:effects_beliefs}, Panel C, Columns 1-3, we test for interaction effects between the learning treatment and the amenity value associated with each job ad, finding the interaction terms to be negative, but not statistically significant.

While the evidence thus far considers the mean values of respondents' pay beliefs, it is useful to consider whether the information intervention also affected the precision of their beliefs. Using the additive structure in Equations \eqref{eq:meanvariancemodel}-\eqref{eq:meanvariancemodel2}, we estimate each respondent $i$'s posterior variance pay belief for job $j$ in logs, $\mathrm{Var}_i[\tilde{w}_{j}]$. We then use this as an outcome in the regression model \eqref{eq:intensity}, to study whether the precision of respondents' beliefs differs across the treatment and control groups. The average variance of respondents' pay beliefs for the control group is about 0.026 log points. In Table \ref{tab:effects_beliefs}, Panel A, Columns 4-6, we show respondents' posterior variance of pay beliefs reduces by around 0.004, meaning that the intervention reduced variance of pay beliefs by around 15\%. As for the mean pay beliefs, however, we find no evidence in Table \ref{tab:effects_beliefs}, Panels B-C, Columns 4-6, that the precision changed differentially for male respondents or for respondents with higher current salaries or affected the precision of pay beliefs associated with jobs offering better amenities differently. As for the baseline pay beliefs, the mean and variance of posterior pay beliefs used as outcomes here rely on the model \eqref{eq:meanvariancemodel}-\eqref{eq:meanvariancemodel2} to correct for uncertainty bias. In Appendix Table~\ref{tab:robustness_effects}, we provide robustness to alternative specifications, such as using measures of expected pay in levels or not correcting for the posterior variance. We also provide robustness to alternative checks for respondent inattention. Our findings remain unchanged across all specifications.   

To summarize, the information intervention on average increased respondents' pay beliefs by about 4\%, reduced the dispersion of pay beliefs, lowered the probability that respondents hold either relatively low or relatively high pay beliefs, and increased the precision of their beliefs. In our Bayesian framework detailed in Appendix~\ref{sec:appendix_belief_model}, the level effect implies that the posterior weight respondents place on the disclosed signals is $\omega_\mu \approx 0.22$ (the ratio of the $4\%$ treatment effect to the $18\%$ baseline gap between mean beliefs and actual pay). This is consistent with non-degenerate but moderately tight priors over the unconditional pay level: signals about pay from ten job ad pairs shift the posterior mean by roughly one-fifth. Meanwhile, we do not find statistically significant treatment effect heterogeneity across respondent- or job characteristics available in our data. We note that the absence of statistically significant heterogeneity in treatment effects partly reflects limited power: with less than 500 respondents per treatment arm, the confidence intervals for these interactions are wide, and we cannot rule out economically meaningful effects within these bounds.

\subsection{Impacts on Job Choices and the Amenity-Pay Tradeoff}\label{subsec:wtp_learning}

The evidence thus far indicates that our control respondents on average expected starting salaries to be 18\% lower than what the actual salaries turned out to be, and the pay information intervention on average raised pay beliefs by 4\%. The evidence provided in Section \ref{sec:beliefs} further showed that the control respondents expected jobs offering better amenities to also pay more, and these associations in the respondents' beliefs were much stronger than the observational associations between actual pay and amenities found in our job ads. While the information intervention on average raised pay beliefs and reduced the dispersion of pay beliefs, we found no evidence that it significantly affected the amenity-pay belief associations.

We now provide evidence on how each information treatment---the short-term pay information intervention (``learning treatment'') and the persistent pay transparency intervention (``full information treatment'')---affected respondents' stated job choices and the implied tradeoffs between pay and workplace amenities. In Table \ref{tab:effects_choices}, we focus on differences between the average characteristics of the control respondents' stated choices and the corresponding characteristics of the stated choices of treated respondents in each treatment arm. As described in Section \ref{subsec:intensity}, we scale the treatment effects by the intensity of treatment. The average characteristics of chosen alternatives for the control group without explicit pay information (Column 1) and the treatment group after having previously received explicit pay information (Column 2) are quite similar, suggesting no economically meaningful changes in job choices due to the learning treatment (Column 3). The only statistically significant change is a reduction in the average actual pay in chosen alternatives, though the impact is economically small. By comparison, when we consider the full information treatment (Column 4), we find evidence that having explicit pay information for choices under consideration raises the average actual pay in chosen alternatives by a statistically and economically significant magnitude of 3.8\% (Column 5). While both standardized revealed-preference measures of job quality have positive coefficients, suggesting improvements in job quality following the full information treatment, we are unable to reject null impacts on these outcomes.

\begin{table}[t!]
\begin{center}
\caption{The Effects of Pay Information Intervention on Stated Job Choices.}
\label{tab:effects_choices}
\renewcommand{\arraystretch}{0.90}
\input{export_tsd/effects_realized_jobs}
\renewcommand{\arraystretch}{1}
\vspace{-0.25cm}
\end{center}
\par \scriptsize{
\emph{Notes:} This table provides estimates of the effects of learning treatment (Column 3) and full information treatment (Column 5), respectively, on characteristics of preferred choice alternatives across choices between real job ads in Stages 3-4 of the survey experiment. The learning treatment effect is the difference in mean values of characteristics of alternatives chosen by the treatment group in Stage 4 (mean values reported in Column 2) and the alternatives chosen by the control group in Stage 3 (mean values reported in Column 1), which captures the effect of previously having been exposed to explicit pay information on similar job ads. The full information treatment is the difference in mean values of characteristics of alternatives chosen by the treatment group in Stage 3 (mean values reported in Column 4) and the alternatives chosen by the control group in Stage 3 (mean values reported in Column 1), capturing the effect of having explicit pay information for choices under consideration. The actual pay is reported in logs of monthly starting salaries for each job ad, while the Sorkin Value and Poaching Index are standardized revealed-preference measures of job quality associated with the employer posting each job ad, respectively, estimated using matched employer-employee data \citep{audoly24}, and the amenity value associated with each job ad is calculated using the WTPs listed in Table \ref{tab:willingness_to_pay}, Column (5), and information on mentions of workplace amenities detected using the text-analysis approach of \cite{audoly24}. Treated group means and treatment effects account for the intensity of treatment as discussed in Section \ref{subsec:intensity}. All specifications include fixed effects for occupation-sector and survey batch. Standard errors are clustered by individual and job. $^{*} p<0.1$, $^{**} p<0.05$, $^{***} p<0.01$.
}
\end{table}

How does pay information provision affect respondents' tradeoffs between pay and workplace amenities? To answer this question, we first consider the treated respondents' tradeoffs when they have explicit pay information for choices under consideration. The implied WTPs for treated respondents under this full information treatment are reported in Column 4 of Table \ref{tab:effects_choices_wtp}. Interestingly, these WTPs are close to the WTPs reported in Table \ref{tab:willingness_to_pay}, Column 3, from stylized full-information choice experiments with hypothetical jobs. Next, we consider respondents' \textit{perceived} tradeoffs between expected pay and amenities in settings without explicit pay information.  Specifically, to evaluate the perceived amenity-pay tradeoffs, we estimate the mixed logit choice model \eqref{eq_utility}-\eqref{eq_prob} using information on mentions of workplace amenities listed in job ad texts, daily travel time between respondent residence and workplace locations, and respondents' stated posterior mean pay belief in logs, $\mathbb{E}_i[\tilde{w}_{j}]$, as a measure of their expected pay in each job (as opposed to the actual pay ${w}_{j}$). We estimate such models, respectively, using stated choices for the control respondents without explicit pay information and for the treated respondents who were previously exposed to explicit pay information. In Appendix~\ref{sec:wtp_omitted_q}, we clarify how one should interpret the WTP estimates from choice models with data on pay beliefs and unobserved job quality. Importantly,  even when respondents hold consistent preferences about workplace amenities, the WTP estimates derived from such choice models reflect a combination of true preferences and belief parameters that capture the inferences that respondents make about unobserved job quality.

The estimates of ``implied'' WTPs derived from the control respondents' stated choices without explicit pay information are reported in Column 1 of Table \ref{tab:effects_choices_wtp}. These results suggest that control respondents are willing to give up some of the expected pay for permanent jobs as well as for jobs offering flexible hours, and must be compensated in terms of higher expected pay for jobs involving shift work and longer commute. Using the treated respondents' stated choices in scenarios without explicit pay information yet having prior exposure to pay information, we perform a similar analysis to assess whether the learning treatment affected their amenity-pay tradeoff. The ``implied'' WTPs for the treated respondents who were previously exposed to explicit pay information are reported in Column 2 of Table \ref{tab:effects_choices_wtp}. Consistent with the absence of significant changes in job choices, we do not find any evidence that the learning treatment affected the implied WTPs in Column 3. Meanwhile, consistent with the evidence suggesting that the full information treatment affected respondents' job choices, we do find meaningful differences in the implied WTPs in Column 5. The chi-square statistic reported in the final rows of Column 5 clearly rejects the joint null hypothesis that all five WTPs for the treated and control respondents across Columns 1 and 4 are identical.

\paragraph{Interpretation} 

Taken together, the evidence in Table \ref{tab:effects_choices_wtp} could be interpreted as showing that while the learning treatment did not affect the post-intervention valuations of workplace amenities, the full information treatment appears to have affected valuations of amenities that could affect choices. In particular, when respondents lack explicit pay information--as the controls in Column 1--they appear to undervalue workplace amenities relative to pay, as compared to the valuations derived from choices under the full information treatment in Column 4. These findings raise two questions. First, how can we rationalize that the learning treatment affected pay beliefs, but did not affect the amenity-pay tradeoffs? Second, why do respondents appear to undervalue workplace amenities when they do not have explicit pay information? We discuss both issues in the following.

The Bayesian model in Appendix~\ref{sec:appendix_belief_model} clarifies how to read the asymmetric pattern of changes in pay beliefs, but no changes in the implied amenity-pay tradeoffs. First, the null learning effect on WTPs provides indirect evidence that the slope of the relationship between pay beliefs and unobserved (to the econometrician) job quality is unaffected by our learning treatment. This result is consistent with our earlier findings that the learning treatment shifted belief levels, but not the amenity-pay belief slope, as we showed in Section \ref{subsec:treat_belief}. Second, the learning effects depend crucially on the dosage of information. In particular, pay signals from ten job ad pairs are much more informative about the average pay level than about the pay-amenity gradient, so even moderately tight priors on the gradient deliver a near-zero slope effect under a one-shot intervention. The null effect of the learning treatment on the amenity-pay tradeoff is consistent with belief-formation models where respondents would substantially revise their slope priors under more sustained, more salient, or higher-variance disclosure. The evidence in Table \ref{tab:effects_choices_wtp}, Column 4, provides a natural benchmark for what a full-scale pay-transparency policy enforcing pay disclosure may achieve.

Why do respondents appear to undervalue workplace amenities when they do not have explicit pay information? As noted above, estimates of ``implied'' WTPs from choice models using data on pay beliefs and unobserved job quality should be interpreted with caution. In particular, even when respondents hold consistent preferences about workplace amenities, the WTP estimates derived from such choice models reflect a combination of true preferences and beliefs that capture the inferences that respondents make about unobserved job quality. This implies that the WTPs reported in Table \ref{tab:effects_choices_wtp}, Columns 1-2, should not be interpreted as measures of true preferences, as the elicited pay beliefs also contain respondents' expectations about unobserved job quality. The apparent finding that respondents undervalue workplace amenities when they do not have explicit pay information thus reflects biases in beliefs about the amenity-pay relationship, as opposed to true differences in their valuations of amenities.

\begin{table}[t!]
\begin{center}
\caption{The Effects of Pay Information Intervention on the Amenity-Pay Tradeoffs.}
\label{tab:effects_choices_wtp}
\renewcommand{\arraystretch}{0.90}
\input{export_tsd/treatment_wtp.tex}
\renewcommand{\arraystretch}{1}
\vspace{-0.25cm}
\end{center}
\par \scriptsize{
\emph{Notes:} This table provides alternative estimates of implied willingness to pay (WTP) for five non-pecuniary workplace attributes (see details in Table \ref{tab:willingness_to_pay}), from estimation of the mixed logit choice model \eqref{eq_utility}-\eqref{eq_prob} fully interacted with pay information status and treatment status (see Figure \ref{fig:surveyDesign}). Column 1 reports mean WTPs for the control group based on their choices between real job ads without explicit pay information in Stage 3, while Column 2 reports mean WTPs for the treatment group based on such choices in Stage 4 after previously being exposed to pay information for similar job ads. Column 4 reports mean WTPs for the treatment group based on their choices between real job ads with pay information in Stage 3. Column 3 reports the effects of learning treatment of previously being exposed to pay information, as measured by the difference between the estimates reported in Column 2 (treatment) and Column 1 (control). Column 5 reports the effects of full information treatment of having explicit pay information for choices under consideration, as measured by the difference between the estimates reported in Column 4 (treatment) and Column 1 (control). Respondents' stated posterior mean pay belief in logs, $\mathbb{E}_i[\tilde{w}_{j}]$, is used as a measure of expected pay in each job in Columns 1 and 2, while the actual pay ${w}_{j}$ is used as the pay measure in Column 3. Posterior mean beliefs are estimated using the two-equation model \eqref{eq:meanvariancemodel}-\eqref{eq:meanvariancemodel2} and corrected for uncertainty bias. Treated group mean WTPs and treatment effects account for the intensity of treatment as discussed in Section \ref{subsec:intensity}. All models additionally control for a set of further job ad attributes detected using the text-analysis approach as in \cite{audoly24}. Standard errors are reported in parenthesis, and clustered at the respondent level. The chi-square statistic (and associated joint $p$-value) tests the joint null that all five WTP differences between the treated and control respondents are zero. $^{*} p<0.1$, $^{**} p<0.05$, $^{***} p<0.01$.
}
\end{table}

\paragraph{Implications} 

More broadly, our findings may provide a potential mechanism for why much of the empirical research using observational data fails to find evidence in favor of compensating differentials, despite individuals expressing strong preferences for workplace amenities as reflected in their hypothetical choices in experimental settings. Within the Bayesian framework provided in Appendix~\ref{sec:appendix_belief_model}, the empirical patterns we document are consistent with respondents holding either an upward-biased prior over the pay-amenity gradient or applying an over-extrapolating weight to amenity signals when forming pay beliefs. Both mechanisms generate a perceived menu of jobs in which higher-amenity jobs are also expected to pay more, and both can therefore drive a wedge between observational pay-amenity correlations and the amenity-pay tradeoffs that job seekers act on. Such mechanisms may be especially relevant for young and inexperienced job seekers, for whom entry-level job choices can have long-lasting consequences (see, e.g., \cite{KAHN2010303, Oreopoulos2012}). 

Our evidence shows that a one-time pay information intervention did not change the pay belief-amenity gradient among our respondents, yet we found significant gaps between the slope of beliefs on advertised amenities and the slope of actual pay on advertised amenities. Within our Bayesian framework, such gaps are inconsistent with rational Bayesian inference for any group with correctly-specified priors, and thus having beliefs data for broader populations may be informative about whether other groups also exhibit slope-prior bias or over-extrapolation. Our learning treatment effects on the level and dispersion of beliefs may, by contrast, be specific to our sample of respondents whose priors are loose and whose anchor (current pay) is markedly below the displayed-pay benchmark. At the same time, we found that the full information treatment affected choices--with respondents selecting into high-paying jobs--and these choices implied valuations of workplace amenities that are more in line with WTPs from stylized full-information choice experiments with hypothetical jobs.

\section{Conclusion}\label{sec:conclusion}
Workers' choices between jobs depend not only on the underlying values they place on pay and amenities, but also on the beliefs they hold about what jobs actually pay. In labor markets where vacancies routinely omit pay information, these beliefs must be formed under considerable uncertainty and from limited public signals, which are themselves correlated with non-pecuniary job attributes. This paper provides direct empirical evidence on the structure of pay beliefs and their interaction with workplace amenities, using a multi-stage incentivized survey experiment that combines hypothetical choice experiments, elicited beliefs over real job ads, and a randomized pay-information intervention.

We document four sets of facts. First, although our respondents exhibit standard, sizable preferences over workplace amenities---closely aligned both with their self-reported subjective WTPs and with estimates from the broader literature---their beliefs about pay are systematically biased downward relative to actual starting salaries, and exhibit systematic dispersion that is attributable relatively more to persistent person components than job components. Second, respondents expect jobs that advertise better amenities to pay \emph{more}, not less, with the slope of pay beliefs on advertised amenity value much steeper than the analogous slope of actual pay. This pattern of perceived augmenting differentials, rather than compensating differentials, helps reconcile the long-standing tension between sizable stated preferences for amenities and the absence of compensating differentials in observational data. Third, a randomized information treatment that displays credible pay information on real job ads raises mean beliefs, reduces both negative and positive biases, and increases the precision of beliefs---but does not alter the strong positive association between perceived pay and advertised amenities, nor the implied amenity-pay tradeoff in stated choices. Finally, respondents' choices between real job ads with explicit pay disclosure imply strong preferences for workplace amenities, closely tracking those found in full-information choice experiments.

In view of recent pay transparency policies in Europe and the US (e.g., the EU Pay Transparency Directive 2023/970 and similar state-level laws in California, Colorado, and New York), our findings suggest that disclosing pay on a subset of vacancies is sufficient to durably move the level and dispersion of beliefs about pay in similar jobs where pay information is not disclosed. Whether such disclosure also moves the slope of beliefs on advertised amenities---and therefore the amenity-pay tradeoff in choices---depends crucially on the dosage and salience of the disclosure relative to the precision of workers' priors over the pay-amenity gradient. A one-shot pay disclosure is, by construction, substantially more informative about the average pay level than about the pay-amenity gradient. Our null on the amenity-pay tradeoff in stated choices after the learning treatment should therefore be read as a lower bound on what a more sustained or salience-enhancing pay-transparency intervention could deliver, rather than as evidence that pay transparency cannot, on its own, restore the textbook compensating differential in workers' actual job choices. Notably, our evidence confirming strong preferences for workplace amenities from real job ads with explicit pay disclosure may provide a benchmark for what a full-scale pay-transparency policy mandating salary disclosure on all vacancies could plausibly achieve. An important caveat for this interpretation is that pay-transparency policies may trigger important demand-side reactions affecting equilibrium outcomes \citep{cullen2024pay,cullenecma2024}, which in turn may affect the amenity-pay tradeoffs that our evidence does not capture.

In terms of methodological contributions, the two-way decomposition of beliefs developed here---combining a mean pay belief equation in the spirit of AKM, a flexible variance equation, a Jensen correction for log-concavity, and empirical Bayes shrinkage---offers a tractable framework for credibly summarizing belief data with both individual and job-level structure. This framework may be applicable beyond the context of pay beliefs, e.g., in settings where researchers are able to elicit expectations data that admit a two-way person-by-object structure. Our paper can also serve as an example of how survey experiments can embed randomized information interventions in discrete choice settings that combine information from job ads and administrative data, and efficiently use Bayesian sampling to dynamically tailor choice scenarios in surveys constrained by limited duration or respondent inattention.

Several questions remain. The persistent positive amenity-pay belief associations we document---which are inconsistent with rational Bayesian inference and pure level bias as the sole mechanisms---raise the question of whether they reflect an upward-biased prior over the pay-amenity gradient, over-extrapolation from amenity signals, or some combination of the two. While our experimental design demonstrates the robustness of the pattern to a one-time pay information intervention, longer or more salient interventions---such as repeated exposure or salience-enhancing presentation formats---may yield different conclusions. Our study also does not measure respondents' actual labor-market behavior; we focus on elicited beliefs and hypothetical choices, not applications, acceptance decisions, or realized wages. Likewise, the long-run consequences of belief-driven amenity-pay tradeoffs for sorting, mobility, and wage growth remain an open question, particularly for younger workers, for whom early-career job choices have been shown to entail lasting effects on later labor market outcomes \citep{KAHN2010303, Oreopoulos2012}. We view these as natural directions for further work.

\bibliographystyle{ecta}
\bibliography{main}

\clearpage
\appendix

\begin{center}
    \section*{Appendix}
\end{center}

\setcounter{page}{1}
\global\long\def\thepage{[Appendix-\arabic{page}]}

\global\long\def\thetable{A.\arabic{table}}%
\setcounter{table}{0}
\global\long\def\thefigure{A.\arabic{figure}}%
\setcounter{figure}{0}
\global\long\def\theequation{A.\arabic{equation}}%
\setcounter{equation}{0}

\section{Additional Figures and Tables}
\label{sec:appendix_material}

\begin{figure}[ht!]
    \centering
    \caption{Job Choice Scenario With Hypothetical Jobs.}
    \label{fig:appendix_hypothetical_jobs}
        \includegraphics[width=1.1\linewidth]{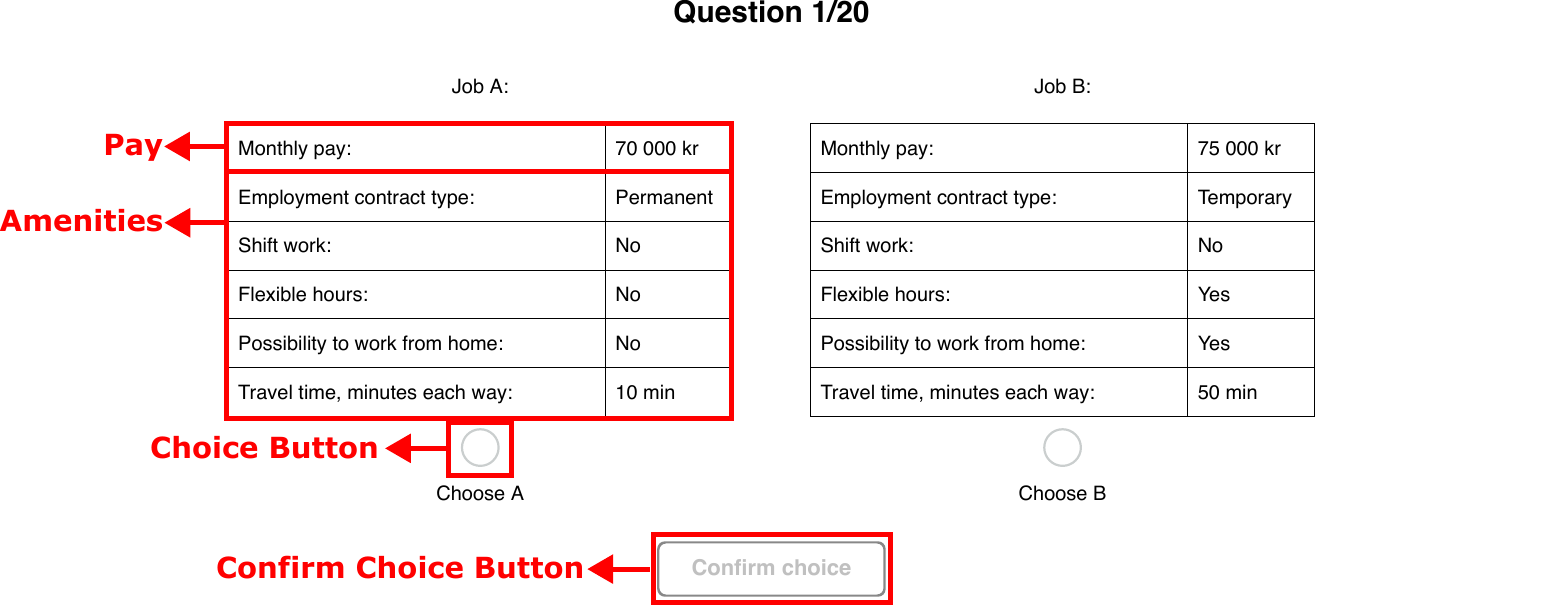}
\end{figure}

\begin{figure}[ht!]
    \centering
    \caption{Information Sheet: Hypothetical Choices}
    \label{fig:information_sheet}

    \begin{subfigure}{\linewidth}
        \includegraphics[width=\linewidth]{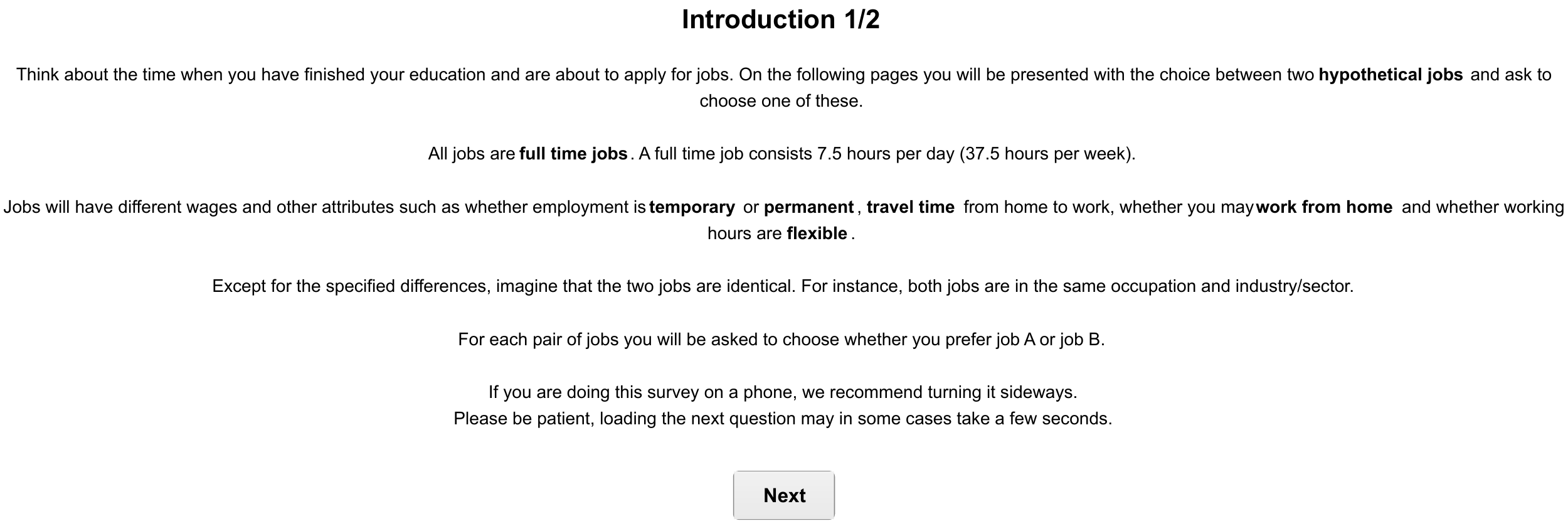}
    \end{subfigure}

    \vspace{1em}

    \begin{subfigure}{\linewidth}
         \includegraphics[width=\linewidth]{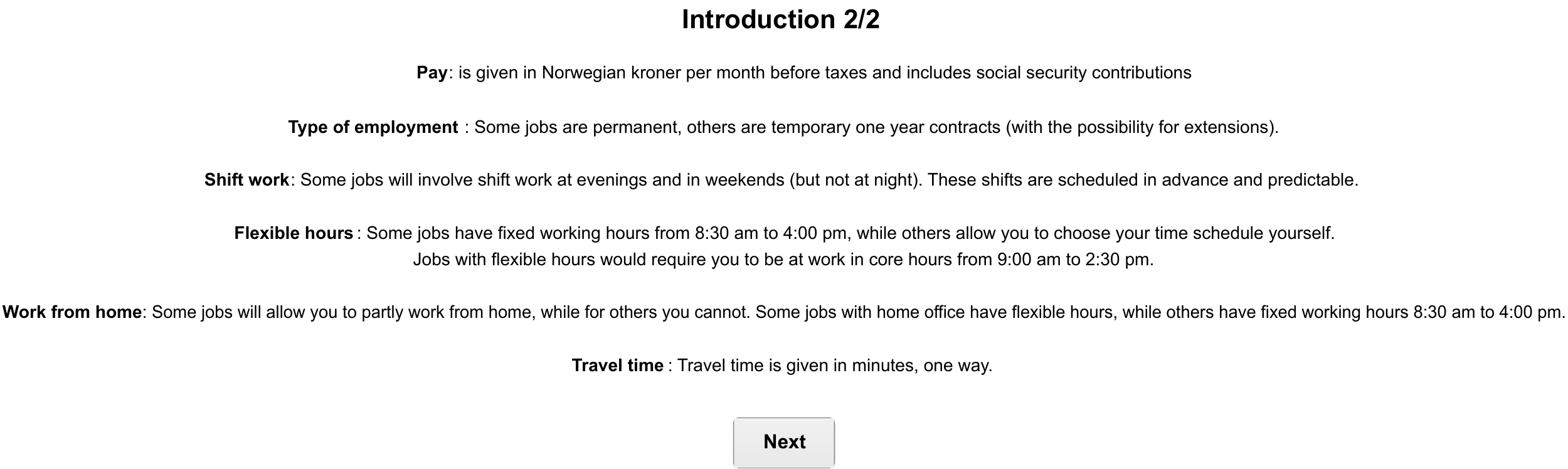}
    \end{subfigure}
\end{figure}

\begin{figure}[ht!]
    \centering
    \caption{Information Sheet: Subjective Willingness to Pay}
    \label{fig:information_sheet_subj}
    \includegraphics[width=0.75\linewidth]{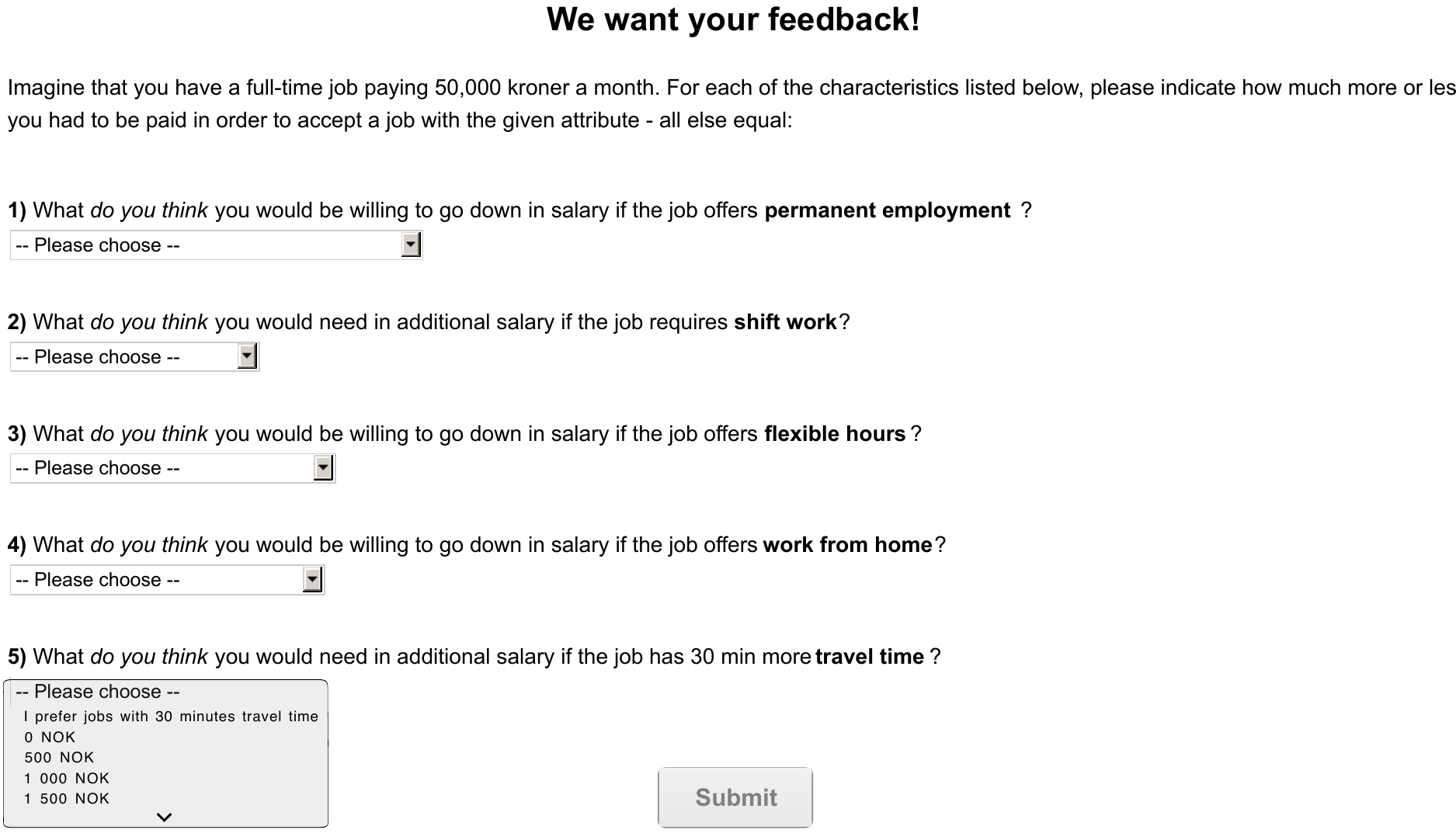}
\end{figure}

\begin{figure}[p!]
    \caption{Experimental Variation in the Provision of Pay Information in Job Ads.} \vspace{-1em}\label{fig:appendix_pay_inf_noinf}
        \begin{center}
        \vspace{1em}
        \subfloat[][Job Choice Scenario With Pay Information ]{\hspace{-1em}\includegraphics[width=1.15\linewidth]{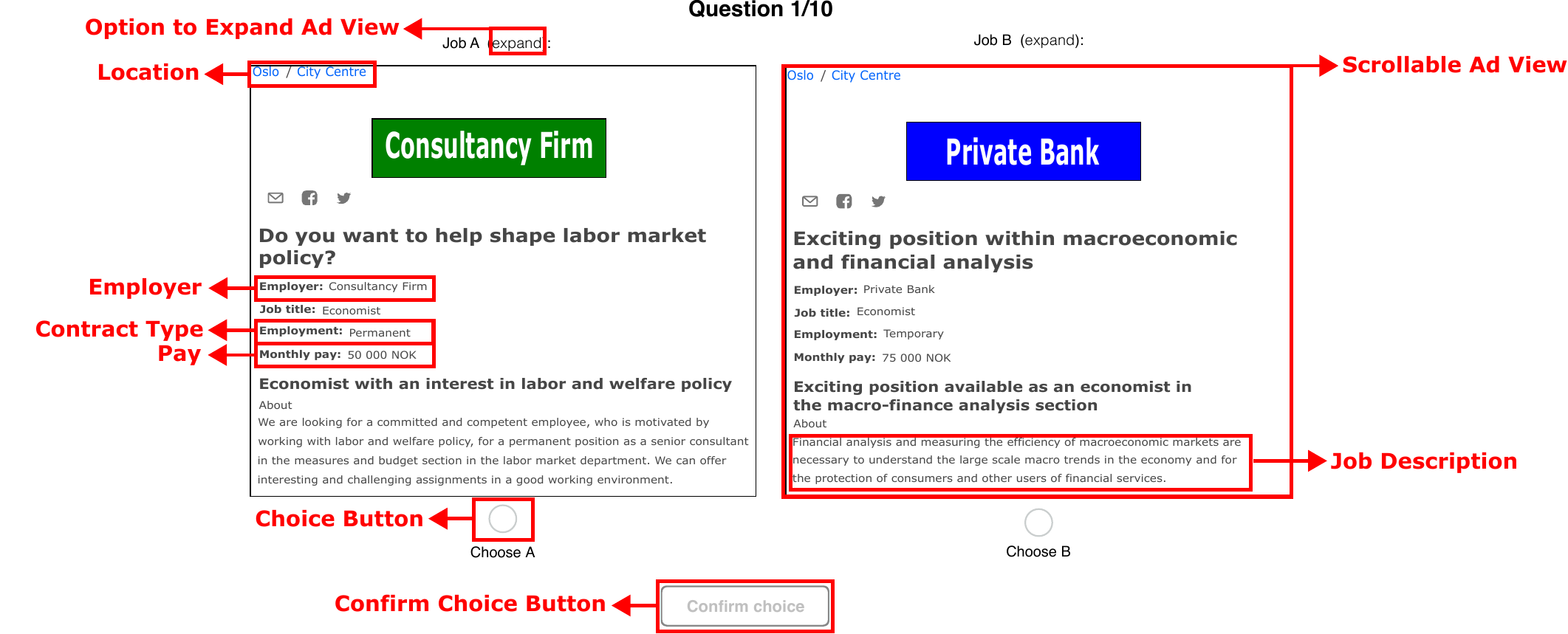}} \\
        \subfloat[][Job Choice Scenario Without Pay Information]{\hspace{-1em}\includegraphics[width=1.15\linewidth]{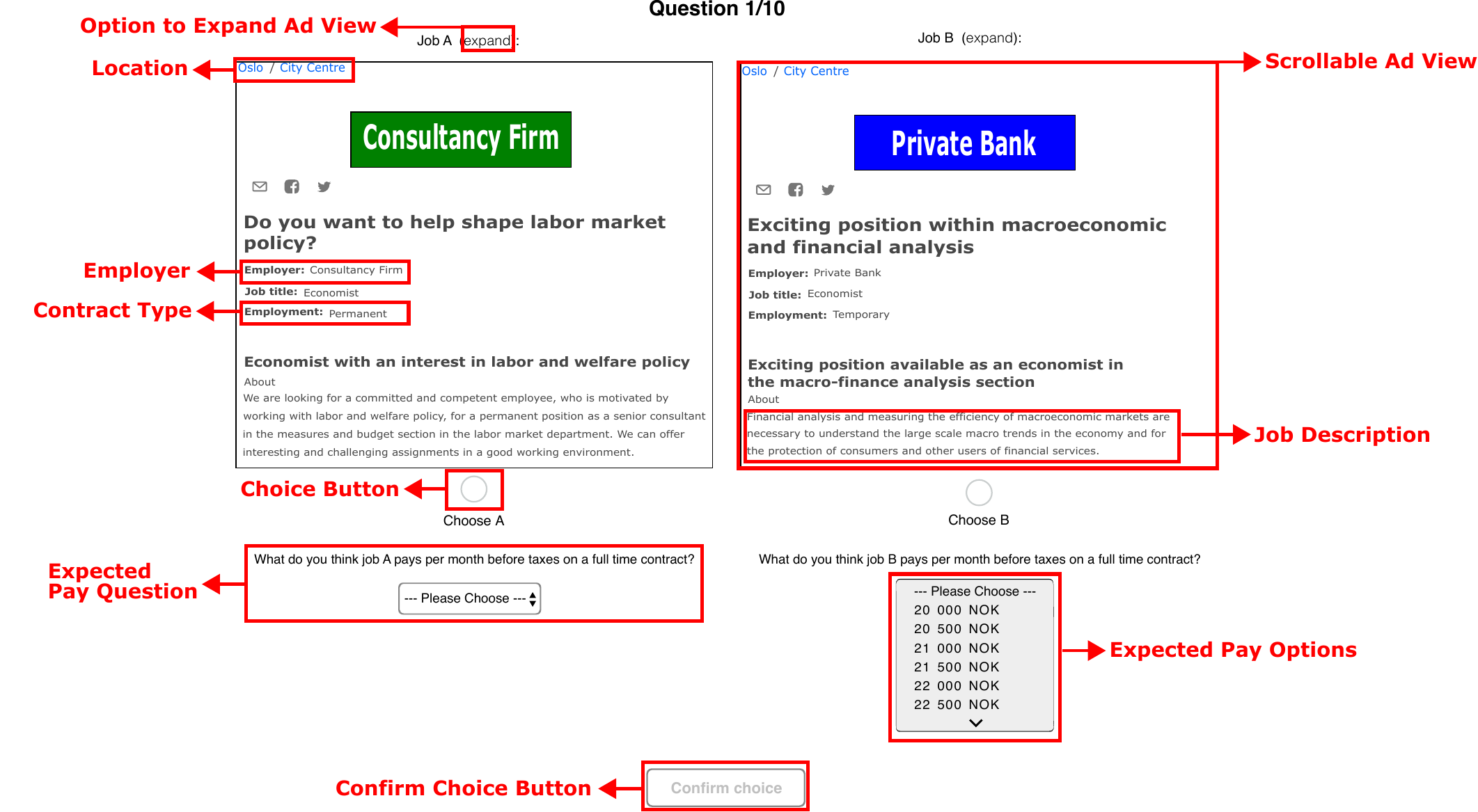}} 
        \end{center}
\end{figure}

\begin{figure}[t!]
        \begin{center}
        \caption{The Distributions of Willingness to Pay for Workplace Amenities.}
        \label{fig:wtp_density}{\includegraphics[width=1\textwidth]{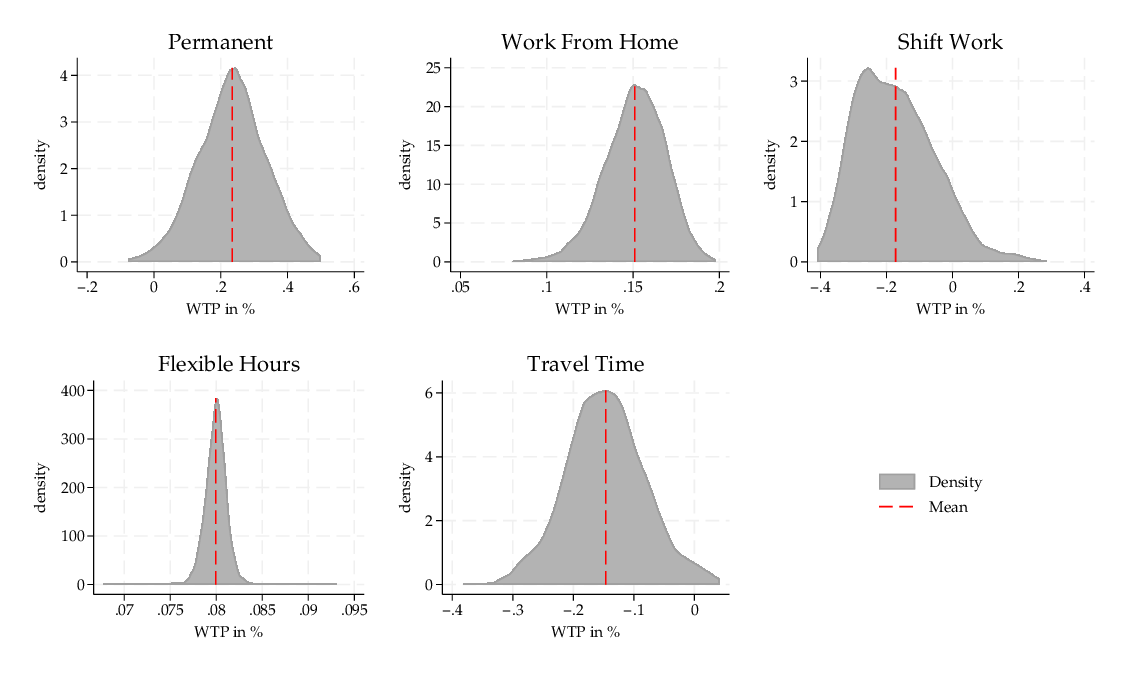}}
        \end{center}
        \par \scriptsize{
\emph{Notes:} This figure shows the distributions of individual-specific expected willingness to pay (WTP) parameters for five non-pecuniary workplace attributes. See notes to Table \ref{tab:willingness_to_pay} for the definition of each attribute. The individual-specific parameters are estimated using a mixed logit model with repeated choices based on the methodology in \cite{revelttrain1998, ReveltTrain2000}, using the combination of choice data collected from choices between hypothetical jobs in Stage 2 and real job ads in Stages 3-4, as illustrated in Figure \ref{fig:surveyDesign}. The corresponding averages of individual-specific expected WTP parameters is provided in Table \ref{tab:willingness_to_pay}, Column 5. Note that y-axis scales differ across the five panels reflecting different concentration of WTP densities.}
\end{figure}

\begin{figure}[t!]
\caption{Illustration: Treatment Intensity by Time Since December 2023.\label{fig:intensity}}
\includegraphics[width=\textwidth]{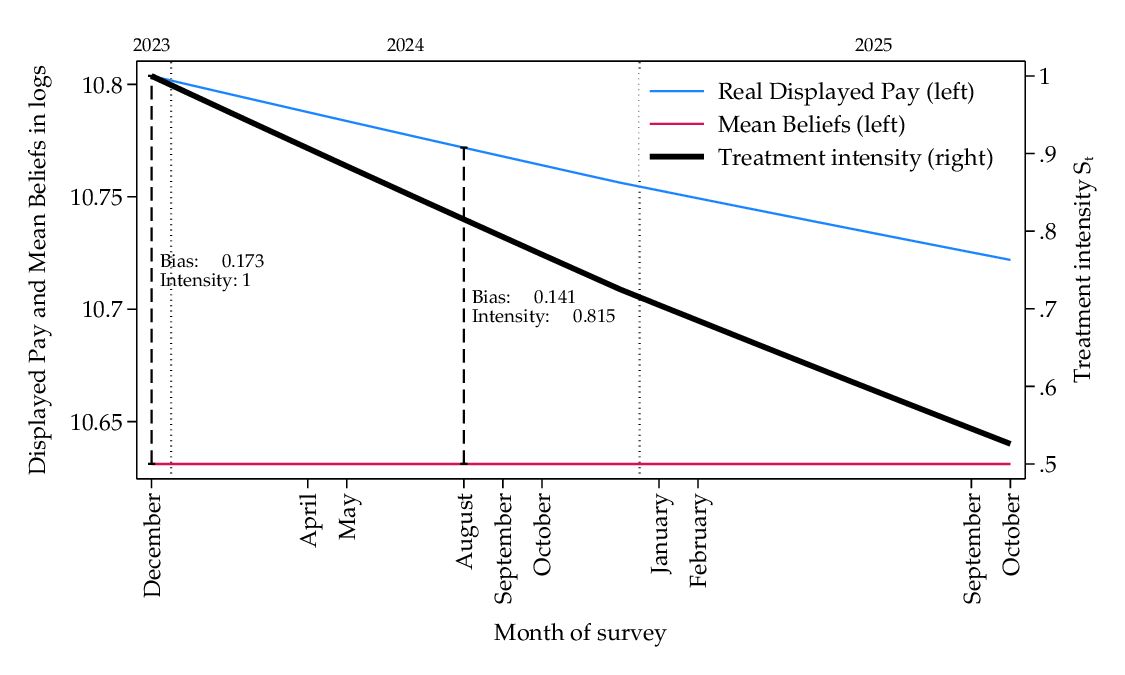}
\scriptsize{\emph{Notes:} This figure displays the variation in treatment intensity in our pay information treatment over the survey period from December 2023 to October 2025. The blue line shows how the average real displayed pay in job ads in our pay information treatment in logs, which was equal to 10.81 in December 2023, depreciated over time due to the aggregate nominal wage growth. The red line captures the average value of posterior pay belief in the control group in logs (deflated to December 2023), which was equal to 10.63. The treatment intensity is $S_{t}=1-\frac{k_t}{\bar{w}-\bar{b}}$, where $\bar{w} \equiv  \overline{\log W_j}$ is the average actual pay in job ads shown to the treatment group (December 2023 salary rates), $\bar{b} \equiv \overline{\mathbb{E}_i[\log\tilde{W}_{j}]}$ is the average value of posterior pay belief in the control group (deflated to December 2023) and $k_t \equiv  \log K_t $ is the general wage index relative to December 2023.}
\end{figure}

\input{results_old/ads_collection}

\begin{table}[tb]
\begin{center}
\caption{Robustness: Differences in Time Use.}
\label{tab:rob-timeuse}
\input{export_tsd/timeuse.tex}
\end{center}
\par \scriptsize{
\emph{Notes:} Table shows regression of time use in minutes for each of the ads modules onto whether the module was the first or second ads module, whether the module contained pay information and a treatment group indicator. Standard errors in parentheses, clustered by individual. $^{*} p<0.1$, $^{**} p<0.05$, $^{***} p<0.01$.
}
\end{table}

\begin{table}[t!]
\centering
\caption{Estimates of Willingness to Pay for Workplace Amenities from the Literature.}
\label{tab:wtpLitA}
\begin{adjustbox}{center}
\scalebox{0.55}{
    \input{results/literatureA}
}
\end{adjustbox}
\end{table}

\begin{table}[t!]
\centering
\caption{Estimates of Relevant WTPs for Workplace Amenities from the Literature.}
\label{tab:wtpLitB}
\begin{adjustbox}{center}
\scalebox{0.7}{
        \input{results/literatureB}
}
\end{adjustbox}
\end{table}

\begin{table}[tb]
\begin{center}
\caption{Correlates of Pay Belief Variance: Actual Pay, Job Quality, and Amenity Value.}
\label{tab:baselinebeliefs_var}
\renewcommand{\arraystretch}{0.95}
\input{export_tsd/baselinebeliefs_var.tex}
\renewcommand{\arraystretch}{1}
\end{center} 
{\scriptsize \emph{Notes:} This table shows the coefficients from regressions of respondent $i$'s posterior variance of pay belief for job $j$ in logs, $\mathrm{Var}_i[\tilde{w}_{j}]$, on job characteristics, as listed in each row header. The dependent variable has a mean value of 0.0264 in the estimation sample. Panels A-C provide coefficients from separate regressions for each job measure, while Panel D provides coefficients from a joint regression including all workplace (dis)amenities as explanatory variables (controlling for travel time). Column (2) adds controls for respondent fixed effects, Column (3) adds controls for occupation-by-sector fixed effects, corresponding to the position listed in the job ad, while Column (4) adds both sets of fixed effects. Standard errors are reported in parentheses, estimated by two-way cluster bootstrap over respondent and job ad, accounting for the estimation of posterior variance using the two-equation model \eqref{eq:meanvariancemodel}-\eqref{eq:meanvariancemodel2}. $^{*} p<0.1$, $^{**} p<0.05$, $^{***} p<0.01$.
}
\end{table}

\begin{table}[tb]
\begin{center}
\caption{Robustness: Willingness to Pay for Workplace Amenities.}
\label{tab:robustness_wtp}
\renewcommand{\arraystretch}{0.95}
\input{export_tsd/robustness_wtp.tex}
\renewcommand{\arraystretch}{1}
\end{center}
{\scriptsize \emph{Notes:} Table shows mean willingness to pay parameters from alternative mixed logit specifications. Column 1 show our baseline results, which includes data from both of our choice experiments on both hypothetical and real jobs. Column 2 allows preferences to be correlated. Column 3 weights individuals with the $1-c_i$, where $c_i$ is the posterior mean estimate that the individual was inattentive (see details in Appendix \ref{sec:appendix_wtp}). Column 4 drops individuals who failed one or more attention checks, defined as scenarios with weakly better amenities and pay, and strongly better for at least one amenity or pay. $^{*} p<0.1$, $^{**} p<0.05$, $^{***} p<0.01$.
}
\end{table}

\begin{table}[tb]
\begin{center}
\caption{Robustness: Correlates of Pay Beliefs.}
\label{tab:robustness_baselinebeliefs}
\renewcommand{\arraystretch}{0.9}
\input{export_tsd/robustness_baselinebeliefs.tex}
\renewcommand{\arraystretch}{1}
\end{center}
\vspace{-0.25cm}
{\scriptsize \emph{Notes:} Table shows robustness of the relationship between posterior mean pay beliefs and job characteristics to accounting for inattention and alternative ways of estimating beliefs. Column 1 show our baseline results. Column 2 weights individuals with the $1-c_i$, where $c_i$ is the posterior mean estimate that the individual was inattentive (see details in Appendix \ref{sec:appendix_wtp}). Column 3 drops individuals who failed one or more attention checks in the hypothetical choice module, defined as scenarios with weakly better amenities and pay, and strongly better for at least one amenity or pay. Column 4 avoids the bias from the lognormal transformation by relating wage beliefs in levels to job characteristics using Poisson regression, while Column 5 simply ignores the problem and relate the log of level wage beliefs to job characteristics.  $^{*} p<0.1$, $^{**} p<0.05$, $^{***} p<0.01$.
}
\end{table}

\begin{table}[tb]
\begin{center}
\caption{Robustness: The Effects of Pay Information on Respondents’ Pay Beliefs.}
\label{tab:robustness_effects}
\renewcommand{\arraystretch}{0.95}
\input{export_tsd/robustness_effects.tex}
\renewcommand{\arraystretch}{1}
\end{center}
{\scriptsize \emph{Notes:} Table shows robustness of the effects of our pay information treatment to accounting for inattention and alternative ways of estimating beliefs. Column 1 shows our baseline results. Column 2 weights individuals with the $1-c_i$, where $c_i$ is the posterior mean estimate that the individual was inattentive (see details in Appendix \ref{sec:appendix_wtp}). Column 3 drops individuals who failed one or more attention checks in the hypothetical choice module, defined as scenarios with weakly better amenities and pay, and strongly better for at least one amenity or pay. Column 4 avoids the bias from the lognormal transformation by relating estimating treatment effects on level wage beliefs using Poisson regression, while Column 5 simply ignores the problem and estimate treatment effects on the log of stated level beliefs.  $^{*} p<0.1$, $^{**} p<0.05$, $^{***} p<0.01$.
}
\end{table}

\begin{table}[tb]
\begin{center}
\caption{Correlates of Pay Belief Variance: Respondent Background Characteristics.}
\label{tab:baselinebeliefs_var_p}
\renewcommand{\arraystretch}{0.95}
\input{export_tsd/baselinebeliefs_var_p.tex}
\renewcommand{\arraystretch}{1}
\end{center}
{\scriptsize \emph{Notes:} This table shows the coefficients from regressions of respondent $i$'s posterior variance of pay belief for job $j$ in logs, $\mathrm{Var}_i[\tilde{w}_{j}]$, on respondent characteristics, as listed in each row. The dependent variable has a mean value of 0.0264 in the estimation sample. Column (2) adds controls for job ad fixed effects, Column (3) adds indicators for the survey batch, capturing the time period and study course that respondent participated in during the survey, while Column (4) adds both sets of fixed effects. Standard errors are reported in parentheses, estimated by two-way cluster bootstrap over respondent and job ad, accounting for the estimation of posterior variance using the two-equation model \eqref{eq:meanvariancemodel}-\eqref{eq:meanvariancemodel2}. $^{*} p<0.1$, $^{**} p<0.05$, $^{***} p<0.01$.
}
\end{table}

\global\long\def\thetable{B.\arabic{table}}%
\setcounter{table}{0}
\global\long\def\thefigure{B.\arabic{figure}}%
\setcounter{figure}{0}
\global\long\def\theequation{B.\arabic{equation}}%
\setcounter{equation}{0}

\global\long\def\thetable{B.\arabic{table}}%
\setcounter{table}{0}
\global\long\def\thefigure{B.\arabic{figure}}%
\setcounter{figure}{0}
\global\long\def\theequation{B.\arabic{equation}}%
\setcounter{equation}{0}

\clearpage

\section{Implementing the BACE Sampling Procedure}
\label{sec:appendix_wtp}

To design the hypothetical choice sets in Stage 2 of our survey experiment, we use the Bayesian
adaptive choice experiments (BACE) methodology proposed by \cite{drake22,drakeetal2023}. This procedure dynamically selects the next scenario to maximize the expected information gain given the posterior distribution of preference parameters using earlier responses. This methodology requires specifying prior distributions over the preference parameters in the population. We calibrate these using a pilot survey we conducted before the main survey and prior evidence from the literature. The priors we used in our BACE implementation are shown in Appendix Figure \ref{fig:priors}. We use truncated log-normal distributions, and intentionally choose priors with somewhat large support to not let the priors restrict the potential scenarios shown too much. We also limit the choice set of the BACE algorithm to avoid certain configurations of the combinations of different job attributes, e.g., jobs that have shift work cannot have flexible hours. Note that these priors affect only the adaptive scenario selection and do not enter the final estimation--our final mixed logit choice model \eqref{eq_utility}-\eqref{eq_prob} is estimated on the realized choice data without imposing any priors.

\begin{figure}[h!]
\begin{center}
\caption{Priors Used in the BACE Adaptive Scenario Selection.}
\includegraphics[width=0.75\textwidth]{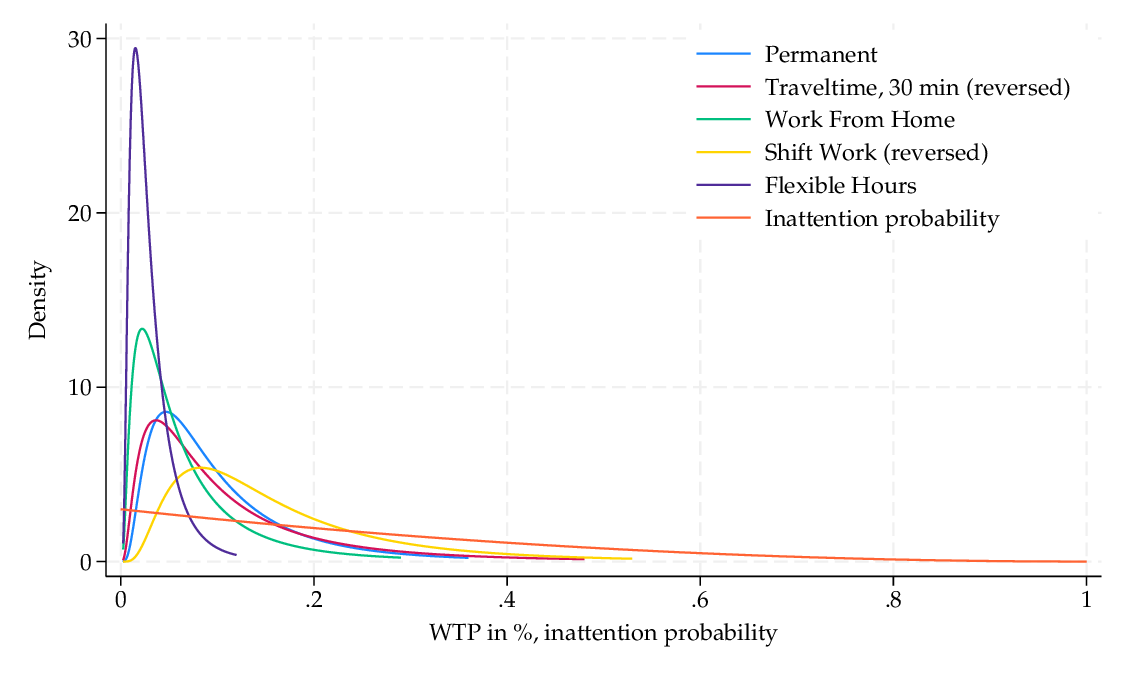}
\label{fig:priors}
\end{center}
\end{figure}

This procedure also requires us to specify a likelihood function for the probability to choose job $j$ over $j'$ from a scenario in order to be able to form posteriors by combining the priors with the choices. We use the same utility specification over amenities and pay as in the mixed logit model \eqref{eq_utility}-\eqref{eq_prob}, but we also want to allow for inattention and generate scenarios that act as attention checks. To this end, we follow in the spirit of \cite{mas2017valuing} and allow individuals to have a probability $c_i$ to be inattentive and simply select a job at random. Unlike \cite{mas2017valuing}, we allow this probability to be individual-specific like all the preference parameters, and give them prior distributions as illustrated in Appendix Figure \ref{fig:priors}.\footnote{An alternative would be to have individual-specific scale parameters in the logit specification.} The result of this is that the probability of choosing job $j$ over $j'$ underlying our scenario selection procedure is: 
\begin{align*}
\Pr(y_{it}=1 \mid c_i,\pi_i,\theta) = (1-c_i)\,\mathbf{1}\left[ \Delta U_{it}>0\right]+\frac{c_i}{2},
\end{align*}
where $\Delta U_{it}=\theta(w_{j(it)}-w_{j'(it)})+\pi_i(z'_{j(it)}-z'_{j'(it)})$ is the difference in utilities between the two jobs.\footnote{Without loss of generality, the adaptive scenario selection procedure can be specified in WTP-space $\pi_i/\theta$.} This is a deterministic rule with a tremble probability $c_i$ that represents the probability that an individual is inattentive and just randomly selects a job. 

The reason why we deviate from the mixed logit utility specification \eqref{eq_utility}-\eqref{eq_prob} for selecting scenarios is that a purely deterministic choice rule without tremble would imply that once the posterior concentrates, some scenarios become fully predictable and generate no information. This means that the BACE algorithm would never present dominated alternatives that could serve as attention checks. Introducing an individual-specific tremble allows the algorithm to both generate scenarios that serve as attention checks and continue to learn about the preference parameters efficiently. Importantly, the tremble parameter is itself learned during the adaptive procedure, so that respondents exhibiting random choice behavior receive more attention check scenarios. In robustness checks, we exploit these attention check scenarios to restrict the sample to attentive individuals.

Between each choice scenario, the posterior over $(\pi_i, p_i)$ is updated. The algorithm then selects the scenario that maximizes the expected information gain. For details on this procedure and the software, see \cite{drake22}.

\global\long\def\thetable{C.\arabic{table}}%
\setcounter{table}{0}
\global\long\def\thefigure{C.\arabic{figure}}%
\setcounter{figure}{0}
\global\long\def\theequation{C.\arabic{equation}}%
\setcounter{equation}{0}

\global\long\def\thetable{C.\arabic{table}}%
\setcounter{table}{0}
\global\long\def\thefigure{C.\arabic{figure}}%
\setcounter{figure}{0}
\global\long\def\theequation{C.\arabic{equation}}%
\setcounter{equation}{0}

\clearpage

\section{A Bayesian Model of Pay-Belief Formation}
\label{sec:appendix_belief_model}
 
This appendix develops a stylized Bayesian model of pay-belief formation that nests rational inference as a benchmark and decomposes deviations from rationality into three economically distinct biases. The model provides predictions for (i) the gap between belief slopes and actual-pay slopes on advertised amenities, and (ii) the effect of a one-shot information treatment on the level and slope of beliefs. We relate our findings to these model predictions. In Appendix \ref{sec:wtp_omitted_q}, we use the belief-formation framework developed here to clarify the interpretation of choice-based estimates of willingness-to-pay for workplace amenities from the job ads experiment when pay is either observed or agents must form pay beliefs and some attributes associated with job quality are unobserved by the econometrician.
 
\subsection{Setup: The Data-Generating Process}
\label{subsec:dgp}
 
Let $w_j$ denote true log starting salary in job $j$ in the relevant labor market. We model $w_j$ as
\begin{equation}
\label{eq:dgp}
w_j \;=\; \mu^{*} \;+\; \phi^{*} z_j \;+\; \kappa^{*} q_j \;+\; e_j,
\qquad e_j \mid z_j, q_j \sim \mathcal{N}(0, \sigma_e^2),
\end{equation}
where $z_j$ are amenities advertised in the job ad and observed by the econometrician, $q_j$ denotes other job attributes that are not observed by the econometrician, and $e_j$ captures factors that are neither observed by the econometrician nor the agent.\footnote{To ease the notation, we let both $z_j$ and $q_j$ be univariate, but the implications carry over to the multivariate case where both $z_j$ and $q_j$ are vectors. For brevity, we also omit other job attributes, such as occupation or sector, that are observed by both the econometrician and the agent. In Section~\ref{sec:beliefs}, we use either a composite amenity or a vector of advertised amenities, and control for occupation and sector.} 

The structural slope $\phi^{*}$ captures the underlying pay-amenity gradient holding fixed unobserved factors $q_j$, while $\kappa^{*}$ captures the true relationship between pay and unobserved factors. For expositional purposes, we define $q_j$ broadly to absorb all components of unobserved job quality that are systematically correlated with $z_j$, so that $e_j$ captures residual variation in pay that is orthogonal to $z_j$, i.e., \(\mathrm{Cov}(e_j,z_j)=0\).   
Throughout, we let
\begin{equation}
\label{eq:betaw}
\phi^{w} \;\equiv\; \frac{\mathrm{Cov}(w_j, z_j)}{\mathrm{Var}(z_j)}
\;=\; \phi^{*} \;+\; \kappa^*\,\frac{\mathrm{Cov}(q_j, z_j)}{\mathrm{Var}(z_j)}
\end{equation}
denote the OLS coefficient from a univariate regression of $w_j$ on $z_j$ in the population of jobs, omitting $q_j$. This is the analogue of the slope coefficients reported in Table \ref{tab:bias}, Column~(3). Equation~\eqref{eq:betaw} makes clear that $\phi^{w}$ need not equal $\phi^{*}$: it is contaminated by the systematic covariance between advertised amenities and unobserved correlated factors $q_j$. 

In theory, we expect the structural slope $\phi^{*}<0$ under classical compensating differentials, while under augmenting-differentials models with search frictions \citep{hwang1998hedonic} or rent sharing \citep{card2018firms}, one may expect $\phi^{*} \geq 0$. In the population of Norwegian high-skilled job ads, it is reasonable to assume $\phi^{w}>0$, consistent with augmenting-differentials patterns between pay and advertised amenities found in \citet{audoly24}.
 
\subsection{The Respondent's Perceived Model}
\label{subsec:perceived}
 
Respondent \(i\) does not directly observe all determinants of pay and forms pay expectations based on $z_j$ and $q_j$. We assume that respondent \(i\) holds a perceived model of log pay as
\begin{equation}
\label{eq:perceived}
w_j \;\sim\; \mathcal{N}\!\left( \mu_i + \phi_i z_j + \kappa_i q_j,\; s_i^2 \right),
\end{equation}
where we let the respondent have the following priors over the parameters $(\mu_i,\phi_i,\kappa_i)$:
\begin{equation}
\label{eq:prior}
\begin{pmatrix}
\mu_i \\ \phi_i \\ \kappa_i
\end{pmatrix}
\sim
\mathcal N\!\left(
\begin{pmatrix}
\mu_{i,0} \\ \phi_{i,0} \\ \kappa_{i,0}
\end{pmatrix},
\begin{pmatrix}
\tau_\mu^{-1} & 0 & 0\\
0 & \tau_\phi^{-1} & 0 \\
0 & 0 & \tau_\kappa^{-1} \\
\end{pmatrix}
\right),
\end{equation}

where $\tau_\mu$ is the precision of the prior over the unconditional mean of pay, while $\tau_\phi$ and $\tau_\kappa$ capture the precision of the slope of pay belief on advertised amenity and unobserved job quality, respectively. We allow $(\mu_{i,0}, \phi_{i,0},\kappa_{i,0})$ to differ from the ``true'' counterparts $(\mu^{*}, \phi^*, \kappa^*)$, and we allow respondents to potentially misweight the amenity signals. The prior mean $\mu_{i,0}$ can be interpreted as the model counterpart to the empirical person component $\alpha_i$ in the two-way decomposition of Equation~\eqref{eq:meanvariancemodel} in person and job components.

Specifically, when respondent $i$ views ad $j$ with advertised amenity $z_j$ and attributes $q_j$ but no explicit pay information, we assume the respondent forms a posterior mean belief
\begin{equation}
\label{eq:beliefmean}
\mathbb{E}_i[\tilde{w}_j \mid z_j,q_j] \;=\; \mu_{i,0} \;+\;
\underbrace{\lambda_i}_{\text{weight on signal}}  \cdot \left( \phi_{i,0} \cdot z_j + \kappa_{i,0} \cdot q_j  \right),
\end{equation}
and exploiting the covariance structure between $z_j$ and $q_j$, we can further derive
\begin{equation}
\label{eq:beliefmean2}
\mathbb{E}_i[\tilde{w}_j \mid z_j] \;=\; \mu_{i,0} \;+\;
\lambda_i \cdot \phi^{RF}_{i,0} \cdot z_j ,
\end{equation}
where $\phi_{i,0}^{RF}$ represents the ``reduced-form'' slope between respondent $i$'s pay belief and advertised amenity $z_j$, including the signal drawn from correlated unobserved job quality $q_j$.
\[
\phi_{i,0}^{RF}
\equiv
\phi_{i,0}
+
\kappa_{i,0}
\frac{\mathrm{Cov}(q_j,z_j)}{\mathrm{Var}(z_j)}.
\]
Given this structure, we let the cross-job slope of beliefs on the advertised amenity  be

\begin{equation}
\label{eq:betab}
\phi_i^b
\equiv
\frac{\mathrm{Cov}(\mathbb{E}_i[\tilde{w}_j \mid z_j],z_j)}{\mathrm{Var}(z_j)}
=
\lambda_i\phi_{i,0}^{RF}.
\end{equation}
This reduced-form representation does not require that respondents ignore job quality unobserved by the econometrician. Rather, any perceived relationship between omitted quality \(q_j\) and pay that is bundled with the advertised amenity is absorbed into \(\phi_{i,0}^{RF}\). The weight $\lambda_i \geq 0$ governs how strongly the respondent extrapolates from the amenity signal: $\lambda_i = 1$ corresponds to standard Bayesian use of the signal, $\lambda_i > 1$ to systematic over-extrapolation (e.g., diagnostic expectations \citep{bordalo2018diagnostic} or representativeness heuristics \citep{tversky1974judgment,bordalo2016stereotypes}), and $\lambda_i < 1$ to under-extrapolation.\footnote{In principle, there could be separate weights on the measured amenity signal and the unobserved job quality signal. While this is straightforward to include, we omit it for notational simplicity.}

Accordingly, we can interpret $\phi^{b}_i$ as the model counterpart to the reduced-form slope coefficient reported in Table \ref{tab:bias}, Column~(1). The difference between the coefficients reported in Table \ref{tab:bias}, Columns (1) and (3), is the relevant counterpart to the model's slope gap \(\phi_i^{b}-\phi^w\), as neither the actual pay regressions nor the belief pay regressions condition on \(q_j\). Thus, the observed belief slope $\phi^{b}_i$ on advertised amenities can be positive either because respondents believe advertised amenities directly predict higher pay, or because advertised amenities are bundled with unobserved job quality that respondents also believe predicts higher pay.

This framework nests three distinct deviations from rational inference. First, the prior mean $\mu_{i,0}$ of unconditional pay can be biased (``level bias''). Second, the prior mean $\phi^{RF}_{i,0}$ of the amenity slope can differ from the true OLS slope $\phi^{w}$ (``slope-prior bias''). Third, the respondent can misweight the signal through $\lambda_i \neq 1$ (``over-extrapolation bias'').

\subsection{Rational Benchmark}
\label{subsec:benchmark}

%The empirical objects we compare are reduced-form, univariate slopes: the slope of actual pay on advertised amenities, \(\phi^w\), and the slope of pay beliefs on advertised amenities, \(\phi_i^b\). Neither slope conditions on omitted job quality ($q_j$). Thus, the rational benchmark is not that respondents recover the structural partial slope ($\phi^*$), but rather that their beliefs reproduce the reduced-form relationship between pay and measured amenities.

If respondent $i$'s priors are unbiased and the signal is weighted correctly,
then $\mu_{i,0}=\mu^*$, $\phi_{i,0}^{RF}=\phi^w$ and $\lambda_i=1$. Under this rational benchmark:
\begin{align}
\mathbb E_i[\tilde w_j\mid z_j]
&=
\mu^*+\phi^w z_j,
\label{eq:benchmark_mean}\\
\phi_i^b
&=
\phi^w.
\label{eq:benchmark_slope}
\end{align}

\noindent\textbf{Implication 1 (Rational Benchmark).}
\emph{Under rational inference, the reduced-form cross-job slope of pay beliefs on advertised amenity equals the reduced-form cross-job slope of actual pay on advertised amenity. The mean of pay beliefs equals the unconditional mean of pay.} \\

This is the benchmark against which we evaluate the evidence shown in Section~\ref{sec:beliefs}.
 
\subsection{Three Bias Mechanisms}
\label{subsec:biases}
 
We now characterize how alternative deviations from the rational benchmark may affect the empirical moments that a researcher can observe in data on beliefs. The key empirical objects are the mean of pay beliefs and the reduced-form cross-job slope of beliefs on amenities.

\paragraph{(i) Biased Prior over the Level ($\mu_{i,0} \neq \mu^{*}$).}
A negative level bias ($\mu_{i,0} < \mu^{*}$) shifts the entire distribution of beliefs downward by $\mu^{*} - \mu_{i,0}$, but does not affect the slope of beliefs on amenities:
\begin{equation*}
\mathbb{E}_i[\tilde{w}_j \mid z_j] - w_j^{\mathrm{rational}}(z_j)
= (\mu_{i,0} - \mu^{*}),
\qquad \phi^{b}_i = \phi^{w},
\end{equation*}

where $w_j^{\mathrm{rational}}(z_j)=\mu^*+\phi^w z_j$ is the rationally expected pay level. This is consistent with our finding that the average control respondent underestimates actual starting salaries by around 18\%. However, level bias \emph{cannot} alone explain why $\phi^{b} > \phi^{w}$.

\paragraph{(ii) Biased Prior over the Reduced-Form Slope ($\phi^{RF}_i \neq \phi^{w}$).}
If respondents hold an upward-biased prior about how much amenity-rich jobs pay, then the belief slope is shifted away from the rational benchmark by a constant. When signals are weighted correctly ($\lambda_i=1$), the gap is
\begin{align*}
\phi_i^b-\phi^w
=
(\phi^{RF}_{i,0}-\phi^w)\qquad \text{when }\lambda_i=1
\end{align*}

Crucially, under additive biased priors, this bias does \emph{not} depend on the value of $\phi^{w}$ itself: the absolute belief-slope bias is the same, independent of whether the true reduced-form slope is small or zero.
 
\paragraph{(iii) Over-extrapolation ($\lambda_i > 1$ ).}
If respondents over-weight the amenity signal relative to the Bayesian benchmark, the belief slope is amplified:
\begin{align*}
\phi^{b}_i - \phi^{w} &= \lambda_i\phi^{RF}_{i,0} - \phi^w \\
&=(\lambda_i-1)\phi^w \qquad \text{when } \phi^{RF}_{i,0}=\phi^w
\end{align*}

The bias is therefore proportional to the underlying reduced-form slope. If
the true reduced-form pay-amenity slope is large because advertised amenities
directly predict pay, or because they are bundled with unobserved job quality that
predicts pay, then over-extrapolation of signals generates a large belief-slope bias. \\
 
\noindent\textbf{Implication 2 (Positive Belief-Slope Gap).}
\emph{\(\phi^{b}>\phi^{w}>0\) rules out rational inference and pure level bias as the sole mechanisms. It is consistent with either an upward-biased prior over the reduced-form pay-amenity slope $\phi^{RF}_i > \phi^{w}$, over-extrapolation $\lambda_i > 1$, or both.}
 
\subsection{Bayesian Updating from the Pay-Information Treatment}
\label{subsec:updating}
 
Treated respondents in our experiment view ten job ad pairs ($M = 10 \times 2=20$) with disclosed pay before reporting their pay beliefs for a separate set of ten ad pairs. Let $\mathcal{S}_M = \{(z_\ell, w_\ell)\}_{\ell=1}^M$ denote the signals received by the treated respondents. Since the empirical signal contains realized pay and advertised amenity, but does not separately identify omitted job quality \(q_j\), the relevant object for updating is the reduced-form pay-amenity slope. Standard Bayesian updating of the prior \eqref{eq:prior} using $\mathcal{S}_M$ gives a posterior that is again Gaussian, with mean
\begin{align}
\mu_{i,1}
&=
\mu_{i,0}
+
\omega_\mu
\left(
\bar w_M-\mu_{i,0}-\phi_{i,0}^{RF}\bar z_M
\right),
\label{eq:postmu}\\
\phi_{i,1}^{RF}
&=
\phi_{i,0}^{RF}
+
\omega_\phi
\left(
\hat\phi_M^{\mathrm{OLS}}-\phi_{i,0}^{RF}
\right).
\label{eq:postbeta}
\end{align}
where $\bar{w}_M, \bar{z}_M$ are the sample means and $\hat{\phi}_M^{\mathrm{OLS}}$ is the within-sample OLS slope of $w$ on $z$. The two posterior weights are

\begin{equation}
\label{eq:weights}
\omega_\mu
=
\frac{M}{M+\sigma_u^2\tau_\mu},
\qquad
\omega_\phi
=
\frac{M s_{z,M}^2}{M s_{z,M}^2+\sigma_u^2\tau^{RF}_\phi}.
\end{equation}
where $s_{z,M}^2 = M^{-1}\sum_\ell (z_\ell - \bar{z}_M)^2$, and $\tau^{RF}_\phi$ is the precision of the prior on the reduced-form slope of pay belief on the advertised amenity (see Equation \eqref{eq:betab}).

The change in the level of treated beliefs is therefore
\begin{equation}
\label{eq:level_effect}
\Delta_{\mathrm{T-C}}\!\big[\bar{b}\big] \;=\;
\omega_\mu \cdot \big(\bar{w}_M - \mu_{i,0} - \phi^{RF}_{i,0}\,\bar{z}_M\big),
\end{equation}

which is positive whenever respondents underestimate actual pay (i.e., $\mu_{i,0} + \phi^{RF}_{i,0}\bar{z}_M < \bar{w}_M$). Likewise, the change in the slope of treated beliefs on amenities is
\begin{equation}
\label{eq:slope_effect}
\Delta_{\mathrm{T-C}}\!\big[\phi^{b}\big] \;=\;
\omega_\phi \cdot \big(\hat{\phi}_M^{\mathrm{OLS}} - \phi^{RF}_{i,0}\big).
\end{equation}
 
\noindent\textbf{Implication 3 (Asymmetric Updating).}
\emph{The Bayesian model predicts that the information treatment moves the level of beliefs by $\omega_\mu$ and the reduced-form slope of beliefs by $\omega_\phi$. The model can rationalize a sizable level effect together with a near-zero slope effect if and only if $\omega_\mu \gg \omega_\phi$, i.e., the prior over the level is relatively loose ($\sigma_u^2 \tau_\mu / M$ small) while the prior over the reduced-form slope is relatively tight ($\sigma_u^2 \tau^{RF}_\phi / (M\,s_{z,M}^2)$ large).} \\
 
The condition $\omega_\mu \gg \omega_\phi$ has a clean interpretation. The level $\mu$ has a precision-loaded prior with $M$ informative observations per unit of $\sigma_u^2 \tau_\mu$. The slope $\phi$ has $M \cdot s_{z,M}^2$ informative observations per unit of $\sigma_u^2 \tau^{RF}_\phi$. Even with relatively loose priors on both parameters, ten job ad pairs contain substantially less information about a slope than about a mean: with $s_{z,M}^2$ on the order of the variance of $z_j$ (around $0.013$ in our data) and $M = 20$, the effective sample size for slope updating is roughly $M \cdot s_{z,M}^2 \approx 0.26$. Thus, the model predicts that one short exposure to twenty pay signals from ten ad pairs is materially informative about the average pay level, but only weakly informative about the reduced-form amenity-pay gradient. \\
 
\noindent\textbf{Implication 4 (Slope Persistence under Information).}
\emph{When $\omega_\phi$ is small, the information treatment leaves $\phi^{b}_i$ nearly unchanged, regardless of whether the slope-prior bias mechanism (ii) or the over-extrapolation mechanism (iii) underlies the gap $\phi^{b} > \phi^{w}$. A null effect of disclosure on the amenity-pay tradeoff in stated choices is therefore not, by itself, evidence against pay transparency policies that provide more sustained or salient information.}
 
\subsection{Discussion and Mapping to the Empirical Findings}
\label{subsec:mapping}
 
Taken together with our empirical evidence, Implications 1--4 yield the following:
 
\begin{enumerate}
\item The 18\% downward bias in mean beliefs (Table~\ref{tab:bias}, final row) implies $\mu_{i,0} < \mu^{*}$. This is a level bias only and would not, on its own, generate a slope discrepancy.
 
\item The positive belief slope $\phi^{b} > \phi^{w} > 0$ (Table~\ref{tab:bias}, Columns (1) and (3)) requires either $\phi^{RF}_{i,0} > \phi^{w}$, $\lambda_i > 1$ or both. It cannot be explained by level bias alone or by rational inference.
 
\item The 4\% increase in mean beliefs and 15\% reduction in their variance produced by the information treatment
(Table~\ref{tab:effects_beliefs}, Panel A) imply $\omega_\mu \in (0,1)$ -- i.e., respondents' priors over the level are loose but not infinitely loose.
 
\item The absence of a measurable slope effect of the information treatment (Table~\ref{tab:effects_choices_wtp}) implies $\omega_\phi$ is small in our context. By Equation~\eqref{eq:weights}, this is consistent with (a) high prior precision $\tau^{RF}_\phi$, (b) limited within-treatment variation $s_{z,M}^2$, or (c) both.
\end{enumerate}
 
The model thus rationalizes our findings under \emph{either} mechanism (ii) or mechanism (iii), and is consistent with our null on the amenity-pay tradeoff under disclosure if the slope-relevant precision in twenty pay signals is dominated by the precision of the prior on the reduced-form slope of pay beliefs on the advertised amenity.

\section{Choices with Pay Beliefs and Omitted Job Quality}
\label{sec:wtp_omitted_q}

In Sections \ref{sec:wtp} and \ref{subsec:wtp_learning}, we provided alternative estimates of respondents' willingness-to-pay (WTP) for workplace amenities, using different portions of choice data collected as part of our survey experiment. Specifically, the WTP estimates provided in Section \ref{sec:wtp} were, respectively, from a benchmark hypothetical choice setting with full information about pay and workplace amenities (Table \ref{tab:willingness_to_pay}, Column (3)) and from hypothetical choices between job ads with full information about pay and amenities advertised  in job ad texts (Table \ref{tab:willingness_to_pay}, Column (4)). Meanwhile, the estimates in Section \ref{subsec:wtp_learning} were derived from respondents' choices between job ads featuring advertised amenities in job ad texts and relied solely on respondents' reported pay beliefs (Table \ref{tab:effects_choices_wtp}). In this appendix, we clarify the interpretation of WTP estimates from choice models when job alternatives may plausibly contain attributes that are partly unobserved by the econometrician (e.g., omitted job quality) and when agents do not hold explicit pay information and must rely on available information to form pay beliefs.

\subsection{Choice Setup}

To simplify notation, we consider a single respondent and suppress individual subscripts. Let \(z_j\) denote the measured amenity and let \(q_j\) denote omitted job attributes or job quality that is not observed by the econometrician. The respondent's utility is given by:
\[
U_j
=
\theta \mathcal{W}_j
+
\pi z_j
+
\chi q_j
+
\xi_j,
\]
so the respondent values both advertised amenity $z_j$ and omitted job quality $q_j$, has idiosyncratic tastes $\xi_j$, and $\mathcal{W}_j$ is a generic monetary value associated with the job choice that enters perceived utility with a weight $\theta$. When the agent has full information about pay, $\mathcal{W}_j=w_j$, while otherwise, if the agent must form pay expectations given the available information $(z_j,q_j)$, then $\mathcal{W}_j=\mathbb{E}_i[\tilde{w}_j\mid z_j,q_j]$, using the notation from Appendix \ref{sec:appendix_belief_model}.  

If job quality \(q_j\) was observed by the econometrician and included in the estimation of the choice model, then the WTP for the advertised amenity \(z_j\), holding \(q_j\) fixed, would be
\[
WTP^{true}
=
\frac{\pi}{\theta}.
\]
In practice, \(q_j\) is omitted. Therefore, choice-model estimates based only on $\mathcal{W}_j$ and \(z_j\) may generally not recover the structural parameters \((\theta,\pi)\). The benchmark hypothetical choice experiments with full information about pay and workplace amenities construct experimental settings where \(q_j\) are equalized across choice alternatives, so the WTP derived from such experiments (Table \ref{tab:willingness_to_pay}, Column (3)) are expected to provide consistent estimates of the structural parameters. In the following, we consider settings where this may not hold.

\subsection{Full Pay Information with Omitted Job Quality}\label{app_choice_omitted}
Under full information about pay displayed in job ads but the researcher potentially not capturing all measures of job quality $q_j$, as in the experimental setup using real job ads in Section  \ref{sec:wtp} (Table \ref{tab:willingness_to_pay}, Column (4)), we can express the respondent's utility as
\[
U_j^F
=
\theta w_j
+
\pi z_j
+
\chi q_j
+ 
\xi^F_j.
\]
Suppose the econometrician estimates the misspecified logit index
\[
\widetilde U_j^F
=
\widetilde\theta^F w_j
+
\widetilde\pi^F z_j
+ 
\widetilde\xi^F_j,
\]
omitting \(q_j\). To see what the omitted job quality does, we can write the linear projection of omitted job quality on the included attributes as
\[
q_j
=
\rho_w^F w_j
+
\rho_z^F z_j
+
\eta_j^F,
\]

where \(\rho_w^F\) and \(\rho_z^F\) are the linear projection coefficients from projecting omitted job quality \(q_j\) on true pay and advertised amenities captured by the researcher in the full information experiment, and \(\eta_j^F\) is the residual from this projection.

Substituting into true utility gives
\[
U_j^F
=
\left(\theta+\chi\rho_w^F\right)w_j
+
\left(\pi+\chi\rho_z^F\right)z_j
+
\chi\eta_j^F
+ 
\xi^F_j.
\]
Thus, if the residual component \(\eta_j^F\) is orthogonal to the included attributes and does not generate additional misspecification, the estimated logit coefficients are approximately
\[
\widetilde\theta^F
\approx
\theta+\chi\rho_w^F,
\qquad
\widetilde\pi^F
\approx
\pi+\chi\rho_z^F.
\]
The full-information WTP estimated from a model omitting \(q_j\) is therefore approximately
\[
WTP^F
=
\frac{\widetilde\pi^F}{\widetilde\theta^F}
\approx
\frac{\pi+\chi\rho_z^F}{\theta+\chi\rho_w^F}.
\]

Hence, omitted job quality can bias the implied WTP derived from agents' choices even under full information whenever \(q_j\) is valued by the respondent and is correlated with actual pay \(w_j\), advertised amenities \(z_j\), or both. When $q_j$ is valued by respondents and is positively correlated with both pay and amenities, both the numerator and the denominator are too large. The implied WTP can therefore be biased in either direction.

Two findings are reassuring in this case. First, we find remarkably similar estimates of the WTPs when we include a larger set of advertised amenities from the job ads, besides those included in the experimental setup using real job ads in Section \ref{sec:wtp} (Table \ref{tab:willingness_to_pay}, Column (4)). Second, the similarity between the WTP estimates derived from the hypothetical choice experiments (Table \ref{tab:willingness_to_pay}, Column (3)) and the real-ad full-information scenarios  (Table \ref{tab:willingness_to_pay}, Column (4)) suggests that omitted job quality does not materially change the estimated pay--amenity tradeoff when pay is observed. This does not require \(q_j\) to be literally uncorrelated with \(w_j\) and \(z_j\); it only requires that its omission does not substantially alter the $\pi/\theta$ ratio.

\subsection{Pay Beliefs with Omitted Job Quality}\label{app_choice_beliefs}

When the respondent does not have full information about pay, the monetary value associated with a job reflects the expected pay, $\mathcal{W}_j=\mathbb{E}_i[\tilde{w}_j\mid z_j,q_j]$. To ease notation, we refer to this object as $\widehat w_j$ in the following. We can thus express the respondent's \textit{perceived} utility as
\[
U_j^N
=
\theta \widehat w_j
+
\pi z_j
+
\chi q_j
+ 
\xi^N_j
\]
Following the exact same steps as in the full information case, the WTP is approximately

\[
WTP^N
=
\frac{\widetilde\pi^N}{\widetilde\theta^N}
\approx
\frac{\pi+\chi\rho_z^N}{\theta+\chi\rho^N_{\widehat{w}}}.
\]
Notably, however, the linear projection coefficients between advertised amenities and unobserved job quality ($\rho^N_z$) and pay ($\rho^N_{\widehat{w}}$) are different from the analogous expressions in the full information case. To see this, we can use \eqref{eq:beliefmean2}-\eqref{eq:betab} from the pay-belief formation model in Appendix \ref{sec:appendix_belief_model}, and express the expected pay (omitting $i$-subscripts) as follows:
\[
\widehat w_j
:=
\mathbb{E}_i[\tilde{w}_j\mid z_j,q_j]
=
\mu_0
+
\lambda \phi z_j
+
\lambda \kappa q_j
+
r_j,
\]
where \(\phi\) is the perceived partial relationship between advertised amenities and pay, \(\kappa\) is the perceived relationship between unobserved job quality and pay, and \(r_j\) captures belief-variation not explained by \(z_j\) or \(q_j\). The reduced-form belief slope between  $\widehat w_j$ and $z_j$ is:
\[
\phi^b
=
\lambda \left( \phi
+
\kappa
\frac{\mathrm{Cov}(q_j,z_j)}{\mathrm{Var}(z_j)}
\right).
\]

Suppose the econometrician estimates a choice model using pay beliefs $\widehat w_j$ and advertised amenities $z_j$, imposing the following utility structure:
\[
\widetilde U_j^N
=
\widetilde\theta^N \widehat w_j
+
\widetilde\pi^N z_j
+ 
\tilde{\xi}^N_j,
\]
where unobserved job quality \(q_j\) is omitted. Since pay beliefs themselves partly reflect omitted job quality, \(\widehat w_j\) can proxy for \(q_j\). Inverting the pay belief equation gives
\[
q_j
=
\frac{1}{\lambda \kappa}
\left(
\widehat w_j
-
\mu_0
-
\lambda\phi z_j
-
r_j
\right),
\]
which substituted into the \textit{perceived}  utility yields
\[
U_j^N
=
\left(
\theta+\frac{\chi}{\lambda \kappa}
\right)\widehat w_j
+
\left(
\pi-\frac{\chi\phi}{\kappa}
\right)z_j
-
\frac{\chi}{\lambda \kappa}r_j
+
\text{constant}.
\]
Thus, abstracting from the residual belief component \(r_j\), the coefficients from the misspecified no-pay information choice model are approximately
\[
\widetilde\theta^N
\approx
\theta+\frac{1}{\lambda \kappa}\chi,
\qquad
\widetilde\pi^N
\approx
\pi-\frac{\phi}{\kappa}\chi.
\]
The implied WTP estimated from the no-pay information model is therefore approximately
\[
WTP^N
=
\frac{\widetilde\pi^N}{\widetilde\theta^N}
\approx
\frac{
\pi-\frac{\phi}{\kappa}\chi
}{
\theta+\frac{1}{\lambda \kappa}\chi
}.
\]

This expression has a different interpretation from the full-information counterpart $WTP^F$. In the no-pay information model, pay belief is not merely a monetary ``attribute'', it also summarizes omitted job quality. If \(\chi>0\), \(\kappa>0\) and \(\lambda>0\), then jobs with higher expected pay also tend to look like better jobs along omitted dimensions. As a result, the coefficient on expected pay can reflect both the value of pay and the value of omitted job quality. At the same time, when $\phi>0$, the residual coefficient on the advertised amenity is net of the component of omitted job quality that pay beliefs absorb. The estimated no-information WTP can therefore be small for two reasons. First, the inferred value of the advertised amenity $\widetilde\pi^N$ is pulled below $\pi$. Second, as expected pay proxies for omitted job quality, the inferred pay valuation is pulled above $\theta$. In this sense, $WTP^N$ is a residual, belief-conditional willingness-to-pay, not the total perceived value of the advertised amenity.

\paragraph{Interpretation.}
The full-information and no-information WTP ratios differ for two reasons. First, the monetary attribute differs: full-information choices use actual pay \(w_j\), whereas no-information choices use pay belief \(\widehat w_j\). Second, pay belief may themselves be a function of advertised amenities and omitted job quality. Thus, WTP estimates from the no-information model should be interpreted as belief-conditional WTPs: marginal rates of substitution between the advertised amenity and expected pay, holding expected pay fixed. They are comparable to full-information WTPs only as direct marginal rates of substitution with respect to the monetary object entering utility, not as total amenity values.

\paragraph{Implication for the full-information versus no-information WTP gap.}
A comparison of the estimates reported in Sections \ref{sec:wtp} and \ref{subsec:wtp_learning} suggest that the WTPs drop considerably from the full-information in Section \ref{sec:wtp} to the no pay information comparisons in Section  \ref{subsec:wtp_learning}. Putting aside the fact that the two settings measure WTPs in real vs. perceived pay, our derivations above can rationalize this finding. The gap between full-information and no-information WTPs is easiest to interpret using analogous omitted-variable projections:
\[
\widetilde WTP^F-WTP^N
\approx
\frac{\pi+\chi\rho_z^F}{\theta+\chi\rho_w^F}
-
\frac{\pi+\chi\rho_z^N}{\theta+\chi\rho_{\widehat w}^N}.
\]
If omitted quality is valued, \(\chi>0\), then
\[
\rho_{\widehat w}^N>\rho_w^F
\quad \text{and} \quad
\rho_z^N<\rho_z^F
\]
are sufficient to push no-information WTP below full-information WTP. This ``bias'' in the WTPs reflects that pay beliefs absorb omitted job quality. Intuitively, expected pay is a tighter proxy for omitted quality than actual pay because beliefs themselves load on observed amenities, while actual pay does not, so conditional on expected pay there is less residual amenity-quality covariance left to attribute to advertised amenities.

The structural belief equation makes the same point. If
\[
\widehat w_j
:=
\mathbb{E}_i[\tilde{w}_j\mid z_j,q_j]
=
\mu_0
+
\lambda \phi z_j
+
\lambda \kappa q_j
+
r_j,
\]
then, abstracting from \(r_j\),
\[
\rho_{\widehat w}^N\approx \frac{1}{\lambda\kappa},
\qquad
\rho_z^N\approx -\frac{\phi}{\kappa}.
\]
Thus, the conditions above correspond to
\[
\frac{1}{\lambda\kappa}>\rho_w^F,
\qquad
-\frac{\phi}{\kappa}<\rho_z^F.
\]
When \(\kappa>0\), \(\phi>0\) and \(\lambda>0\), expected pay proxies positively for omitted quality, while the residual association between advertised amenities and omitted job quality is reduced after conditioning on expected pay. This rationalizes
\[
\widetilde WTP^N<\widetilde WTP^F
\]
even if the underlying preference parameters \((\theta,\pi,\chi)\) are unchanged.

\paragraph{Implication for our findings of treatment effects on no-information WTP.}
In Section  \ref{subsec:wtp_learning}, we showed that the tradeoff between amenities and pay did not change following the information treatment. In Section \ref{subsec:treat_belief}, we showed that the information treatment increased level beliefs ($\mu_{i,0}$), but did not change the amenity-pay slope in beliefs ($\phi^{b}_i$). The formula above allows us to interpret this as indirect evidence that the information treatment did not affect the perceived slope between unobserved job quality and expected pay:

\[
WTP^N
\approx
\frac{
\pi-\frac{\phi}{\kappa}\chi
}{
\theta+\frac{1}{\lambda \kappa}\chi
}.
\]

Holding preferences fixed, a change in $WTP^N$ would indicate that the treatment changed how beliefs load on advertised amenities or omitted job quality. The empirical finding of little or no treatment effect on no-information WTP is consistent with the belief results: the information treatment raises average wage beliefs but does not materially change the way respondents use amenities or omitted job quality to infer relative pay. In this sense, the null result provides indirect evidence that the treatment does not substantially change \(\kappa\), the perceived relationship between omitted job quality and pay, nor the reduced-form belief slope \(\phi^b\), which structurally depends on all belief parameters \(\kappa\), \(\phi\) and \(\lambda\); see Equation \eqref{eq:betab}.
\end{document}

%% file: export_tsd/macros.tex
\newcommand{\vartotzero}{0.089865656654729}
\newcommand{\varpersonzero}{0.034288498148662}
\newcommand{\varjobzero}{0.023932391882883}
\newcommand{\varunexpzero}{0.030128826569175}
\newcommand{\covzero}{0.001515940054010}

%% file: export_tsd/attrition.tex
{
\def\sym#1{\ifmmode^{#1}\else\(^{#1}\)\fi}
\begin{tabular}{l*{4}{c}}
\toprule
            &\multicolumn{1}{c}{Gave Consent}&\multicolumn{1}{c}{Participated}&\multicolumn{1}{c}{Participated}&\multicolumn{1}{c}{Participated}\\
            &\multicolumn{1}{c}{and}&\multicolumn{1}{c}{in}&\multicolumn{1}{c}{in}&\multicolumn{1}{c}{and}\\
            &\multicolumn{1}{c}{Filled}&\multicolumn{1}{c}{Hypothetical}&\multicolumn{1}{c}{Pay}&\multicolumn{1}{c}{Completed}\\
            &\multicolumn{1}{c}{Background}&\multicolumn{1}{c}{Choice}&\multicolumn{1}{c}{Information}&\multicolumn{1}{c}{the Full}\\
            &\multicolumn{1}{c}{Information}&\multicolumn{1}{c}{Experiment}&\multicolumn{1}{c}{Experiment}&\multicolumn{1}{c}{Survey}\\
            &\multicolumn{1}{c}{in Stage 1}&\multicolumn{1}{c}{in Stage 2}&\multicolumn{1}{c}{in Stages 3-4}&\multicolumn{1}{c}{in Stages 1-4}\\
            &\multicolumn{1}{c}{(1)}&\multicolumn{1}{c}{(2)}&\multicolumn{1}{c}{(3)}&\multicolumn{1}{c}{(4)}\\
\midrule
Age         &       21.79&     -0.0203         &     -0.0610         &     -0.0626         \\
            &     [3.173]         &    (0.0144)         &    (0.0389)         &    (0.0391)         \\
Male        &       0.301         &     0.00225         &    0.000589         &    0.000183         \\
            &     [0.459]         &   (0.00172)         &   (0.00421)         &   (0.00428)         \\
Immigrant Background&       0.101         &    0.000128         &    -0.00621\sym{*}  &    -0.00602\sym{*}  \\
            &     [0.302]         &   (0.00128)         &   (0.00341)         &   (0.00343)         \\
Year in Program&       2.540         &    -0.00161         &    -0.00776         &    -0.00564         \\
            &     [2.003]         &   (0.00907)         &    (0.0187)         &    (0.0188)         \\
Years Studied&       1.717         &   0.0000461         &    -0.00917         &    -0.00977         \\
            &     [1.738]         &   (0.00676)         &    (0.0157)         &    (0.0159)         \\
Currently Works&       0.653         &   -0.000278         &    -0.00569         &    -0.00641         \\
            &     [0.476]         &   (0.00206)         &   (0.00418)         &   (0.00421)         \\
Ever Worked Full Time&       0.355         &    -0.00147         &    -0.00585         &    -0.00513         \\
            &     [0.479]         &   (0.00216)         &   (0.00454)         &   (0.00458)         \\
\midrule
Respondents &        1,063         &        1,042         &         975         &         973         \\
Completion Rate&                     &       0.980         &       0.917         &       0.915         \\
Joint $ F $-statistic&                     &       1.325         &       1.471         &       1.442         \\
Joint $ p $-value&                     &       [0.235]         &       [0.174]         &       [0.185]         \\
\bottomrule
\end{tabular}
}

%% file: export_tsd/willingness_to_pay.tex
{
\def\sym#1{\ifmmode^{#1}\else\(^{#1}\)\fi}
\begin{tabular}{l*{5}{c}}
\toprule
                    &\multicolumn{1}{c}{WTP}&\multicolumn{1}{c}{Subjective}&\multicolumn{3}{c}{WTP Estimates from}\\
                    &\multicolumn{1}{c}{Estimates}&\multicolumn{1}{c}{WTP}&\multicolumn{3}{c}{Hypothetical Choice Experiments:}\\
                    &\multicolumn{1}{c}{from the}&\multicolumn{1}{c}{Stated by}&\multicolumn{1}{c}{\textit{Hypothetical}}&\multicolumn{1}{c}{\textit{Real}}&\multicolumn{1}{c}{}\\
                    &\multicolumn{1}{c}{Literature}&\multicolumn{1}{c}{Respondents}&\multicolumn{1}{c}{\textit{Jobs}}&\multicolumn{1}{c}{\textit{Job Ads}}&\multicolumn{1}{c}{\textit{Combined}}\\
                    &\multicolumn{1}{c}{(1)}&\multicolumn{1}{c}{(2)}&\multicolumn{1}{c}{(3)}&\multicolumn{1}{c}{(4)}&\multicolumn{1}{c}{(5)}\\
\midrule
Permanent           &       0.173\sym{***}&       0.107\sym{***}&       0.161\sym{***}&       0.145\sym{***}&       0.233\sym{***}\\
                    &    (0.0240)         &   (0.00238)         &    (0.0101)         &    (0.0161)         &    (0.0104)         \\
Work From Home      &       0.130\sym{***}&      0.0742\sym{***}&       0.105\sym{***}&     0.00462         &       0.151\sym{***}\\
                    &    (0.0120)         &   (0.00211)         &   (0.00698)         &    (0.0211)         &   (0.00817)         \\
Shift Work          &      -0.127\sym{***}&      -0.183\sym{***}&     -0.0769\sym{***}&     -0.0833\sym{***}&      -0.171\sym{***}\\
                    &   (0.00700)         &   (0.00436)         &    (0.0115)         &    (0.0206)         &    (0.0127)         \\
Flexible Hours      &       0.184\sym{***}&      0.0645\sym{***}&      0.0644\sym{***}&      0.0802\sym{***}&      0.0799\sym{***}\\
                    &    (0.0140)         &   (0.00178)         &   (0.00517)         &    (0.0134)         &   (0.00614)         \\
Travel Time &      -0.150\sym{***}&      -0.131\sym{***}&      -0.169\sym{***}&     -0.0767\sym{***}&      -0.147\sym{***}\\
                    &    (0.0190)         &   (0.00302)         &   (0.00523)         &   (0.00846)         &   (0.00537)         \\
\midrule
Respondents         &                     &         965         &         973         &         973         &         973         \\
Choice Alternatives &                     &                     &       38,920         &       19,460         &       58,380         \\
\bottomrule
\end{tabular}
}

%% file: export_tsd/baselinebeliefs_mean_p.tex
\begin{tabular}{lcccc}
\toprule
& (1) & (2) & (3) & (4) \\
\midrule
Male        &      0.0559\sym{***}&      0.0560\sym{***}&      0.0571\sym{**} &      0.0593\sym{***}\\
            &    (0.0202)         &    (0.0185)         &    (0.0226)         &    (0.0207)         \\
Age         &     0.00159         &    0.000690         &    0.000724         &   -0.000203         \\
            &   (0.00606)         &   (0.00551)         &   (0.00601)         &   (0.00557)         \\
Immigrant Background&      0.0239         &      0.0112         &      0.0366         &      0.0230         \\
            &    (0.0329)         &    (0.0301)         &    (0.0341)         &    (0.0311)         \\
Year in Program&    -0.00380         &    -0.00473         &    -0.00176         &    -0.00292         \\
            &   (0.00603)         &   (0.00534)         &    (0.0106)         &   (0.00956)         \\
Years Studied&     0.00409         &     0.00551         &     0.00126         &     0.00173         \\
            &   (0.00982)         &   (0.00883)         &    (0.0111)         &   (0.00997)         \\
Currently Works&     -0.0234         &     -0.0234         &     -0.0164         &     -0.0164         \\
            &    (0.0183)         &    (0.0164)         &    (0.0189)         &    (0.0172)         \\
Ever Held Full-Time Job&     -0.0288         &     -0.0277         &     -0.0240         &     -0.0246         \\
            &    (0.0212)         &    (0.0185)         &    (0.0212)         &    (0.0189)         \\
Current Job: Pay&       0.116\sym{***}&       0.117\sym{***}&       0.114\sym{***}&       0.114\sym{***}\\
            &    (0.0337)         &    (0.0308)         &    (0.0336)         &    (0.0309)         \\
Current Job: Permanent&      0.0152         &      0.0106         &      0.0124         &     0.00780         \\
            &    (0.0216)         &    (0.0186)         &    (0.0213)         &    (0.0188)         \\
Current Job: Work From Home&     0.00388         &    -0.00900         &      0.0214         &     0.00610         \\
            &    (0.0292)         &    (0.0261)         &    (0.0308)         &    (0.0277)         \\
Current Job: Shift Work&     0.00295         &    0.000582         &    0.000599         &    -0.00228         \\
            &    (0.0202)         &    (0.0179)         &    (0.0203)         &    (0.0183)         \\
Current Job: Flexible Hours&     -0.0328         &     -0.0243         &     -0.0338         &     -0.0258         \\
            &    (0.0204)         &    (0.0181)         &    (0.0219)         &    (0.0195)         \\
\midrule
Job Ad FE&                     &$\checkmark$         &                     &$\checkmark$         \\
Survey Batch FE    &                     &                     &$\checkmark$         &$\checkmark$         \\
Observations&        9,446         &        9,446         &        9,446         &        9,446         \\
Respondents &         479         &         479         &         479         &         479         \\
Job Ads        &         994         &         994         &         994         &         994         \\
\bottomrule
\end{tabular}

%% file: export_tsd/baselinebeliefs_mean.tex
\begin{tabular}{lcccc}
\toprule
& (1) & (2) & (3) & (4) \\
\midrule
\addlinespace
\multicolumn{5}{c}{\textbf{A: Job Quality}} \\
\midrule
Sorkin Value&      0.0616\sym{***}&      0.0576\sym{***}&      0.0453\sym{***}&      0.0407\sym{***}\\
            &   (0.00872)         &   (0.00730)         &   (0.00878)         &   (0.00744)         \\
Poaching Index&      0.0493\sym{***}&      0.0456\sym{***}&      0.0374\sym{***}&      0.0330\sym{***}\\
            &   (0.00960)         &   (0.00826)         &   (0.00874)         &   (0.00750)         \\
\addlinespace
\multicolumn{5}{c}{\textbf{B: Actual Pay}} \\
\midrule
Employer Pay Premium&       0.622\sym{***}&       0.616\sym{***}&       0.222\sym{*}  &       0.209\sym{*}  \\
            &     (0.144)         &     (0.134)         &     (0.119)         &     (0.110)         \\
Actual Job Pay  &       0.205\sym{***}&       0.197\sym{***}&       0.120\sym{***}&       0.108\sym{***}\\
            &    (0.0593)         &    (0.0532)         &    (0.0426)         &    (0.0366)         \\
\addlinespace
\multicolumn{5}{c}{\textbf{C: Amenity Value}} \\
\midrule
Literature  &       0.453\sym{***}&       0.458\sym{***}&       0.391\sym{***}&       0.394\sym{***}\\
            &    (0.0676)         &    (0.0597)         &    (0.0629)         &    (0.0521)         \\
Subjective  &       0.749\sym{***}&       0.759\sym{***}&       0.643\sym{***}&       0.648\sym{***}\\
            &    (0.0880)         &    (0.0777)         &    (0.0844)         &    (0.0723)         \\
Choice Model&       0.477\sym{***}&       0.488\sym{***}&       0.409\sym{***}&       0.420\sym{***}\\
            &    (0.0652)         &    (0.0578)         &    (0.0585)         &    (0.0492)         \\
\addlinespace
\multicolumn{5}{c}{\textbf{D: Specific Amenities}} \\
\midrule
Permanent   &      0.0851\sym{***}&      0.0879\sym{***}&      0.0770\sym{***}&      0.0804\sym{***}\\
            &    (0.0230)         &    (0.0205)         &    (0.0187)         &    (0.0155)         \\
Work From Home&     0.00435         &      0.0162         &     0.00691         &      0.0209         \\
            &    (0.0340)         &    (0.0275)         &    (0.0300)         &    (0.0222)         \\
Shift Work  &      -0.117\sym{***}&      -0.114\sym{***}&      -0.111\sym{***}&      -0.107\sym{***}\\
            &    (0.0225)         &    (0.0195)         &    (0.0228)         &    (0.0193)         \\
Flexible Hours&      0.0519\sym{**} &      0.0432\sym{**} &      0.0372\sym{*}  &      0.0279\sym{*}  \\
            &    (0.0212)         &    (0.0193)         &    (0.0197)         &    (0.0162)         \\
%Travel Time &     -0.0380\sym{***}&     -0.0466\sym{***}&     -0.0254\sym{***}&     -0.0322\sym{***}\\
%            &    (0.0108)         &    (0.0105)         &   (0.00939)         &   (0.00850)         \\
\midrule
Respondent FE   &                     &$\checkmark$                     &                      &$\checkmark$                     \\
Occupation $\times$ Sector FE&                     &         &$\checkmark$                     &$\checkmark$         \\
Observations&        9,446         &        9,446         &        9,446         &        9,446         \\
Respondents &         479         &         479         &         479         &         479         \\
Job Ads        &         994         &         994         &         994         &         994         \\
\bottomrule
\end{tabular}

%% file: export_tsd/bias.tex
\begin{tabular}{lcccc}
\toprule
& \multicolumn{2}{c}{Mean Pay Beliefs} & \multicolumn{2}{c}{Actual Pay} \\
& (1) & (2) & (3) & (4)  \\
\midrule
\addlinespace
\multicolumn{5}{c}{\textbf{A: Pay Levels and Gaps}} \\
\midrule
Pay Level &       10.63\sym{***}&       10.63\sym{***}&       10.81\sym{***}&       10.81\sym{***}\\
            &    (0.0123)         &    (0.0123)         &    (0.0130)         &    (0.0123)         \\
Pay Gap (\textit{Within Scenario})&      0.0115         &      0.0115         &       0.209\sym{***}&       0.209\sym{***}\\
            &    (0.0135)         &    (0.0115)         &    (0.0199)         &    (0.0179)         \\
\addlinespace
\multicolumn{5}{c}{\textbf{B: Pay Correlation with Amenity Values}} \\
\midrule
Literature&       0.453\sym{***}&       0.391\sym{***}&       0.144\sym{**} &      0.0931         \\
            &    (0.0702)         &    (0.0681)         &    (0.0634)         &     (0.102)         \\
Subjective&       0.749\sym{***}&       0.643\sym{***}&       0.287\sym{***}&       0.182         \\
            &    (0.0926)         &    (0.0919)         &     (0.103)         &     (0.115)         \\
Choice Model&       0.477\sym{***}&       0.409\sym{***}&       0.159\sym{**} &      0.0905         \\
            &    (0.0706)         &    (0.0667)         &    (0.0664)         &    (0.0698)         \\
\addlinespace
\multicolumn{5}{c}{\textbf{C: Pay Correlation with Specific Amenities}} \\
\midrule
Permanent   &      0.0851\sym{***}&      0.0770\sym{***}&      0.0218         &     0.00272         \\
            &    (0.0238)         &    (0.0203)         &    (0.0250)         &    (0.0205)         \\
Work From Home&     0.00435         &     0.00691         &     -0.0168         &    -0.00659         \\
            &    (0.0355)         &    (0.0311)         &    (0.0290)         &    (0.0320)         \\
Shift Work  &      -0.117\sym{***}&      -0.111\sym{***}&     -0.0754\sym{**} &     -0.0545\sym{*}  \\
            &    (0.0231)         &    (0.0242)         &    (0.0330)         &    (0.0309)         \\
Flexible Hours&      0.0519\sym{**} &      0.0372\sym{*}  &      0.0185         &      0.0139         \\
            &    (0.0207)         &    (0.0196)         &    (0.0228)         &    (0.0438)         \\
\midrule
Occupation $\times$ Sector FE&                     &$\checkmark$         &                     &$\checkmark$         \\
Observations&        9,446         &        9,446         &        9,446         &        9,446         \\
Respondents &         479         &         479         &         479         &         479    \\
Job Ads        &         994         &         994         &         994         &         994         \\
\bottomrule
\end{tabular}

%% file: export_tsd/randomization.tex
\begin{tabular}{lcccc}

\toprule
& Control & Treatment & Difference & Balance \\
& (1) & (2) & (3) & (4) \\
\midrule
\addlinespace
\multicolumn{5}{c}{\textbf{A: Respondent Characteristics}} \\
\midrule
Age                &        21.7&        21.8&        .133&      .00211\\
                    &      (2.92)&      (3.16)&      (.195)&    (.00706)\\
Male                &        .322&        .283&      -.0388&      -.0577\\
                    &      (.467)&       (.45)&     (.0294)&     (.0351)\\
Immigrant Background&       .0939&       .0993&      .00535&      .00526\\
                    &      (.292)&      (.299)&     (.0189)&     (.0542)\\
Year in Program   &         2.5&        2.56&       .0582&      .00944\\
                    &      (2.03)&      (1.97)&      (.128)&     (.0102)\\
Years Studied       &        1.71&        1.71&     -.00214&      -.0089\\
                    &       (1.8)&      (1.68)&      (.112)&     (.0146)\\
Currently Works   &        .656&        .637&      -.0182&      -.0385\\
                    &      (.475)&      (.481)&     (.0307)&     (.0347)\\
Ever Worked Full Time&        .317&        .383&       .0657&         .08\\
                    &      (.465)&      (.486)&     (.0304)&     (.0355)\\
\addlinespace
\multicolumn{5}{c}{\textbf{B: Job Characteristics}} \\
\midrule
Permanent           &        .868&        .867&     -.00108&     -.00781\\
                    &      (.339)&       (.34)&    (.00364)&    (.00532)\\
Work From Home      &       .0528&       .0568&      .00402&     -.00169\\
                    &      (.224)&      (.231)&    (.00261)&    (.00567)\\
Shift Work          &       .0819&       .0796&     -.00231&       .0033\\
                    &      (.274)&      (.271)&    (.00245)&    (.00535)\\
Flexible Hours      &        .219&         .22&      .00132&       .0119\\
                    &      (.413)&      (.414)&    (.00397)&    (.00999)\\
Travel Time &        1.05&        1.07&       .0267&     -.00156\\
                    &      (.793)&      (.804)&     (.0212)&    (.00789)\\
On the Job Training &         .64&        .633&     -.00726&       .0202\\
                    &       (.48)&      (.482)&    (.00466)&      (.012)\\
Good Colleagues     &        .721&        .718&     -.00238&      -.0118\\
                    &      (.449)&       (.45)&    (.00429)&    (.00867)\\
Central Location    &        .613&        .616&      .00239&       .0036\\
                    &      (.487)&      (.486)&    (.00482)&    (.00709)\\
Other Minor Perks   &       .0813&       .0826&      .00131&       .0143\\
                    &      (.273)&      (.275)&    (.00269)&    (.00908)\\
\midrule
Observations        &       19,160&       19,740&       38,900&       38,860\\
Respondents         &         479&         494&         973&         973\\
Job Ads                 &        1,172&        1,170&        1,175&        1,175\\
Joint $ F$-statistic&            &            &            &        1.16\\
$ p$-value          &            &            &            &         [.29]\\
\bottomrule
\end{tabular}

%% file: export_tsd/effects_beliefs.tex
\begin{tabular}{lcccccc}
\toprule
& \multicolumn{3}{c}{Posterior Mean} & \multicolumn{3}{c}{Posterior Variance}  \\
& (1) & (2) & (3) & (4) & (5) & (6) \\
\midrule
\addlinespace
\multicolumn{7}{c}{\textbf{A: Baseline Treatment Effects}} \\
\midrule
Learning Treatment&      0.0386\sym{**} &      0.0359\sym{**} &      0.0408\sym{**} &    -0.00442\sym{**} &    -0.00410\sym{**} &    -0.00414\sym{**} \\
            &    (0.0183)         &    (0.0182)         &    (0.0178)         &   (0.00175)         &   (0.00180)         &   (0.00179)         \\
Control Group Mean&       10.63\sym{***}&       10.63\sym{***}&       10.63\sym{***}&      0.0264\sym{***}&      0.0263\sym{***}&      0.0263\sym{***}\\
            &    (0.0120)         &    (0.0121)         &    (0.0118)         &  (0.000974)         &  (0.000996)         &  (0.000985)         \\
\addlinespace
\multicolumn{7}{c}{\textbf{B: Treatment Interaction Effects w/ Gender and Current Pay}} \\
\midrule
%Information Treatment&      0.0428\sym{**} &      0.0399\sym{**} &      0.0410\sym{**} &    -0.00452\sym{***}&    -0.00417\sym{**} &    -0.00411\sym{**} \\
%            &    (0.0171)         &    (0.0170)         &    (0.0169)         &   (0.00171)         &   (0.00174)         &   (0.00177)         \\
%Male        &      0.0596\sym{***}&      0.0504\sym{**} &      0.0487\sym{**} &     0.00106         &     0.00225         &     0.00205         \\
%            &    (0.0196)         &    (0.0213)         &    (0.0208)         &   (0.00188)         &   (0.00196)         &   (0.00200)         \\
%Current Pay&       0.106\sym{***}&       0.106\sym{***}&       0.115\sym{***}&    -0.00628         &    -0.00608         &    -0.00663\sym{*}  \\
%            &    (0.0319)         &    (0.0316)         &    (0.0315)         &   (0.00396)         &   (0.00394)         &   (0.00386)         \\
\multicolumn{7}{l}{Learning Treatment  $\times$ ...} \\
... Male&     -0.0224         &     -0.0233         &     -0.0187         &    -0.00344         &    -0.00326         &    -0.00321         \\
            &    (0.0368)         &    (0.0367)         &    (0.0360)         &   (0.00321)         &   (0.00324)         &   (0.00328)         \\
... Current Pay&     -0.0569         &     -0.0617         &     -0.0630         &    -0.00152         &   -0.000481         &   0.0000236         \\
            &    (0.0621)         &    (0.0619)         &    (0.0626)         &   (0.00688)         &   (0.00665)         &   (0.00650)         \\
\addlinespace
\multicolumn{7}{c}{\textbf{C: Treatment Interaction Effects w/ Amenity Value}} \\
\midrule
\multicolumn{3}{l}{Learning Treatment  $\times$ ...} \\
...  Literature&     -0.0479         &     -0.0490         &     -0.0609         &    -0.00227         &   -0.000780         &    -0.00141         \\
            &    (0.0839)         &    (0.0809)         &    (0.0758)         &    (0.0105)         &    (0.0104)         &    (0.0103)         \\
...  Subjective&      -0.116         &      -0.111         &      -0.118         &      0.0107         &      0.0129         &      0.0114         \\
            &     (0.128)         &     (0.124)         &     (0.119)         &    (0.0169)         &    (0.0168)         &    (0.0167)         \\
...  Choice Model&     -0.0552         &     -0.0565         &     -0.0647         &     0.00294         &     0.00449         &     0.00367         \\
            &    (0.0896)         &    (0.0867)         &    (0.0806)         &    (0.0117)         &    (0.0116)         &    (0.0116)         \\
\midrule
Job Ad FE      &                     &$\checkmark$         &$\checkmark$         &                     &$\checkmark$         &$\checkmark$         \\
Respondent Controls    &                     &                     &$\checkmark$         &                     &                     &$\checkmark$         \\
Survey Batch FE    &                     &                     &$\checkmark$         &                     &                     &$\checkmark$         \\
Observations &       19,219         &       19,219         &       19,219         &       19,219         &       19,219         &       19,219         \\
Respondents &         973         &         973         &         973         &         973         &         973         &         973         \\
Job Ads        &        1,125         &        1,125         &        1,125         &        1,125         &        1,125         &        1,125         \\
\bottomrule
\end{tabular}

%% file: export_tsd/effects_realized_jobs.tex
\begin{tabular}{lccccc}
\toprule
& \textbf{} & \multicolumn{2}{c}{\textbf{}} & \multicolumn{2}{c}{\textbf{Full Information}} \\
& \textbf{Control Group:} & \multicolumn{2}{c}{\textbf{Learning Treatment:}} & \multicolumn{2}{c}{\textbf{Treatment:}} \\
& \textit{Mean} & \textit{Mean} & \textit{Difference} & \textit{Mean} & \textit{Difference} \\
& (1) & (2) & (3) & (4) & (5)  \\
\midrule
\multicolumn{5}{l}{Characteristics of Preferred Choice Alternative: } \\
\hspace{.1cm} Actual Pay&       10.82\sym{***}&       10.81\sym{***}&    -0.0080\sym{**} &       10.86\sym{***}&      0.0383\sym{***}\\
            &   (0.0024)         &   (0.0025)         &   (0.0036)         &   (0.0026)         &   (0.0056)         \\
\hspace{.1cm}  Amenity Value&      0.0141\sym{***}&      0.0146\sym{***}&    0.0002         &      0.0121\sym{***}&    -0.0020         \\
            &   (0.0013)         &   (0.0015)         &   (0.0025)         &   (0.0014)         &   (0.0027)         \\
\hspace{.1cm}  Sorkin Value&      0.0581\sym{***}&      0.0555\sym{***}&      0.0111         &      0.0756\sym{***}&      0.0324         \\
            &    (0.0130)         &    (0.0137)         &    (0.0213)         &    (0.0140)         &    (0.0238)         \\
\hspace{.1cm} Poaching Index&      0.0596\sym{***}&      0.0495\sym{***}&     0.00204         &      0.0796\sym{***}&      0.0381         \\
            &    (0.0125)         &    (0.0134)         &    (0.0206)         &    (0.0132)         &    (0.0239)         \\
\midrule
Observations&        4,790         &        4,940         &        9,730         &        4,940      & 9,730   \\
Respondents &         479         &         494         &         973         &         494      & 973   \\
Job Ads        &         971         &         984         &        1,126         &         975      & 1,117   \\
\bottomrule
\end{tabular}

%% file: export_tsd/treatment_wtp.tex
\begin{tabular}{lccccc}
\toprule
%& (1) & (2) & (3) & (4) & (5)  \\
%& \shortstack{Control \\ Group, \\ no info} & \shortstack{Treatment \\ Group, \\ no info} & \shortstack{Learning \\ Treatment} & \shortstack{Treatment \\ Group, \\ info} & \shortstack{Pay \\ Transparency \\ Treatment} \\
& \textbf{} & \multicolumn{2}{c}{\textbf{}} & \multicolumn{2}{c}{\textbf{Full Information}} \\
& \textbf{Control Group:} & \multicolumn{2}{c}{\textbf{Learning Treatment:}} & \multicolumn{2}{c}{\textbf{Treatment:}} \\
& \textit{WTP} & \textit{WTP} & \textit{Difference} & \textit{WTP} & \textit{Difference} \\
& (1) & (2) & (3) & (4) & (5)  \\
\midrule
Permanent   &      0.0721\sym{***}&      0.0371\sym{*}  &     -0.0350         &       0.198\sym{***}&       0.126\sym{***}\\
            &    (0.0172)         &    (0.0207)         &    (0.0302)         &    (0.0335)         &    (0.0381)         \\
Work From Home&     -0.0486\sym{*}  &      0.0226         &      0.0711         &      0.0230         &      0.0715         \\
            &    (0.0253)         &    (0.0390)         &    (0.0514)         &    (0.0453)         &    (0.0518)         \\
Shift Work  &     -0.0386\sym{*}  &     -0.0631\sym{**} &     -0.0245         &    -0.00957         &      0.0290         \\
            &    (0.0234)         &    (0.0284)         &    (0.0410)         &    (0.0424)         &    (0.0488)         \\
Flexible Hours&      0.0652\sym{***}&      0.0475\sym{**} &     -0.0177         &       0.115\sym{***}&      0.0496         \\
            &    (0.0168)         &    (0.0202)         &    (0.0296)         &    (0.0293)         &    (0.0342)         \\
Travel Time&     -0.0352\sym{***}&     -0.0526\sym{***}&     -0.0174         &     -0.0823\sym{***}&     -0.0471\sym{**} \\
            &   (0.00808)         &    (0.0120)         &    (0.0159)         &    (0.0176)         &    (0.0195)         \\
\midrule
Choice Alternatives&        4,790         &        4,940         &        9,730         &        4,940         &        9,730         \\
Respondents &         479         &         494         &         973         &         494         &         973         \\
Job Ads        &         971         &         984         &        1097         &         975         &        1117         \\
Joint $ \chi2$ statistic&                     &                     &       4.992         &                     &       20.34         \\
$ p$ value  &                     &                     &       [0.417]         &                     &     [0.00108]         \\
\bottomrule
\end{tabular}

%% file: results_old/ads_collection.tex
\begin{table}[!h]
\caption{Overview of Job Ads Used in the Survey.}
\label{tab:jobAdsCollection}
\begin{center}
 \vspace{-1em}
 \begin{adjustbox}{center}
 \scalebox{0.9}{
 \renewcommand{\arraystretch}{0.8}
\begin{tabular}{>{\raggedright\arraybackslash}p{5cm}>{\centering\arraybackslash}p{2.5cm}>{\centering\arraybackslash}p{2.5cm}>{\centering\arraybackslash}p{2.5cm}}
\toprule
 & \textbf{Initial Sample of Job Ads} & \textbf{Linked to Employer, Occupation and Salary} & \textbf{Final Sample of Job Ads}\\
\midrule
 & (1) & (2) & (3) \\
\midrule
Sector:\\
\hspace{.1cm} Private Sector Job & 68.3\% & 72.5\% & 66.7\%\\
\hspace{.1cm} Public Sector Job & 26.3\% & 23.0\% & 28.5\%\\
\midrule
Workplace Attributes:\\
\hspace{.1cm} Permanent & 76.4\% & 79.2\% & 87.7\%\\
\hspace{.1cm} Work From Home & 10.3\% & 13.3\% & 15.4\%\\
\hspace{.1cm} Shift Work & 10.9\%  & 8.9\% & 5.0\%\\
\hspace{.1cm} Flexible Hours & 11.9\% & 13.7\% & 17.3\%\\
\hspace{.1cm} On-the-Job Training & 56.9\% & 55.0\% & 61.1\%\\
\hspace{.1cm} Good Colleagues & 37.7\% & 36.5\% & 40.0\%\\
\hspace{.1cm} Central location & 55.5\%  & 58.4\% & 59.2\%\\
\hspace{.1cm} Other Minor Perks & 64.5\%  & 67.7\% & 67.2\%\\
%Teaching &  &  & 3.9\% & 12.6\%\\
%Advisors within Finance, \hspace*{1em}Administration and Sales &  & 21.0\% & 54.3\%\\
%Legal, Social Science and \hspace*{1em}Humanities Occupations &  & 2.3\% & 5.8\%\\
%Employees in Finance, \hspace*{1em}Administration and Sales &  & 8.1\% & 21.6\%\\
%Finance and Logistics Employees &  & 2.6\% & 5.8\%\\
\midrule
Number of Job Ads & 11,393 & 5,017 & 1,234\\
\bottomrule
\end{tabular}
}
\end{adjustbox}
\end{center}
\par \scriptsize {
\emph{Notes:} This table shows the composition of job ads used in the survey in terms of the sector of posting employer and offered workplace attributes in the job ad texts. Column (1) shows the initial sample of job ads for full-time positions posted by employers based in Oslo or the surrounding labor market regions between June and December 2023. Column (2) considers the subset of job ads with information on establishment identifiers, assigned occupation and information on expected starting salary. Column (3) considers the final subset of high-skilled service sector job ads, including advisors and professionals in consulting, sales, teaching, finance, public administration, logistics, and legal occupations, with an expected monthly starting salary between \$2,000 and \$8,000, excluding job ads posted by staffing agencies, and excluding job ads that explicitly provided information on salary compensation as part of the publicly posted information.
}
\end{table}

\begin{comment}
\begin{table}[!h]
\caption{Overview of Relevant Job Ads Used in the Survey.}
\label{tab:jobAdsCollection}
\centering
 \begin{adjustbox}{center}
 \scalebox{0.9}{
 \renewcommand{\arraystretch}{0.8}
\begin{tabular}{>{\raggedright\arraybackslash}p{5cm}>{\centering\arraybackslash}p{2.5cm}>{\centering\arraybackslash}p{2.5cm}>{\centering\arraybackslash}p{2.5cm}>{\centering\arraybackslash}p{2.5cm}}
\toprule

 & \textbf{All Scraped Ads} & \textbf{Matched to Employer ID} & \textbf{Matched to Actual Pay Information} & \textbf{Relevant Sample of Ads}\\
\midrule
Private Sector Job & 68.3\% & 61.8\% & 72.5\% & 66.7\%\\
Public Sector Job & 26.3\% & 34.4\% & 23.0\% & 28.5\%\\
Permanent Job & 76.4\% & 79.9\% & 79.2\% & 87.7\%\\
Work from Home & 10.3\% & 12.0\% & 13.3\% & 15.4\%\\
Shift Work & 10.9\% & 11.4\% & 8.9\% & 5.0\%\\
Flexible Hours & 11.9\% & 15.0\% & 13.7\% & 17.3\%\\
On-the-Job Training & 56.9\% & 57.9\% & 55.0\% & 61.1\%\\
Good Colleagues & 37.7\% & 40.1\% & 36.5\% & 40.0\%\\
Central location & 55.5\% & 58.7\% & 58.4\% & 59.2\%\\
Other Minor Perks & 64.5\% & 68.8\% & 67.7\% & 67.2\%\\
%Teaching &  &  & 3.9\% & 12.6\%\\
%Advisors within Finance, \hspace*{1em}Administration and Sales &  &  & 21.0\% & 54.3\%\\
%Legal, Social Science and \hspace*{1em}Humanities Occupations &  &  & 2.3\% & 5.8\%\\
%Employees in Finance, \hspace*{1em}Administration and Sales &  &  & 8.1\% & 21.6\%\\
%Finance and Logistics Employees &  &  & 2.6\% & 5.8\%\\
\midrule
Number of Job Ads & 11,393 & 6,619 & 5,017 & 1,234\\
\bottomrule
\end{tabular}
}
\end{adjustbox}
\end{table}
\end{comment}

%% file: export_tsd/timeuse.tex
{
\def\sym#1{\ifmmode^{#1}\else\(^{#1}\)\fi}
\begin{tabular}{l*{2}{c}}
\toprule
            &\multicolumn{1}{c}{(1)}         &\multicolumn{1}{c}{(2)}         \\
\midrule
Pay Information Module&      -3.702\sym{***}&      -3.702\sym{***}\\
            &     (0.228)         &     (0.231)         \\
\addlinespace
Second Module Taken&      -3.157\sym{***}&      -3.157\sym{***}\\
            &     (0.228)         &     (0.231)         \\
\addlinespace
Treatment Group&      -0.306         &      -0.306         \\
            &     (0.577)         &     (0.600)         \\
\addlinespace
Constant    &       10.60\sym{***}&       10.60\sym{***}\\
            &     (0.452)         &     (0.454)         \\
\midrule
Observations&        1,938         &        1,938         \\
Respondents &         969         &         969         \\
Survey Round FE&                     &$\checkmark$         \\
\bottomrule
\end{tabular}
}

%% file: results/literatureA.tex
\newcommand{\tabOne}{\input{raw_files/literatureA.tex}}% table.tex > \mytable
{
\def\sym#1{\ifmmode^{#1}\else\(^{#1}\)\fi}
\renewcommand{\arraystretch}{0.8}
\begin{tabular}{lllccp{8cm}ccccp{4cm}}

    \midrule   
    Paper & Method & Country & Obs. & Scale & Amenity & Mean & SE & Scaled Mean & Scaled SE & Source \\
    \midrule

\input{raw_files/literatureB.tex}
\hline\hline
\addlinespace
\multicolumn{11}{p{2\linewidth}}{\footnotesize \emph{Notes:} This table contains reported estimates on willingness-to-pay (WTP) for workplace amenities in the surveyed literature that are relevant for our study. Column ``Scale'' indicates the scaling factor we apply to give estimates a relative interpretation when earnings of jobs are specified in levels, and correspond to baseline wages in the study in the local currency. For the travel time estimates, the ``Scale'' column contains the scaling necessary for the estimate to reflect the willingness-to-pay for an hour longer daily commute (30 minutes of additional travel time each way), in order to be comparable to our survey.}\\

\end{tabular}
}

%% file: raw_files/literatureA.tex
\addlinespace\addlinespace
\multicolumn{10}{l}{\textbf{Permanent}} \\
\cite{datta2019} & Mixed logit & UK & 2013 & 13 & Permanent position  vs. 1 month & 9.69 & 0.313 & 0.745 & 0.024 & Table 6 \\
\cite{datta2019} & Mixed logit & US & 1871 & 18.4971 & Permanent position  vs. 1 month & 11.13 & 0.422 & 0.602 & 0.023 & Table 6 \\
\cite{datta2019} & Mixed logit & UK & 2013 & 13 & 1 year position vs. 1 month & 6.54 & 0.196 & 0.503 & 0.015 & Table 6 \\
\cite{datta2019} & Mixed logit & US & 1871 & 18.4971 & 1 year position vs. 1 month & 7.88 & 0.275 & 0.426 & 0.015 & Table 6 \\
\cite{mahmud} & Conditional Logit & BD & 23568 & 100 & Permanent position  vs. No contract & 43.529 & 2.369 & 0.435 & 0.024 & Table 5 (col 4) \\
\cite{mahmud} & Conditional Logit & BD & 23568 & 100 & 1 year contract  vs. No contract & 26.863 & 1.511 & 0.269 & 0.015 & Table 5 (col 4) \\
\cite{mahmud} & Conditional Logit & BD & 23568 & 100 & 6 month contract  vs. No contract & 19.02 & 1.27 & 0.19 & 0.013 & Table 5 (col 4) \\
%\cite{valet2021} & Conditional Logit & DE & 2659 &  & Permament contract vs 2-year & 0 & 0 & 0 & 0 & S2 Table \\
\addlinespace\addlinespace
\multicolumn{10}{l}{\textbf{Work From Home}} \\
\cite{datta2019} & Mixed logit & UK & 2013 & 13 & 50\% work from home vs. 0 & 1.87 & 0.199 & 0.232 & 0.015 & Table 6 \\
\cite{datta2019} & Mixed logit & US & 1871 & 18.4971 & 50\% work from home vs. 0 & 3.97 & 0.291 & 0.268 & 0.016 & Table 6 \\
\cite{datta2019} & Mixed logit & UK & 2013 & 13 & 100\% work from home vs. 0 & 3.02 & 0.215 & 0.232 & 0.017 & Table 6 \\
\cite{datta2019} & Mixed logit & US & 1871 & 18.4971 & 100\% work from home vs. 0 & 4.95 & 0.321 & 0.268 & 0.017 & Table 6 \\
\cite{drakeetal2023} & Bayesian mixed logit & US & 1500 & 1 & Flexible location & 0.139 & 0.003 & 0.139 & 0.003 & Figure 4 \\
\cite{mas2017valuing} & Mixed logit & US & 608 & 17 & Work from home & 1.33 & 0.29 & 0.078 & 0.017 & Table 5 \\
\cite{nagleretal2022} & Logit & DE & 3307 & 1 & Working from home up to 2 days & 0.055 & 0.004 & 0.055 & 0.004 & Figure 1 \\
\cite{nagleretal2022} & Logit & DE & 3307 & 1 & Working from home up to 5 days & 0.075 & 0.005 & 0.075 & 0.005 & Figure 1 \\
\cite{lewandowski2023} & Logit & PL & 11116 & 1 & WFH 2-3 days a week & 0.051 & 0.003 & 0.051 & 0.003 & Table 7 \\
\cite{lewandowski2023} & Logit & PL & 11116 & 1 & WFH 5 days a week & 0.613 & 0.003 & 0.613 & 0.003 & Table 7 \\
\cite{maestas2023value}  & Mixed logit & US & 1738 & 1 & Telecommute & 0.041 & 0.007 & 0.041 & 0.007 & Table 2 \\
\cite{kesternich} & Logit & NL & 778 & 1 & Telecommute & 0.071 & 0.008 & 0.071 & 0.008 & Table 3 (control) \\
\cite{kesternich} & Logit & NL & 1317 & 1 & Telecommute & 0.039 & 0.005 & 0.039 & 0.005 & Table 4 (control) \\
\cite{derbyshire} & Multinomial logit & UK & 109 & 1 & Working from home ("job involves working from home and does not require travel") & 0.512 & 0.257 & 0.512 & 0.257 & Table 5 (no-disability) \\
\cite{landeghem} & Conditional logit & BE & 20396 & 1 & Work from home (at least 1 day) & 0.15 & 0.027 & 0.15 & 0.027 & Table 3 \\
\cite{landeghem} & Conditional logit & BE & 20396 & 1 & Work from home (at least 2 days) & 0.176 & 0.027 & 0.176 & 0.027 & Table 3 \\
\cite{landeghem} & Conditional logit & BE & 20396 & 1 & Work from home ("as much as I want") & 0.197 & 0.027 & 0.197 & 0.027 & Table 3 \\
\addlinespace\addlinespace
\multicolumn{10}{l}{\textbf{Shift Work}} \\
\cite{desiere2023} & Nonparametric & BE & 709 & 15 & Morning  shift & -0.56 & 0.22 & -0.037 & 0.015 & Table 4, median \\
\cite{desiere2023} & Nonparametric & BE & 709 & 15 & Evening shift & -3.36 & 0.26 & -0.224 & 0.017 & Table 4, median \\
\cite{desiere2023} & Nonparametric & BE & 709 & 15 & Night shift & -5.42 & 0.49 & -0.361 & 0.033 & Table 4, median \\
\cite{desiere2023} & Nonparametric & BE & 709 & 15 & Rotating - predictable & -3.36 & 0.28 & -0.224 & 0.019 & Table 4, median \\
\cite{desiere2023} & Nonparametric & BE & 709 & 15 & Rotating - unpredictable & -5.71 & 0.6 & -0.381 & 0.04 & Table 4, median \\
\cite{kesternich} & Logit & NL & 778 & 1 & Evening, night or weekend shifts & -0.179 & 0.01 & -0.179 & 0.01 & Table 3 (control) \\
\cite{kesternich} & Logit & NL & 1317 & 1 & Evening, night or weekend shifts & -0.108 & 0.005 & -0.108 & 0.005 & Table 4 (control) \\
\addlinespace\addlinespace
\multicolumn{10}{l}{\textbf{Flexible Hours}} \\
\cite{nagleretal2022} & Logit & DE & 3307 & 1 & Flexible schedule & 0.054 & 0.003 & 0.054 & 0.003 & Figure 1 \\
\cite{drakeetal2023} & Bayesian mixed logit & US & 1500 & 1 & Flexible schedule & 0.09 & 0.003 & 0.09 & 0.003 & Figure 4 \\
\cite{mas2017valuing} & Mixed logit & US & 640 & 17 & Flexible schedule & 0.48 & 0.24 & 0.028 & 0.014 & Table 5 \\
\cite{maestas2023value}  & Mixed logit & US & 1738 & 1 & Set own schedule & 0.088 & 0.007 & 0.088 & 0.007 & Table 2 \\
\cite{kesternich} & Logit & NL & 778 & 1 & Full schedule freedom & 0.109 & 0.012 & 0.109 & 0.012 & Table 3 (control) \\
\cite{kesternich} & Logit & NL & 1317 & 1 & Full schedule freedom & 0.051 & 0.008 & 0.051 & 0.008 & Table 4 (control) \\
\cite{derbyshire} & Multinomial logit & UK & 109 & 1 & Flexible schedule & 0.093 & 0.06 & 0.093 & 0.06 & Table 5 (no-disability) \\
\cite{bustelo} & Conditional logit & CO & 36651 & 284 & Flexible schedule & 56.65 & 6.454 & 0.199 & 0.023 & Table 7 (text above for CI) \\
\cite{landeghem} & Conditional logit & BE & 20396 & 1 & Various fixed schedules & 0.223 & 0.018 & 0.223 & 0.018 & Table 3 \\
\cite{landeghem} & Conditional logit & BE & 20396 & 1 & Can ask changes & 0.213 & 0.018 & 0.213 & 0.018 & Table 3 \\
\cite{landeghem} & Conditional logit & BE & 20396 & 1 & Complete flexibility & 0.192 & 0.017 & 0.192 & 0.017 & Table 3 \\
\cite{jost2023} & Linear probability & CH & 1796 & 100 & Flexible schedule & 2.76 & 0.087 & 0.028 & 0.001 & Table S7 \\
\addlinespace\addlinespace
\multicolumn{10}{l}{\textbf{Travel Time}} \\
\cite{nagleretal2022} & Logit & DE & 3307 & 0.5 & Commuting time (30 vs 15) & -0.053 & 0.005 & -0.106 & 0.01 & Figure 1 \\
\cite{nagleretal2022} & Logit & DE & 3307 & 1 & Commuting time (45 vs 15) & -0.13 & 0.005 & -0.13 & 0.005 & Figure 1 \\
\cite{nagleretal2022} & Logit & DE & 3307 & 1.5 & Commuting time (60 vs 15) & -0.22 & 0.006 & -0.147 & 0.004 & Figure1 \\
\cite{willen} & Nonparametric & NO & 10008 & 0.6667 & Doubling of commuting time, men & -0.112 & 0.003 & -0.168 & 0.005 & Correspondence based on Figure A2 \\
\cite{landeghem} & Conditional logit & BE & 20396 & 1 & Respondent baseline + 15 min, daily & -0.116 & 0.015 & -0.116 & 0.015 & Table 3 \\
\cite{landeghem} & Conditional logit & BE & 20396 & 1 & Respondent baseline - 15 min, daily & 0.09 & 0.015 & 0.09 & 0.015 & Table 3 \\
\cite{labarbanchon} & Admin data & FR & 75071 & 1 & CES of wage and commuting time & -0.135 & 0.048 & -0.135 & 0.048 & Table V \\

%% file: raw_files/literatureB.tex
\addlinespace
\cite{bustelo} &  &  &  & 0.199 &  & 36651 &  \\
 &  &  &  & (0.023) &  &  &  \\
\cite{willen} &  &  &  &  & -0.168 & 10008 &  \\
 &  &  &  &  & (0.005) &  &  \\
\cite{datta2019}, UK & \shortstack{0.242 \\ (0.028)} & \shortstack{0.232 \\ (0.015)} &  &  &  & 2013 & Permanent: Difference between permanent and 1 year \\
\cite{datta2019}, US & \shortstack{0.176 \\ (0.027)} & \shortstack{0.268 \\ (0.016)} &  &  &  & 1871 & Home office: Option to work from home 50\% \\
\cite{derbyshire}  &  & 0.512 &  & 0.093 &  & 109 &  \\
 &  & (0.257) &  & (0.060) &  &  &  \\
\cite{desiere2023} &  &  & \shortstack{-0.104 \\ (0.022)} &  &  & 709 & Predictable rotating shift - 1/3 night shift \\
\cite{drakeetal2023} &  & 0.139 &  & 0.090 &  & 1500 & Flexible schedule, location \\
 &  & (0.003) &  & (0.003) &  &  &  \\
\cite{jost2023} &  &  &  & 0.028 &  & 1796 &  \\
 &  &  &  & (0.001) &  &  &  \\
\cite{kesternich}, study A &  & 0.071 & -0.179 & 0.109 &  & 778 &  \\
 &  & (0.008) & (0.010) & (0.012) &  &  &  \\
\cite{kesternich}, study B &  & 0.039 & -0.108 & 0.051 &  & 1317 &  \\
 &  & (0.005) & (0.005) & (0.008) &  &  &  \\
\cite{labarbanchon} &  &  &  &  & \shortstack{-0.064 \\ (0.023)} & 75071 & Average commuting distance 18.6 km, average speed is 3.4 min/km. Percent estimate from paper scaled accordingly. \\
\cite{lewandowski2023} &  & 0.051 &  &  &  & 11116 & Average of WTP at all margins \\
 &  & (0.003) &  &  &  &  &  \\
\cite{maestas2023value} &  & 0.041 &  &  &  & 1738 & Telecommute \\
 &  & (0.007) &  &  &  &  &  \\
\cite{mahmud} & \shortstack{0.167 \\ (0.028)} &  &  &  &  & 23568 & Difference in WTP(permanent vs no contract) and WTP(1 year vs no contract) \\
\cite{mas2017valuing} &  & 0.078 &  & 0.028 &  & 608 & Work from home, flexible schedule \\
 &  & (0.017) &  & (0.014) &  &  &  \\
\cite{nagleretal2022} &  & 0.055 &  &  & -0.128 & 3307 & Work from home up to 2 days \\
 &  & (0.004) &  &  & (0.004) &  &  \\
\cite{landeghem} &  & \shortstack{0.176 \\ (0.027)} &  & \shortstack{0.192 \\ (0.017)} & \shortstack{-0.464 \\ (0.060)} & 20396 & Multiplying by 4 to get 30 mins each way \\
\midrule
Average & 0.195 & 0.151 & -0.130 & 0.099 & -0.206 &  &  \\
 & (0.016) & (0.024) & (0.008) & (0.009) & (0.016) &  &  \\
Weighted average & 0.173 & 0.130 & -0.127 & 0.184 & -0.150 &  &  \\
 & (0.024) & (0.012) & (0.007) & (0.014) & (0.019) &  &  \\

%% file: results/literatureB.tex
\newcommand{\tabOne}{\input{raw_files/literatureB.tex}}% table.tex > \mytable
{
\def\sym#1{\ifmmode^{#1}\else\(^{#1}\)\fi}
\renewcommand{\arraystretch}{0.8}
\begin{tabular}{lp{2cm}p{1.5cm}p{1.5cm}p{1.5cm}p{1.5cm}p{1.5cm}p{7cm}}
\toprule
      Paper & Permanent & Work from Home & Shift Work & Flexible Hours & Travel Time & Obs. & Source from Table \ref{tab:wtpLitA}\\
        \midrule

\input{raw_files/literatureB.tex}
\hline\hline
\addlinespace
\multicolumn{8}{p{1.5\linewidth}}{\scriptsize \emph{Notes:} This table contains our preferred estimates of willingness-to-pay (WTP) from the surveyed papers in Table \ref{tab:wtpLitA}, where the relevant parameters are chosen to most closely match the definition of amenities in our survey and the Norwegian labor market. For each paper, we take the amenity measure that most closely resembles the amenity of interest in our survey, or construct one from multiple WTP parameters reported in the papers as detailed in the ``Source from Table \ref{tab:wtpLitA}'' column. The weighted averages reported at the bottom of this table use the relative share of observations as weights. Variance estimates ignore covariances between parameter estimates within paper, which are typically not reported in the original studies.}\\
\end{tabular}
}

%% file: export_tsd/baselinebeliefs_var.tex
\begin{tabular}{lcccc}
\toprule
& (1) & (2) & (3) & (4) \\
\midrule
\addlinespace
\multicolumn{5}{c}{\textbf{A: Job Quality}} \\
\midrule
Sorkin Value&     0.00130         &     0.00132         &     0.00153\sym{*}  &     0.00151\sym{*}  \\
            &  (0.000840)         &  (0.000818)         &  (0.000905)         &  (0.000871)         \\
Poaching Index&    0.000505         &    0.000584         &    0.000885         &     0.00100         \\
            &  (0.000818)         &  (0.000797)         &  (0.000918)         &  (0.000896)         \\
\addlinespace
\multicolumn{5}{c}{\textbf{B: Actual Pay}} \\
\midrule
Employer Pay Premium&      0.0194\sym{**} &      0.0181\sym{*}  &      0.0151         &      0.0124         \\
            &   (0.00985)         &   (0.00929)         &    (0.0108)         &    (0.0101)         \\
Actual Job Pay  &     0.00235         &     0.00195         &     0.00100         &   0.0000278         \\
            &   (0.00364)         &   (0.00372)         &   (0.00364)         &   (0.00367)         \\
\addlinespace
\multicolumn{5}{c}{\textbf{C: Amenity Value}} \\
\midrule
Literature  &     -0.0180\sym{***}&     -0.0172\sym{***}&     -0.0205\sym{***}&     -0.0199\sym{***}\\
            &   (0.00677)         &   (0.00606)         &   (0.00736)         &   (0.00664)         \\
Subjective  &     -0.0270\sym{**} &     -0.0259\sym{***}&     -0.0286\sym{**} &     -0.0281\sym{***}\\
            &    (0.0108)         &   (0.00970)         &    (0.0116)         &    (0.0104)         \\
Choice Model&     -0.0197\sym{***}&     -0.0178\sym{***}&     -0.0215\sym{***}&     -0.0199\sym{***}\\
            &   (0.00712)         &   (0.00653)         &   (0.00752)         &   (0.00686)         \\
\addlinespace
\multicolumn{5}{c}{\textbf{D: Specific Amenities}} \\
\midrule
Permanent   &    -0.00591\sym{***}&    -0.00432\sym{*}  &    -0.00661\sym{***}&    -0.00502\sym{**} \\
            &   (0.00225)         &   (0.00223)         &   (0.00232)         &   (0.00227)         \\
Work From Home&   -0.000379         &    -0.00128         &   -0.000104         &   -0.000755         \\
            &   (0.00265)         &   (0.00267)         &   (0.00255)         &   (0.00257)         \\
Shift Work  &     0.00305         &     0.00362         &     0.00253         &     0.00334         \\
            &   (0.00273)         &   (0.00253)         &   (0.00294)         &   (0.00273)         \\
Flexible Hours&    -0.00322\sym{*}  &    -0.00361\sym{**} &    -0.00384\sym{*}  &    -0.00413\sym{**} \\
            &   (0.00193)         &   (0.00180)         &   (0.00203)         &   (0.00194)         \\
%Travel Time &    -0.00224\sym{***}&    -0.00243\sym{***}&    -0.00173\sym{*}  &    -0.00177\sym{**} \\
%            &  (0.000868)         &  (0.000859)         &  (0.000886)         &  (0.000875)         \\
\midrule
Respondent FE   &                     &$\checkmark$                     &                     &$\checkmark$                     \\
Occupation $\times$ Sector FE&                     &        &$\checkmark$                     &$\checkmark$         \\
Observations&        9,446         &        9,446         &        9,446         &        9,446         \\
Respondents &         479         &         479         &         479         &         479         \\
Job Ads        &         994         &         994         &         994         &         994         \\
\bottomrule
\end{tabular}

%% file: export_tsd/robustness_wtp.tex
{
\def\sym#1{\ifmmode^{#1}\else\(^{#1}\)\fi}
\begin{tabular}{l*{4}{c}}
\toprule
            &\multicolumn{1}{c}{(1)}&\multicolumn{1}{c}{(2)}&\multicolumn{1}{c}{(3)}&\multicolumn{1}{c}{(4)}\\
            &\multicolumn{1}{c}{Baseline}&\multicolumn{1}{c}{\shortstack{Correlated \\ Preferences}}&\multicolumn{1}{c}{\shortstack{Attention- \\ Weighted}}&\multicolumn{1}{c}{\shortstack{No Failed \\ Attention Check}}\\
\midrule
Permanent   &       0.233\sym{***}&       0.235\sym{***}&       0.243\sym{***}&       0.238\sym{***}\\
            &    (0.0104)         &    (0.0104)         &    (0.0109)         &    (0.0104)         \\
\addlinespace
Work From Home&       0.151\sym{***}&       0.152\sym{***}&       0.156\sym{***}&       0.150\sym{***}\\
            &   (0.00817)         &   (0.00845)         &   (0.00856)         &   (0.00819)         \\
\addlinespace
Shift Work  &      -0.171\sym{***}&      -0.175\sym{***}&      -0.195\sym{***}&      -0.173\sym{***}\\
            &    (0.0127)         &    (0.0129)         &    (0.0139)         &    (0.0127)         \\
\addlinespace
Flexible Hours&      0.0799\sym{***}&      0.0826\sym{***}&      0.0835\sym{***}&      0.0801\sym{***}\\
            &   (0.00614)         &   (0.00616)         &   (0.00624)         &   (0.00604)         \\
\addlinespace
Travel Time &      -0.147\sym{***}&      -0.142\sym{***}&      -0.145\sym{***}&      -0.144\sym{***}\\
            &   (0.00537)         &   (0.00536)         &   (0.00546)         &   (0.00543)         \\
\midrule
Respondents &         973         &         973         &         973         &         907         \\
Choice Alternatives&       58,380         &       58,380         &       58,380         &       54,420         \\
\bottomrule
\end{tabular}
}

%% file: export_tsd/robustness_baselinebeliefs.tex
\begin{tabular}{lccccc}
\toprule
& (1) & (2) & (3) & (4) & (5) \\
& Baseline & \shortstack{Attention- \\Weighted} & \shortstack{No Failed \\ Attention \\ Check} & \shortstack{Wage in \\ Levels \\ (PPML)} & \shortstack{Naive \\ No \\Correction} \\ 
\midrule
\addlinespace
\addlinespace
\multicolumn{6}{c}{\textbf{A: Job Quality}} \\
\midrule
Sorkin Value&      0.0616\sym{***}&      0.0608\sym{***}&      0.0619\sym{***}&      0.0618\sym{***}&      0.0622\sym{***}\\
            &   (0.00804)         &   (0.00812)         &   (0.00827)         &   (0.00714)         &   (0.00687)         \\
Poaching Index&      0.0493\sym{***}&      0.0480\sym{***}&      0.0486\sym{***}&      0.0489\sym{***}&      0.0496\sym{***}\\
            &   (0.00881)         &   (0.00878)         &   (0.00913)         &   (0.00884)         &   (0.00809)         \\
\addlinespace
\multicolumn{6}{c}{\textbf{B: Actual Pay}} \\
\midrule
Employer Pay Premium&       0.622\sym{***}&       0.613\sym{***}&       0.612\sym{***}&       0.645\sym{***}&       0.632\sym{***}\\
            &     (0.144)         &     (0.145)         &     (0.140)         &     (0.147)         &     (0.140)         \\
Actual Job Pay  &       0.205\sym{***}&       0.208\sym{***}&       0.214\sym{***}&       0.210\sym{***}&       0.206\sym{***}\\
            &    (0.0576)         &    (0.0581)         &    (0.0587)         &    (0.0629)         &    (0.0601)         \\
\addlinespace
\multicolumn{6}{c}{\textbf{C: Amenity Value}} \\
\midrule
Literature  &       0.453\sym{***}&       0.449\sym{***}&       0.455\sym{***}&       0.419\sym{***}&       0.444\sym{***}\\
            &    (0.0639)         &    (0.0710)         &    (0.0682)         &    (0.0615)         &    (0.0584)         \\
Subjective  &       0.749\sym{***}&       0.754\sym{***}&       0.762\sym{***}&       0.732\sym{***}&       0.736\sym{***}\\
            &    (0.0889)         &    (0.0950)         &    (0.0913)         &    (0.0789)         &    (0.0719)         \\
Choice Model&       0.477\sym{***}&       0.479\sym{***}&       0.488\sym{***}&       0.459\sym{***}&       0.467\sym{***}\\
            &    (0.0660)         &    (0.0721)         &    (0.0725)         &    (0.0615)         &    (0.0574)         \\
\addlinespace
\multicolumn{6}{c}{\textbf{D: Specific Amenities}} \\
\midrule
Permanent   &      0.0851\sym{***}&      0.0847\sym{***}&      0.0881\sym{***}&      0.0795\sym{***}&      0.0821\sym{***}\\
            &    (0.0227)         &    (0.0244)         &    (0.0244)         &    (0.0217)         &    (0.0203)         \\
Work From Home&     0.00435         &     0.00666         &      0.0112         &     0.00985         &     0.00416         \\
            &    (0.0335)         &    (0.0351)         &    (0.0358)         &    (0.0341)         &    (0.0316)         \\
Shift Work  &      -0.117\sym{***}&      -0.121\sym{***}&      -0.120\sym{***}&      -0.121\sym{***}&      -0.116\sym{***}\\
            &    (0.0217)         &    (0.0222)         &    (0.0222)         &    (0.0178)         &    (0.0167)         \\
Flexible Hours&      0.0519\sym{**} &      0.0467\sym{**} &      0.0480\sym{**} &      0.0405\sym{**} &      0.0503\sym{***}\\
            &    (0.0204)         &    (0.0219)         &    (0.0209)         &    (0.0202)         &    (0.0191)         \\
%Travel Time &     -0.0380\sym{***}&     -0.0418\sym{***}&     -0.0384\sym{***}&     -0.0445\sym{***}&     -0.0391\sym{***}\\
%            &    (0.0109)         &    (0.0108)         &    (0.0109)         &    (0.0101)         &   (0.00939)         \\
\midrule
Observations&        9,773         &        9,446         &        8,825         &        9,446         &        9,446         \\
Respondents &         479         &         479         &         449         &         479         &         479         \\
Job Ads        &         994         &         994         &         963         &         994         &         994         \\
\bottomrule
\end{tabular}

%% file: export_tsd/robustness_effects.tex
\begin{tabular}{lccccc}
\toprule
& (1) & (2) & (3) & (4) & (5)  \\
& Baseline & \shortstack{Attention- \\Weighted} & \shortstack{No Failed \\ Attention \\ Check} & \shortstack{Wage in \\ Levels \\ (PPML)} & \shortstack{Naive \\ No \\Correction} \\
\midrule
\addlinespace
\multicolumn{6}{c}{\textbf{A: Posterior Mean of Pay Beliefs}} \\
\midrule
Pay Information Effect&      0.0386\sym{**} &      0.0325\sym{*}  &      0.0368\sym{**} &      0.0280         &      0.0385\sym{**} \\
            &    (0.0173)         &    (0.0183)         &    (0.0176)         &    (0.0173)         &    (0.0167)         \\
Control Group Mean&       10.63\sym{***}&       10.63\sym{***}&       10.64\sym{***}&       10.69\sym{***}&       10.64\sym{***}\\
            &    (0.0115)         &    (0.0127)         &    (0.0119)         &    (0.0116)         &    (0.0116)         \\
\addlinespace
\multicolumn{6}{c}{\textbf{B: Posterior Variance of Pay Beliefs}} \\
\midrule
Pay Information Effect&    -0.00442\sym{**} &    -0.00471\sym{***}&    -0.00511\sym{***}&                     &                     \\
            &   (0.00173)         &   (0.00182)         &   (0.00179)         &                     &                     \\
Control Group Mean&      0.0264\sym{***}&      0.0262\sym{***}&      0.0259\sym{***}&                     &                     \\
            &  (0.000969)         &   (0.00106)         &  (0.000987)         &                     &                     \\
\midrule
Observations&       19,219         &       19,219         &       17,858         &       19,219         &       19,219         \\
Respondents &         973         &         973         &         907         &         973         &         973         \\
Job Ads        &        1125         &        1125         &        1105         &        1125         &        1125         \\
\bottomrule
\end{tabular}

%% file: export_tsd/baselinebeliefs_var_p.tex
\begin{tabular}{lcccc}
\toprule
& (1) & (2) & (3) & (4) \\
\midrule
Male        &    0.000791         &    0.000673         &     0.00267         &     0.00294         \\
            &   (0.00186)         &   (0.00195)         &   (0.00214)         &   (0.00219)         \\
Age         &   -0.000261         &   -0.000194         &   -0.000141         &   -0.000129         \\
            &  (0.000483)         &  (0.000487)         &  (0.000490)         &  (0.000504)         \\
Immigrant Background&     0.00145         &     0.00106         &   -0.000615         &   -0.000870         \\
            &   (0.00300)         &   (0.00306)         &   (0.00307)         &   (0.00316)         \\
Year in Program&  -0.0000870         &  -0.0000335         &   -0.000303         &    0.000212         \\
            &  (0.000594)         &  (0.000616)         &   (0.00100)         &   (0.00105)         \\
Years Studied&    0.000381         &    0.000234         &    0.000898         &    0.000740         \\
            &  (0.000850)         &  (0.000875)         &  (0.000992)         &   (0.00102)         \\
Currently Works&   -0.000118         &   -0.000698         &    0.000502         &   -0.000188         \\
            &   (0.00174)         &   (0.00178)         &   (0.00186)         &   (0.00188)         \\
Ever Held Full-Time Job&     0.00301         &     0.00329         &     0.00265         &     0.00288         \\
            &   (0.00218)         &   (0.00217)         &   (0.00225)         &   (0.00227)         \\
Current Job: Pay&    -0.00678\sym{*}  &    -0.00732\sym{*}  &    -0.00709\sym{*}  &    -0.00742\sym{*}  \\
            &   (0.00375)         &   (0.00377)         &   (0.00379)         &   (0.00386)         \\
Current Job: Permanent&    -0.00157         &    -0.00181         &    -0.00207         &    -0.00208         \\
            &   (0.00184)         &   (0.00188)         &   (0.00191)         &   (0.00195)         \\
Current Job: Work From Home&    0.000692         & -0.00000989         &  -0.0000321         &   -0.000502         \\
            &   (0.00281)         &   (0.00283)         &   (0.00288)         &   (0.00290)         \\
Current Job: Shift Work&    0.000674         &    0.000515         &     0.00174         &     0.00146         \\
            &   (0.00180)         &   (0.00182)         &   (0.00189)         &   (0.00193)         \\
Current Job: Flexible Hours&     0.00315         &     0.00334\sym{*}  &     0.00358\sym{*}  &     0.00361\sym{*}  \\
            &   (0.00196)         &   (0.00195)         &   (0.00206)         &   (0.00208)         \\
\midrule
Job Ad FE&                     &$\checkmark$         &                     &$\checkmark$         \\
Survey Batch FE    &                     &                     &$\checkmark$         &$\checkmark$         \\
Observations&        9,446         &        9,446         &        9,446         &        9,446         \\
Respondents &         479         &         479         &         479         &         479         \\
Job Ads        &         994         &         994         &         994         &         994         \\
\bottomrule
\end{tabular}